# The orbital dynamics of the LMC and SMC about the Milky Way

**Gurtina Besla[1] · Nitya Kallivayalil[2]**

**Abstract**
This review summarizes the changing observational and theoretical landscape that has led to rapid developments in our understanding of the orbital history of the Milky Way's most massive satellite galaxies, the LMC and SMC. The determination of high precision center-of-mass proper motions for the Clouds and their environs, coupled with new insights into the structure of dark matter halos from cosmological simulations, indicates that the Clouds are recent additions to the Milky Way's satellite system, where the LMC was sufficiently massive at infall to induce distortions in the Milky Way's dark matter halo, and bring in satellites of its own. To gain insights into the orbital history of the Clouds it is, therefore, necessary to integrate orbits in a time-evolving Milky Way + LMC potential. The proper motion measurements further indicate that the LMC and SMC have had a recent strong encounter with each other. This leaves open the impact of this encounter on their internal structures, and the degree of correspondence between their stellar, HI and dark matter centers. The choice of kinematic center in turn affects the center of mass proper motions; this is the biggest observational unknown, even in the high precision *Gaia* era. Given the expected mass loss of the SMC, the distortions it induces in the LMC halo, and the recent LMC–SMC close encounter, understanding the orbits of objects about the LMC (including the SMC) and interpreting observational data of the Clouds requires full N-body simulations that are constrained by their present-day structure and kinematics.

Gurtina Besla and Nitya Kallivayalil have contributed equally to this work.

✉ Gurtina Besla
gbesla@arizona.edu

✉ Nitya Kallivayalil
njk3r@virginia.edu

[1] Department of Astronomy, Steward Observatory, University of Arizona, 933 North Cherry Avenue, Tucson, AZ 85721, USA

[2] Department of Astronomy, University of Virginia, 530 McCormick Road, Charlottesville, VA 22904, USA

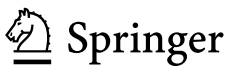

## Contents

# 1 Introduction

Our understanding of the orbital history of the two most massive satellite galaxies of the Milky Way (MW) has evolved considerably. The canonical view, wherein these galaxies have completed multiple orbits about the MW over a Hubble time (e.g., Murai and Fujimoto 1980), has changed to one, where they are recent additions, just completing their first or second passage about our Galaxy (e.g., Besla et al. 2007). This dramatic change has been driven by two factors: (1) high-precision astrometric measurements; and (2) an improved understanding of the mass and structure of galactic dark matter halos. These new observational and theoretical tools are the landscape that motivates this review.

This review is long; we have provided summary statements at the end of each section to highlight key takeaway points.

## 1.1 Historical context: the cranes of the southern night sky

The two nearest star forming galaxies to the Milky Way are readily observable to the naked eye in the southern hemisphere. Consequently, these two galaxies have been known to humans, since there have been humans in those regions of the planet. Just like today, the origin and purpose/meaning behind their presence have been the subject of discussion and conjecture, as reflected in the mythology of most civilizations in the Southern Hemisphere. For example, the Mapuche people of Chile referred to these objects as "water ponds" (*Rüganko* or *Menoko*). The Clouds also feature in the oral histories of most Aboriginal groups in Australia. For example, in the Lower Murray they are known as *Prolggi*, which has been translated as "Cranes".

The earliest written records of the Clouds are possibly traced to the work of Persian astronomer Abd-al-Rahman al-Sufi (around 964 AD) in his treatise *Kitāb suwar al-kawākib* ("The Book of Fixed Stars").

In western astronomy, these galaxies have been known collectively as the Magellanic Clouds, and individually as the Large Magellanic Cloud (LMC), for the larger of the two, and the Small Magellanic Cloud (SMC), for the smaller. The term "Magellanic" for the Large and Small Cloud was not coined by Magellan himself or his chronicler, and Magellan did not directly refer to the LMC and SMC themselves in his writings (Dennefeld 2020).

In this review, we will refer to these two galaxies collectively as "the Clouds" and separately as the LMC and SMC. As a field, Astronomy must adopt naming conventions thoughtfully and be ready to modify historical precedent (de los Reyes M 2023). As with the story of our understanding of the dynamics of the Clouds themselves, conventional wisdom cannot dictate modern astrophysical discourse. If we wish to advance our physical understanding of the universe, we must build an inclusive community that can support and sustain a diversity of scholars who will bring new ideas.

**Summary:** We encourage the field of Astronomy to adopt a new naming convention for the Clouds that recognizes their global significance as objects that have fascinated humans broadly, both across the world and across time.

## 1.2 The perturbed structure and kinematics of the clouds

Owing to their proximity to our Galaxy, the Clouds have been observed in wavebands spanning almost the entire electromagnetic spectrum, allowing us to study the interstellar medium (ISM) and stellar content of two entire galaxies in unprecedented detail (e.g., Trümper et al. 1991; Snowden and Petre 1994; Kim et al. 2003; Fukui et al. 2008; Meixner et al. 2006; Skrutskie et al. 2006; Israel et al. 2010; Gordon et al. 2011; Mao et al. 2012; Meixner et al. 2013; Ackermann et al. 2016; Gaia Collaboration et al. 2021). The LMC is also the first external galaxy whose internal stellar dynamics have been measured using proper motions, with HST (van der Marel and Kallivayalil 2014), Gaia (Vasiliev 2018; Gaia Collaboration et al. 2021), and ground-based facilities (Niederhofer et al. 2022). This has also been possible recently for the SMC (Zivick et al. 2018; Niederhofer et al. 2018). These data open new avenues to understand how galactic dynamics, turbulence, stellar feedback, stellar sub-structure (bars, warps, spiral arms), and star formation all work together to affect galactic structure on small to large scales. There is no other pair of star forming galaxies, where the entirety of the system can be observed with such detail. Future data sets (*Roman, Rubin*) will continue to advance this landscape.

However, the Clouds are subject to environmental factors that can alter their structure, making it difficult to disentangle secular processes that drive ISM physics from external perturbations. At a distance of 50–60 kpc from the MW, the internal dynamics of the Clouds must be affected by the gravitational field of the MW, but the degree to which this matters depends on the duration over which the Clouds have been in the vicinity of the MW (Besla et al. 2007). Furthermore, at this distance, the Clouds are embedded within the circumgalactic medium (CGM) of the MW. The impact of the MW's CGM on the ISM of the Clouds depends on their orbital speed and trajectory (Salem et al. 2015; Mastropietro et al. 2005). Finally, the Clouds are also in close proximity to each other ($\sim$ 20 kpc), meaning their mutual orbital history will dictate the degree to which tidal and hydrodynamic interactions between the two galaxies influence their structure, star formation histories, and dynamical state.

These three environmental factors (the MW's gravity, motion through the MW CGM, and LMC–SMC interactions) modify the current morphological and kinematic structure of the Clouds. For example, the Clouds are surrounded by a massive complex of HI gas in the form of a $\sim$150 degree long stream trailing behind them (the Stream), a gaseous bridge connecting them (the Bridge), and an HI complex that leads them (the Leading Arm) (e.g., Putman et al. 2003; Nidever et al. 2010). Mirabel and Turner (1973) originally searched for this gaseous complex based on an unpublished prediction by Alar Toomre that the outskirts of the Clouds should have been disrupted during a close passage with the MW, creating a gaseous tidal bridge connecting the Clouds and the MW. While this search did not lead to a detection, there were indications in earlier HI maps that such gaseous structures exist (Dieter 1965). Later, a high-velocity, elongated HI feature was discovered (van Kuilenburg 1972; Wannier and Wrixon 1972), and eventually identified as the Stream (Mathewson et al. 1974). This gaseous HI structure is believed to have once resided within the Clouds and was likely removed by a combination of gravitational forces and hydrodynamic processes (see reviews by D'Onghia and Fox 2016; Lucchini 2024).

The stellar content of the Clouds is also perturbed (e.g., de Vaucouleurs and Freeman 1972; Mackey et al. 2018). The LMC is a proto-type of the Magellanic Irregular galaxy classification, wherein galaxies possess one-armed spirals and photometrically off-center stellar bars (de Vaucouleurs and Freeman 1972). Recent work has demonstrated significant perturbations to the LMC's outer stellar disk (e.g., van der Marel 2001; Olsen and Salyk 2002; Mackey et al. 2016; Choi et al. 2018; Jiménez-Arranz et al. 2025; Oden et al. 2026) and stellar bar (e.g., Subramaniam and Subramanian 2009; Haschke et al. 2012; Rathore et al. 2025b; Jiménez-Arranz and Roca-Fàbrega 2025). The SMC's stellar body has a very large line-of-sight depth (∼20 kpc; Nidever et al. 2013; Subramanian and Subramaniam 2012; Ripepi et al. 2017), with complex structure, both internal and in its periphery (Nidever et al. 2011; Niederhofer et al. 2018; Tatton et al. 2021; Dias et al. 2022; Cullinane et al. 2023). The internal kinematics of the SMC display radially outward motions, indicating that it is in a state of disruption (Zivick et al. 2018; De Leo et al. 2020; Niederhofer et al. 2021; Zivick et al. 2021; Rathore et al. 2026; Vijayasree et al. 2026).

In order to realize the potential of existing and future observational data sets of the Clouds to study/test/improve models of ISM physics and star formation in low mass galaxies, theoretical models for the Clouds are needed that accurately capture the external gravitational and hydrodynamic forces that alter their structure over time. These forces in turn depend on the orbital history of these galaxies about the MW and also about each other. This orbital history is the subject of this review.

**Summary:** The Clouds are interacting with each other and the MW, causing the LMC and SMC to be structurally and kinematically perturbed. Understanding the relative role of environmental, secular, or tidal processes or dark matter physics on the observed system requires understanding the orbital history of the Clouds. This is the motivation for this review.

### 1.3 Knowledge of the orbit of the clouds as of 2006

Prior to the first high-precision proper motion measurements of the Clouds in 2006 (Kallivayalil et al. 2006b), the orbits of the Clouds were constrained using radial velocity measurements of the Clouds and along the Stream. When combined with theoretical modeling, such data revealed two key constraints on the orbits of the Clouds.

- *The LMC is at pericenter, not apocenter:* Because the radial velocities of the Clouds are lower than those along the Stream, the Clouds are either near their closest (pericenter) or furthest (apocenter) approach to our MW (Lin et al. 1995). Several authors have explored an apocenter model (Hunter and Toomre 1969; Lin and Lynden-Bell 1977; Fujimoto and Sofue 1976). In this scenario, Alar Toomre theorized that the Stream would be a tidal bridge connecting the Clouds to the MW. However, with an apocenter at $\sim 50$ kpc, the pericenter of the orbit would be $<$20 kpc. At such distances the tidal field of the MW would prevent the Clouds from being so closely located on the sky, disrupting any binary state (Lin and Lynden-Bell 1982; Gardiner et al. 1994). As such, theoretical models favored a high angular momentum, high eccentricity ($> 0.6$) orbit, where the Clouds currently had non zero transverse motion ($>150$ km/s). The first measurements of the current tangential speeds of the Clouds determined that the tangential speeds far exceeded their radial speeds (van der Marel et al. 2002), effectively settling any debate: the Clouds are currently near pericenter (see, e.g., Fig. 20).

- *The Clouds are moving in a polar orbit, leading the Stream on the sky:* Feitzinger et al. (1977) were the first to observationally determine the sense in which the Clouds move across the sky. They measured an offset in the velocity of the line of nodes of several objects in the LMC that they attributed to the fact that the LMC subtends a large angle in the sky, such that "a purely transverse motion of its center of mass would give radial velocity contributions at other points" (Lin and Lynden-Bell 1982). This is a compelling argument that proved that the Clouds lead the Stream on the sky and have a substantial tangential velocity component with respect to the Galactic center. The Clouds are thus moving in a polar orbit, counter-clockwise as seen from the Sun, looking toward the Galactic center (Murai and Fujimoto 1980). This orbital geometry is illustrated later in this review (see Sect. 5: e.g., Figs. 3 and 7). This orbital geometry is supported by all theoretical models regardless of whether the Stream has a tidal (Gardiner et al. 1994; Yoshizawa and Noguchi 2003) or hydrodynamic (Moore and Davis 1994; Mastropietro et al. 2005) origin. The measurement of the proper motion of the Clouds ultimately confirmed this finding.

In the absence of accurate proper motion measurements, several assumptions were made by theorists about the present-day 3D velocity vectors and past orbits of the Clouds in order to construct models of the gaseous Stream and Bridge. The following outlines two facts that were long-assumed to be the opposite by theoretical studies but have now been confirmed by recent high-precision proper motion measurements.

- *The past orbital trajectory of the LMC is not aligned with the Stream on the sky.* Prior to 2006, theoretical models assumed that the entirety of the tangential motion of the LMC was aligned with the Stream, which requires the north component of the proper motion vector to be negligible. In Kallivayalil et al. (2006b), and subsequent proper motion measurements, the north component of the LMC was measured to be non-zero. This means that the past orbit of the LMC does not align with the Stream (see Figs. 7, 18, and 23). This offset poses a challenge for a purely hydrodynamic (ram pressure stripping) LMC origins for the Stream, where gas removed in this manner would closely trace the orbit of the LMC. Conversely, tidal streams gain energy, allowing them to deviate markedly from the past orbit.

- *The SMC is not in a circular orbit about the LMC.* Early theoretical studies generally assume that the SMC is in a circular orbit about the LMC, maintaining a steady separation over time. This idea was first challenged by Gingold (1984), who argued that the SMC would have been disrupted by now if it were on a circular orbit about the LMC. Bekki and Chiba (2005) explored the importance of dynamical friction on the orbit of the SMC, illustrating that the SMC must be in a decaying orbit and, therefore, could not sustain a circular orbit about the LMC (see Sect. 5 and Fig. 5). With high-precision proper motions, the relative speed of the SMC with respect to the LMC is measured to be of order 100 km/s (Kallivayalil et al. 2006b), requiring the SMC to be on a highly eccentric orbit about the LMC, if it is bound to the LMC at all (see Figs. 10, 20, 21, 22). This fact is critical to understanding the orbital history of the system, as a highly eccentric binary is more easily disrupted by MW tides, requiring a recent infall in order to explain their observed proximity today.

The last point begs the question: Have the Clouds always been a binary pair? There are four primary arguments in favor of keeping the Clouds a binary system for at least 5 Gyr.

First, the majority of theoretical studies of the origin of the Stream that invoke tidal stripping require that the SMC has completed at least two orbits about the LMC. Given the high relative velocity of the SMC with respect to the LMC, this stipulation would require the Clouds to be a binary for at least 4–5 Gyr. The binarity of the Clouds is assumed in most studies prior to and after 2006 (when the first high precision SMC proper motion was measured; Kallivayalil et al. 2006b). With the recent discovery of a stellar counterpart to the gaseous Stream at Galactocentric distances of $\sim$100 kpc (Chandra et al. 2023), it is now clear that repeated tidal interactions between the Clouds are a viable, and needed, physical process to explain the Stream (e.g., Murai and Fujimoto 1980; Gardiner and Noguchi 1996; Yoshizawa and Noguchi 2003; Bekki and Chiba 2005; Besla et al. 2010; Diaz and Bekki 2011; Besla et al. 2013; Pardy et al. 2018; Lucchini et al. 2020).

Second, the majority of proper motion measurements, both prior to and after 2006, force a recent ($<$ 200 Myr ago) close encounter between the Clouds (Růžička et al. 2010). The Clouds have interacted at least once in the recent past.

Third, similarities in the star formation histories of both Clouds (e.g., Noël et al. 2009; Harris and Zaritsky 2009; Weisz et al. 2013; Massana et al. 2022; Burhenne et al. 2026) and their stellar populations (Irwin et al. 1990) point to a shared orbital history for at least 5 Gyr. To significantly affect the star formation histories of either galaxy, the separation between the Clouds at pericenters needs to be less than 50 kpc (e.g. see Fig. 16 in Besla et al. 2012). The star formation rates of both Clouds are elevated relative to their average star formation rate (SFR) over the past $\sim$3.5 Gyr (Weisz et al. 2013; Massana et al. 2022; Cohen et al. 2024a; Burhenne et al. 2026). Isolated dwarf pairs with separations less than 50 kpc have SFRs elevated by a factor of $\geq$2.3 relative to that of isolated dwarfs (Stierwalt et al. 2015). The elevated SFR in the Clouds thus requires them to have maintained a close separation for at least $3.5 - 4$ Gyr.

Fourth, as argued in Gardiner et al. (1994), “the sparse distribution of the outer satellites of our Galaxy make the capture of the SMC by the LMC a rather improbable event”. It is more probable that the Clouds were captured together as a galaxy group (D’Onghia and Lake 2008).

Together, the above suggest that the Clouds have maintained a long-lived binary state ($>$ 5 Gyr). However, different assumptions about the masses of the MW and LMC can dramatically alter the orbit of the SMC and longevity of the LMC–SMC binary.

Meaning that the longevity of the LMC–SMC binary may be the critical discriminant in identifying viable orbital solutions for the Clouds about the MW.

**Summary:** Prior to 2006, it was already established that the LMC is just past its closest approach to the MW, where the Clouds are moving in a polar orbit that leads the gaseous Stream. Theoretical models prior to 2006 assumed that: (1) the past orbital trajectory of the Clouds is aligned with Stream; and (2) the SMC is in a circular orbit about the LMC. These two assumptions were proven false by high-precision proper motion measurements in 2006. The Clouds are likely a long-lived binary, where the SMC is on an eccentric orbit about the LMC.

## 2 New tools to understand the dynamics of the LMC and SMC: precision astrometry

A sound measurement of the proper motion (PM) of the Clouds requires all of the following key factors: (1) an instrument that can perform the astrometry with adequate precision, (2) an inertial reference frame that provides wide coverage, (3) secure determination of the membership of stars in the Clouds, and (4) a reliable kinematic model of the internal kinematics of the Clouds, including their kinematic centers. As facilities with ever increasing precision, such as HST and Gaia have come online, it is the final point (4) that now limits our understanding of the past orbits of both galaxies.

This review provides a brief summary of past measurements of the PMs of both galaxies, but eventually focuses on modeling the most recent independent results from HST and Gaia. The following conventions are adopted in this review. The PM in the west ($\mu_W$ ) and north ($\mu_N$ ) directions in terms of the change in right ascension ($\alpha$) and declination ($\delta$) on the sky are defined as

$$\mu_W = -(d\alpha/dt)\cos(\delta),\ \ \mu_N = d\delta/dt. \tag{1}$$

The motions in angular units can be converted to physical units if the distance is known, using the following relation:

$$V_T = 4.74\ \mu\ D, \tag{2}$$

where $V_T$ is the transverse velocity in $\mathrm{kms}^{-1}$, $\mu$ is the PM in mas $\mathrm{yr}^{-1}$ and $D$ is the distance in kpc.

### 2.1 Overview of ground-based efforts to measure the PMs of the clouds

Given the fundamental importance of the Clouds as the brightest and most massive galaxies in the vicinity of the Milky Way (MW), many groups have attempted to measure their motions, first using ground-based approaches starting in the early 1990s. For the LMC, Kroupa et al. (1994) used stars from the Positions and Proper Motions (PPM) Catalogue; Jones et al. (1994) used photographic plates with a 14-yr epoch span; Drake et al. (2001), used data from the Massive Compact Halo Objects (MACHO) project; Anguita et al. (2000) and Pedreros et al. (2002) used CCD frames with an 11-yr epoch span; and Momany and Zaggia (2005) used the USNO CCD Astrograph all-sky Catalog (UCAC2).

Of these pioneering ground-based studies, Anguita et al. (2000) and Momany and Zaggia (2005) presented particularly large PMs with $\mu_N \sim 4$ mas/yr, which would translate to a velocity of $\sim$ 950 km/s at a distance of 50 kpc. The weighted average of the remaining measurements implied $V_T$ for the LMC of the order of 280 km/s (van der Marel et al. 2002), which has been used in several interpretive works about the LMC. Ground-based efforts continued with Costa et al. (2009) who determined the LMC PM relative to background quasars using the Iréneé du Pont 2.5 m telescope; and Vieira et al. (2010), who undertook an ambitious program involving analysis of photographic plates, the Yale/San Juan Southern Proper Motion program, which targeted a million objects and spanned a baseline of 40 years.

Historically, there have been fewer ground-based studies of the PM of the SMC, probably due to the fact that it is further away. Irwin et al. (1996) used AAT (Anglo-Australian Telescope) and CTIO (Cerro Tololo Inter-American Observatory) 4 m photographic plates covering a baseline of $15 - 20$ yr and centered on $\sim$ 1000 background galaxies. A value for the PM from this study is quoted in Irwin (1999), but the analysis of these data is unpublished. The next big effort was that of Vieira et al. (2010) and the Yale/San Juan Southern Proper Motion program.

At this point in the history of ground-based efforts, several large surveys of the Clouds began in earnest, such as the near-infrared survey VMC (VISTA survey of the Magellanic Clouds). The main VMC survey ran between November 2009 and October 2018 but additional observations were taken between August 2021 and January 2023 to increase the time baseline for deriving PMs. This survey remains an active powerhouse for large-scale investigations of the Clouds, and firmly brings the study of the Clouds into the $\sim$ millions of sources regime.

The first paper on the LMC PM tied VMC data to Two Micron All Sky Survey (2MASS) positions over a 10-year baseline and centered on background galaxies (Cioni et al. 2014). The first SMC PM study utilized multi-epoch VMC data itself (Cioni et al. 2016). Of course the general time frame for VMC is also coincident with the main *Gaia* data releases (DR2, eDR3 and DR3), which similarly affords an all-sky view of the Clouds. The more recent VMC and other ground-based efforts have tied to *Gaia*, which if not explicitly used as an epoch of data, is still used in source selection, as it allows to more cleanly distinguish between sources in the Clouds and MW stars. Moving forward, tying large ground-based data sets to *Gaia* in some way is bound to become the dominant approach by which PMs in the Clouds' vicinity, and indeed many parts of the MW, are determined.

It is, therefore, a little artificial to separate, as we have in this review, ground-based and space-based efforts to measure Cloud PMs, but we keep this structure to give an idea of the direction the field is taking.

Coming back to VMC, Niederhofer et al. (2018) utilize more epochs of data, and enhanced source selection with *Gaia*, presenting stellar PMs within the central $3.1 \times 2.4$ kpc of the SMC. This is expanded to some 40 deg$^2$ while also getting a better sense of systematic errors in Niederhofer et al. (2021). The COM results are consistent with previous works but target $\sim$ 700, 000 sources. Turning to mapping the PMs as a function of position within the SMC, the authors find a nonuniform velocity pattern indicative of a tidal feature behind the main body of the SMC and a flow of stars in the south–east moving predominantly along the line-of-sight. This is attributed mainly to the presence of a “counter Bridge”, which in the lexicon of tidal theory, is a tidal tail emanating from the SMC that formed from a tidal interaction with the LMC (Toomre and Toomre 1972). According to N-body simulations by Diaz and Bekki (2012), it is expected to originate behind the SMC and accompany the formation of the Bridge.

While not focused on PMs perse, surveys dedicated to the low surface brightness stellar content of the Clouds and to their outskirts, have separately uncovered these substructures as well as revealed others, interpreted as outcomes of the interaction history of the Clouds with each other. These include the Survey of the MAgellanic Stellar History (SMASH), a deep survey of the Clouds using the Dark Energy Camera (Nidever et al. 2017, 2021), the MAgellanic Periphery Survey (MAPS), utilizing a filter set that is especially sensitive to metal poor giants (Nidever et al. 2011), and the VIsible Soar photometry of star Clusters in tApii and Coxi HuguA (VISCACHA) survey targeting star clusters in the outer regions of the SMC (Dias et al. 2014, 2021). These surveys reveal a very complex structure for the SMC. MAPS uncovers a large ($\sim$ 20 kpc) line-of-sight depth in the SMC (Nidever et al. 2013), while VISCACHA reveals stellar counterparts of the Bridge and counter Bridge, including SMC Bridge clusters that are moving toward the LMC (Dias et al. 2022).

Wan et al. (2020) identify $\sim$ 3500 candidate LMC carbon stars from their extremely red $(g - r)$ colors in SkyMapper, a 20, 000 sq. degree survey of the southern sky. Coupled with Gaia DR2 astrometry, they present a PM of the LMC, as well as a derivation of a dynamical center for different stellar populations. They find that the center for young stars is significantly offset from the older populations, in keeping with findings from smaller FOV studies, such as HST (van der Marel and Kallivayalil 2014).

Niederhofer et al. (2022) present VMC PMs for the inner 28 deg$^2$ of the LMC, obtaining center-of-mass PMs, a dynamical center, and rotation amplitude differences for young vs. intermediate/old stellar populations again consistent with HST studies. Schmidt et al. (2022) present VMC PMs in the outer regions of the LMC utilizing a machine learning approach to source selection with Gaia Early Data Release 3 (EDR3) data. They find that the south-eastern region of the LMC shows a slow rotational speed compared to the overall rotation. N-body simulations suggest that this could be caused by a fraction of stripped SMC stars located in that particular region that move opposite to the expected rotation.

Vijayasree et al. (2025) present LMC PMs from the VMC extension through 2023. The extended time baseline improves the precision of VMC PMs from 6 mas/yr to 1.5 mas/yr. Also to give a sense of scale, this latest work estimates PMs for approximately 5 million unique sources ($\sim$ 3 million in Niederhofer et al. 2022 and 2.6 million in Schmidt et al. 2022). The tangential rotation curve reveals an asymmetric drift between young and old stars. The internal rotation map confirms the clockwise rotation around the dynamical center of the LMC, consistent with previous predictions (e.g., van der Marel and Kallivayalil 2014). A significant residual motion is detected toward the north–east of the LMC, directed away from the center, suggesting a possible tidal influence from the MW and the SMC.

An enduring theme that is being solidified by these works, and which we will pick up again in the space-based results, is that the measured PMs of the Clouds are dependent on the location, where they are measured, and also on the targeted stellar population, due to the interactions between the Clouds and with the MW.

Of course it is difficult for ground-based studies to compete with the exceptional precision of *Gaia* despite the longer time baseline of 10 years compared to 3 years for eDR3. However, the near-IR VMC offers substantial completeness across a range of stellar populations and complements *Gaia* in regions of high extinction. The large number of sources can be used to investigate not just center-of-mass motions but the large-scale interactions of the Clouds with each other, providing valuable context.

**Summary:** We have firmly entered the $\sim$ million source regime for studies of the Clouds thanks to large-scale ground-based surveys, such as VMC, SMASH, MAPS, VISCACHA, and SkyMapper. While the per-source accuracy can be limited, these surveys provide important context on the overall structures resulting from the ongoing interactions between the Clouds and with the MW thanks to their broad coverage and completeness; and complement *Gaia* in regions of high extinction or by augmenting with chemistry and radial velocities. Notable discoveries include the large line of sight depth of the SMC, the counter-bridge features, and the confirmation, writ large, that determined centers vary as a function of stellar population.

### 2.2 Overview of space-based efforts

Given that astrometry is greatly aided by instrument stability, it was inevitable to look to space-based methods, where the destabilizing effects of the Earth's atmosphere are curtailed. A first attempt using space-based observatories was made by Kroupa and Bastian (1997) using Hipparcos data, who obtained errors in the $\sim$ 0.3 mas/yr regime. With the advent of HST, however, the field of Local Group astrometry came into its own. Despite the fact that HST was not commissioned with astrometry as the primary aim, it has made pioneering contributions to the field of Local Group astrometry due to its remarkable stability. The workhorse instruments that have typically been used are the Advanced Camera for Surveys (ACS) and the Wide Field Planetary Camera 3 (WFPC3), which both have high resolution, are well-calibrated, and have been shown to be very stable for astrometry (Anderson and King 2003a, 2004, 2006).

Anderson and King (2003b, c) first measured a very accurate relative proper motion between the SMC and the globular cluster 47 Tuc using *HST*'s Wide Field Planetary Camera 2 (WFPC2). However, at the time there was no suitable inertial reference frame for SMC stars. To get to an absolute motion, they had to combine this relative motion with an estimate of the absolute PM of 47 Tuc by Freire et al. (2003) using milli-second pulsars, which, while pioneering in its own right, had relatively large errors.

Subsequently Geha et al. (2003) found a set of quasars behind both Clouds from their optical variability in the Massive Compact Halo Objects (MACHO) database, suitable to be used as an inertial reference frame. Using the High Resolution Camera (HRC) on ACS and a two epoch snapshot program with a baseline of only approximately 2 years centered on 21 background quasars, Kallivayalil et al. (2006a) were able to measure a PM for the LMC to better than 5% accuracy for the first time. This work capitalized on several technological advances. The HRC provided the highest resolution available (the average pixel scale was 28.27 mas/pixel) and, in the pioneering calibration of Anderson and King (2006), it was shown to be well-sampled even in the bluest filters, meaning that a star's integrated flux does not depend strongly on where it lands in a pixel. This enabled *sub pixel* accuracy, necessary to obtain reasonable velocity errors at halo distances. The F606W filter, a broad $V$-type filter with high throughput was chosen for the main astrometric goals. The F814W, a wide $I$-type filter, allowed the construction of color-magnitude diagrams (CMDs) for membership selection and troubleshooting purposes. The fact that the reference frame (quasars) had similar profiles to the target LMC stars, enabled an "apples to apples" comparison. The empirically constructed stable HRC PSF, and the well-calibrated geometric distortion on the camera, all carried out using a well-populated globular cluster field by Anderson and King (2004), contributed to the high accuracy of the program despite the relatively short time baseline.

Several further steps were necessary. The conversion of measured PMs as a function of a given field to a center-of-mass motion is not trivial. There are two main categories of contributions to be taken into account: a geometric component and those coming from internal motions. The geometric component comes from the fact that both Clouds span a significant area in the sky, leading to differing contributions from the components of the three-dimensional velocity vector as a function of position ("viewing perspective"), first discussed in van der Marel (2001). Internal motions may be made up of rotation, contributions from the precession and nutation of the disk, and from streaming due to the bar or tidal interactions (van der Marel et al. 2002). We know that essentially all of these are in play for both galaxies, and the degree of the effect depends on what is chosen as the "center" position. Thus, the dynamical center of the galaxy is a key input. Since the adopted center in turn affects the magnitude of the geometric or internal motion-based correction at a given position in the galaxy, the center, the corrections, and the center-of-mass PM are ideally all best solved for together in iterative fashion. Indeed, the wealth of information on the internal motions of the Clouds coming from ongoing large-scale investigations (e.g., VMC and *Gaia*) continues to drive home this point quite clearly.

Recognizing this, the Kallivayalil et al. (2006a) study included the effect of viewing perspective, which is calculable given a line-of-sight velocity, center-of-mass PM, and the galaxy distance. The systemic line-of-sight velocity was well-known from an analysis of 1041 carbon stars (van der Marel et al. 2002), and the distance modulus from Freedman et al. (2001). The removal of this term allowed for the search for residual motions. Kallivayalil et al. (2006a) attempted to constrain the dynamical center and internal motions, specifically rotation, but did not have the required precision. They recovered the expected magnitude and sign of rotation from a model fit to the carbon stars in van der Marel et al. (2002), but the per-field PM errors were still significant. Similarly, since they were unable to independently fit the center, they used the center fit to the line-of-sight velocities of carbon stars from van der Marel et al. (2002).

Kallivayalil et al. (2006b) presented a PM for the SMC from the same observational program, centered on 5 background quasars from Geha et al. (2003). Although this was the best measurement of the SMC PM at the time, the errors were still larger than those for the LMC by a factor of 3 ($\sim 0.06$ vs. $\sim 0.02$ mas/yr for the LMC), mostly driven by the lack of knowledge of the internal kinematics and structure of the SMC, and by the limited number of background quasars with which to probe these internal kinematics (and derive a center-of-mass motion). Nevertheless, the errors for both Clouds were now small enough to open up a new set of questions, specifically related to the velocities of the Clouds about the MW and relative to each other.

To convert a measured PM and line-of-sight velocity into a 3D Galactocentric velocity, $v_{tot}$, it is necessary to correct for the reflex motion of the Sun. For the parameters adopted in the Kallivayalil et al. (2006a, b) studies, which were the standard IAU values for the Solar radius, $R_0$, and the circular speed at the solar radius, $V_0$ (8.5 kpc and 220 km/s Kerr and Lynden-Bell 1986), and the solar peculiar velocity from Dehnen and Binney (1998), a high LMC $v_{tot} \sim 380$ km/s was obtained for the LMC. A value this high had only been considered in one theoretical work to model the production of the Stream (Heller and Rohlfs 1994), with the others ranging from 250 km/s–350 km/s. This was also higher than a weighted average of the observational estimates at the time, compiled in van der Marel et al. (2002), which was 280 km/s. This led to the question of whether the Clouds are indeed bound to the MW, and the more specific and perhaps well-posed aspect of this question which is whether they were on first infall (Besla et al. 2007). Furthermore, the relative 3D velocity between the Clouds, now obtainable, implied that they could be unbound from each other, depending on the masses adopted (Kallivayalil et al. 2006b).

Given the surprising nature of the implications for the Clouds–MW system implied by this set of PM measurements, Piatek et al. (2008) conducted a reanalysis of this data set. Their study also corrected for the effects caused by the degrading charge transfer efficiency of the HRC, which led to smaller random errors but otherwise consistent results with the Kallivayalil et al. studies.

Subsequently, Kallivayalil et al. (2013) were successful in obtaining a third epoch of HST data. This third epoch was executed on the WFPC3/UVIS channel, which had the advantage of being a completely independent instrument, and the increased time baseline of $\sim 7$ years provided better control of systematic as well as random errors. The three-epoch data yielded PM random errors of just 1–2% per field for both Clouds. For the LMC this was sufficient to constrain the internal dynamics, including the determination of a PM rotation curve and a dynamical center (van der Marel and Kallivayalil 2014) for the first time. This led to several now empirically discernible insights. One was that the choice of dynamical center affects the center-of-mass motion. The other was that the choice of solar motion affects the Galactocentric velocities, with $V_0$ being a direct additive to the Galactocentric Y-direction of the LMC velocity. The newly constrained dynamical center brought down the center-of-mass PM of the LMC in both directions (by $\sim 0.2$ mas/yr), and in concert with a revised understanding of the solar motion ($V_0$ went from 220 km/s to 239 km/s; McMillan 2011) brought down the Galactocentric velocity from 378 km/s to 321 km/s (see also Shattow and Loeb 2009 and Růžička et al. 2010).

The center-of-mass PM uncertainties were now dominated by the limitations in our understanding of the internal kinematics and geometry of the Clouds. The velocity uncertainties were now dominated by distance errors, if the choice of solar motion were held constant. Nonetheless, the increased precision led to several new insights about the Clouds, including that first-infall orbits are preferred if one imposes the requirement that the LMC and SMC must have been a bound pair for at least several Gyr (Kallivayalil et al. 2013).

The increased precision in turn allowed for a renewed exploration of the internal motions and dynamical center of the LMC. After subtraction of the center-of-mass PM, the PM residuals showed a clear signature of clockwise rotation in the plane of the sky, around a dynamical center that could now be fit. van der Marel and Kallivayalil (2014) combined the PMs of stars in the 22 fields from Kallivayalil et al. (2013) with existing radial velocity (RV) measurements for 6790 individual stars, making the first estimate of rotation of any galaxy based on full three-dimensional velocity measurements. Interestingly, van der Marel and Kallivayalil (2014) found that the center of the stellar PM rotation field is consistent with the position of the HI dynamical center. At the time this made some sense, since presumably the stars and gas are orbiting in the same gravitational potential. However, this result differs by 1.12 degrees from the photometric center from near-IR star counts in van der Marel (2001), which is consistent with the brightest part of the LMC bar. Whether the LMC center should agree with the gas or the stars is still one of the most important open questions in this field, especially given the now mounting evidence that the LMC and SMC have had a head-on collision (reminiscent of the Bullet Cluster). van der Marel and Kallivayalil (2014) further found population-dependent kinematics in the LMC, with the young (red supergiant) stars rotating faster than the old (red and asymptotic giant branch) stars due to asymmetric drift.

With the start of *Gaia* operations, van der Marel and Sahlmann (2016) analyzed PMs for eight individual stars in the SMC from the TGAS Catalog (Lindegren et al. 2016). They attempted to investigate the internal structure of the SMC, and the resulting residual motions were not indicative of any coherent motion. While this study foreshadowed the future potential impact of *Gaia*, TGAS itself offered a relatively small sample size.

Motivated by the still sparse PM coverage of the SMC, and our lack of understanding of the internal structure of this morphologically complex dwarf, Zivick et al. (2018) presented a PM based on a larger set of 30 fields containing background quasars and spanning a $\sim$3 year baseline, using *HST's* WFC3/UVIS channel. These quasars, obtained from Kozłowski et al. (2013), focused on the outer regions. Combining this data with the previous five HST fields from Kallivayalil et al. (2013), and an additional eight measurements from van der Marel and Sahlmann (2016), brought the total to 43 SMC fields. The systemic PM was obtained at the 1% level. After subtraction of the systemic motion, an ordered motion in the outer parts of the SMC moving radially away from the galaxy was found, indicating that the SMC is in the process of tidal disruption. The relative velocity between the Clouds was found to be $103 \pm 26$ km/s, indicating that they have experienced a close encounter in the recent past ($147 \pm 33$ Myr ago, with a mean impact parameter of $7.5 \pm 2.5$ kpc; see 9.2 for more details). This picture of a close interaction between the Clouds has been backed up by many subsequent observational studies, both ground-based and space-based, utilizing different tracers (e.g., Vijayasree et al. 2025; Navarrete et al. 2023; Cullinane et al. 2022a) and is also consistent with star formation enhancements (Massana et al. 2022; Sakowska et al. 2024).

Of course, the internal structure and kinematics of the SMC affect the center-of-mass motion and vice versa. There is a clear velocity gradient present in the H I gas that has been interpreted as rotation. The dynamical center of this signal is located in the northeast of the SMC (Stanimirović et al. 2004). This center is offset from the center inferred from the structure of the Cepheid population (Ripepi et al. 2017) and the near-IR populations (Cioni et al. 2000). The red giants studied in Harris and Zaritsky (2006) show a more spheroidal structure with weak if any rotation and high dispersion, indicating that the stars and gas are decoupled from each other. Subsequent line-of-sight studies of red giants and young stars (Dobbie et al. 2014; Evans and Howarth 2008; Lamb et al. 2016) find some evidence for weak rotation, but along a different axis than the gas. While it is plausible that the old stellar populations could be dynamically decoupled from the gas, it is hard to understand how the young stellar populations could be. Indeed Murray et al. (2019) show that a rotating disk model for the gas cannot reproduce the observed motions of young O and B stars, typically used as gas tracers.

In the HST Zivick et al. (2018) study, which is agnostic as to stellar population, the authors attempt to constrain the rotation velocity and the dynamical center of the SMC stars by leaving them as free parameters, but the fit does not converge, preferring a center closer to the H I center but with very large error bars. They, therefore, consider the effects of adopting either the photometric center or the HI center in the determination of the center-of-mass motion and the subsequent orbital modeling.

With the advent of Data Release 2 (DR2) and onward, Gaia has been used to target millions of sources in the Clouds, enabling precise center-of-mass motions as well as internal motions, starting with Helmi et al. (2018). They derive a PM rotation curve for the LMC, now using many orders of magnitude more sources than the HST studies, but with consistent amplitude. Helmi et al. (2018) do not attempt to fit a dynamical center for either Cloud, but rather present results for two centers, the HI centers (Kim et al. 1998; Luks and Rohlfs 1992 for the LMC and Stanimirović et al. 2004 for the SMC) and the photometric centers from near IR star counts (van der Marel 2001 for the LMC and Gonidakis et al. 2009 for the SMC[1]). They do not present a PM for the LMC dynamical center found in van der Marel and Kallivayalil (2014), but this is close to the HI center. For a consistent choice in center, e.g., say HI, the $\mu_N$ is consistent between the DR2 and HST studies, and the $\mu_W$ component is $2\sigma$ lower in DR2. For a consistent center choice, the SMC PMs in both studies are consistent with each other. As mentioned, consistency between studies aside, the choice of center affects all PM values. For example, a choice of photometric center for the LMC (instead of HI) acts to slightly increase the DR2 $\mu_W$ (which would result in higher tangential velocities) and the $\mu_N$ component more significantly by $3\sigma$. This would act to displace the LMC's past orbit in the plane of the sky with respect to the location of the Stream (which needs $\mu_N \sim 0$, and is insensitive to the value of $\mu_W$; Besla et al. 2007). This is discussed further in the review—Figs. 7 and 18. These statements pertaining to the adopted center would hold for all PM determinations (not just the *Gaia* ones).

[1] Note that (Helmi et al. 2018) wrongly attribute their adopted SMC photometric center to Cioni et al. (2000) and also misquote the value from Gonidakis et al. (2009) slightly; see Table 1.

Several subsequent DR2-based studies aim to better characterize the internal dynamics of the Clouds and some attempt to refine the centers. Vasiliev (2018) explore the PMs of $\sim$ 500, 000 red giant stars in the LMC, in concert with a dynamical model, finding that the circular velocity is $\sim$ 90 km/s at 5 kpc, and that the velocity dispersion ranges from 30–40 km/s, consistent with previous studies (e.g., van der Marel et al. 2002; van der Marel and Kallivayalil 2014). They find non-negligible systematics in the data that prevent them from determining a dynamical center. Zivick et al. (2021) explore the red giant population of the SMC and cross-match with publicly available radial velocity catalogs and RR Lyrae-based distance distributions. They use a forward modeling approach, applying kinematic models with varying rotation properties and a prescription for tidal expansion, to generate mock PM catalogs, which they compare with the *Gaia* data. They find evidence for moderate rotation (with a magnitude of $\sim 10 - 20$ km/s at 1 kpc from the SMC center), and tidal expansion ($\sim$ 10 km/s/kpc). This is in contrast to De Leo et al. (2020) who also use Gaia PM and RV information for SMC red giants, but do not find rotation, arguing that it is buried in tidal expansion. Zivick et al. (2021) fit a 2D Gaussian to the red-giant distribution, which they adopt as the center of their models. Both these studies underscore the clear need now for more RV measurements in the SMC, valuable to constrain the rotation and also to identify substructure that can aid in interpreting the structure of the Cloud.

Bolstering the number of available RV's has been the thrust of new efforts with *Gaia* DR3 data (note that astrometry does not change between eDR3 and DR3, the changes are in RVs). *Gaia* eDR3 PMs are generally consistent with DR2 if the same center is adopted (Gaia Collaboration et al. 2021). Gaia Collaboration et al. (2021) do not infer geometric properties from the eDR3 astrometry, assuming that the local parallax zero point of the LMC in this data set distorts most of the 3D structure. Jiménez-Arranz et al. (2023) (DR3) address to what extent the RV's benefit the structural inferences of the LMC. They present a formalism to transform observed space to an LMC frame, which allows inclusion of the line-of-sight component when deriving internal LMC velocities. They do not fit a new center.

Choi et al. (2022) combine eDR3 PMs for red clump stars with $\sim$ 1000 RV's from the literature to investigate the residual velocity field of the LMC. They find strong evidence in these residuals for a close (impact parameter $<$ 10 kpc) and recent ($<$ 250 Myr) interaction with the SMC, in keeping with previous works (e.g., Zivick et al. 2018). They present a kinematic center for the LMC from their analysis of red clump stars. For the SMC, Almeida et al. (2024) conduct a joint analysis of DR3 and APOGEE data, showing mounting evidence for several kinematic clumps and a distance bimodality, seen in previous studies (e.g., Nidever et al. 2013), that make it challenging to even define what a center means. The latest H I investigations similarly show dynamic complexity, indicating that the SMC's ISM is made up of two distinct star-forming systems, with similar gas mass, separated by $\sim$ 5 kpc along the line of sight (Murray et al. 2024a).

As pointed out in lessons from ground-based data, while not focused on PMs per se, there are also powerful space-based surveys, such as Scylla, a multicycle pure-parallel HST campaign that is measuring resolved star formation histories and age gradients in the LMC and SMC, a complementary probe of LMC–SMC interaction (Murray et al. 2024b; Cohen et al. 2024a, b), which also points to repeated past interactions between the Clouds. The *Gaia* XP data are also being used increasingly in this regard to add context from the metallicity structure of the Clouds. Massana et al. (2024) find that, in contrast to the idea that the periphery of the Clouds must be dominated by stripped SMC debris, it may be dominated by LMC debris instead.

**Summary:** The high precision PMs afforded by space-based studies have contributed to a paradigm shift in the interpretation of the orbits of the Clouds about the MW and each other, including that they may be on first infall, and have had a direct collision with each other in the past. In detail, the observations have led to several nuanced insights, including that the choice of LMC and SMC center affects the center-of-mass PMs, and that the choice of solar motion affects the Galactocentric velocities. Notable discoveries include the LMC's clockwise rotation in the plane of the sky, that both the determined LMC center and rotation amplitude are population dependent, and the tidal disruption of the SMC (radial expansion in PM space) with little residual rotation in the stars.

### 2.3 Lessons from current measurements and outlook to the future

We have entered an era of large scale surveys of the Clouds with much to look forward to with LSST and Roman. With high quality 3D velocities, metallicities, and abundances increasingly available for large samples of stars, as well as detailed ISM properties, we are now in many aspects not limited by precision, but by challenges in interpretation. It is worth taking stock of what limits progress in our understanding of the orbits of the Clouds with respect to the MW and with each other. It seems increasingly clear that the Clouds have had a strong encounter, if not head on collision, with each other in the recent past. This leaves open the impact on their internal structures, and specifically the relationship between the stellar, HI and dark matter centers. As discussed above and shown elegantly in the review by Vasiliev (2024), the choice of center affects the center-of-mass PMs, and this is the biggest observational unknown, even in the high precision *Gaia* era.

Figures 1 and 2 show the range of centers that have been determined for the Clouds. There are also recent interesting modeling works aimed at determining centers independently of these observational works, which similarly indicate the dynamical disequilibrium of the Clouds. Lucchini and Han (2025) use a large ensemble of orbit realizations to determine the present-day ejection sites of a set of hypervelocity stars originating from the LMC (Han et al. 2025; Brown et al. 2014), presumably tracing the location of the supermassive blackhole (and located approximately 1.5 deg north of the Wan et al. 2020 center).

Furthermore, depending on the timing and impact parameter of the recent LMC–SMC encounter, there can be a displacement of ∼1.5–2.5 kpc[2] between the density center of the LMC's outer stellar disk and its dynamical center (the location, where

[2] At LMC distance 1 kpc ∼ 1 degree.

**Table 1** Positions and proper motions used in this review

| LMC | RA | DEC | $\mu_W$ | $\mu_N$ | $X$ | $Y$ | $Z$ |
|---|---|---|---|---|---|---|---|
| HI center orbits (K13) | 78.76 | −69.19 | $-1.910 \pm 0.020$ | $0.229 \pm 0.047$ | −1 | −41 | −28 |
| Phot center orbits (Gaia DR3) | 81.28 | −69.78 | $-1.858 \pm 0.020$ | $0.385 \pm 0.020$ | −1 | −41 | −27 |
| **SMC** | **RA** | **DEC** | $\mu_W$ | $\mu_N$ | $X$ | $Y$ | $Z$ |
| HI center orbits (Z18) | 16.25 | −72.42 | $-0.820 \pm 0.02$ | $-1.210 \pm 0.01$ | 15 | −38 | −44 |
| Phot center orbits (Gaia DR3) | 12.80 | −73.15 | $-0.686 \pm 0.02$ | $-1.237 \pm 0.02$ | 16 | −38 | −44 |

The positions and PMs used in this review. RA/DEC is in degrees, PMs are in mas/yr, and the Galactocentric positions are in kpc. In this and the following tables, we use the following conventions. The distance modulus for the LMC in the HI orbits is taken as 18.5 (Freedman et al. 2001), and for Phot orbits is 18.48 (Pietrzyński et al. 2019). That for the SMC in both the HI and Phot orbits is taken as 18.99 (Cioni et al. 2000). These choices are made to be consistent with literature choices for each case. For the solar parameters in all tables we use $R_0 = 8.29$ kpc, and $V_0 = 239$ km/s (McMillan 2011) and the peculiar velocity is from Schönrich et al. (2010). Note that the RA/DEC for Phot center orbits for the SMC is mis-attributed to Cioni et al. (2000) in the Gaia and subsequent papers, but is instead from Gonidakis et al. (2009). The value in this table is taken from the Gaia papers to be consistent with the literature, but that value is also slightly misquoted in the literature. For completeness the RA/DEC from Gonidakis et al. (2009) is (12.75, −73.12) degrees the galaxy potential well is the deepest; Pardy et al. 2016). Even after a collision, the bar center remains coincident with the dynamical center due to the bar's mass; this may favor the photometric center rather than the HI center in the LMC for kinematic studies (Pardy et al. 2016). However, there is a displacement between the LMC dark matter center and the center of the stellar bar, by ~1 kpc (e.g., Fig. 6 in Rathore et al. 2025a, see also Athanassoula et al. 1997; Berentzen et al. 2003; Pardy et al. 2016). The disequilibrium of the LMC is clearly a challenge for understanding the appropriate choice of center.

Since PM determinations for the Clouds are largely in agreement if a consistent choice of center is adopted, in this review we pick two centers that span a representative range and that have also been used in published studies: the HI centers and the photometric centers, labeled "HI" and "Phot" in the Figures. For the LMC, in detail, we do not pick the center labeled "HI" in the figure but rather the center labeled "Proper Motions", which is consistent with the HI center, but is the dynamical center obtained from the PM field in van der Marel and Kallivayalil (2014). The corresponding Galactocentric velocities are shown in Table 2. To span the parameter space allowed by independent PM estimates, we use the combination of HST PMs (specifically Kallivayalil et al. 2013 for LMC and Zivick et al. 2018 for SMC) plus PM/HI centers, and *Gaia* DR3 PMs (specifically Gaia Collaboration et al. 2021) plus photometric centers. In this review, the corresponding orbits (velocities) are referred to as "HI Center orbits (velocities)" and "Phot Center orbits (velocities)".

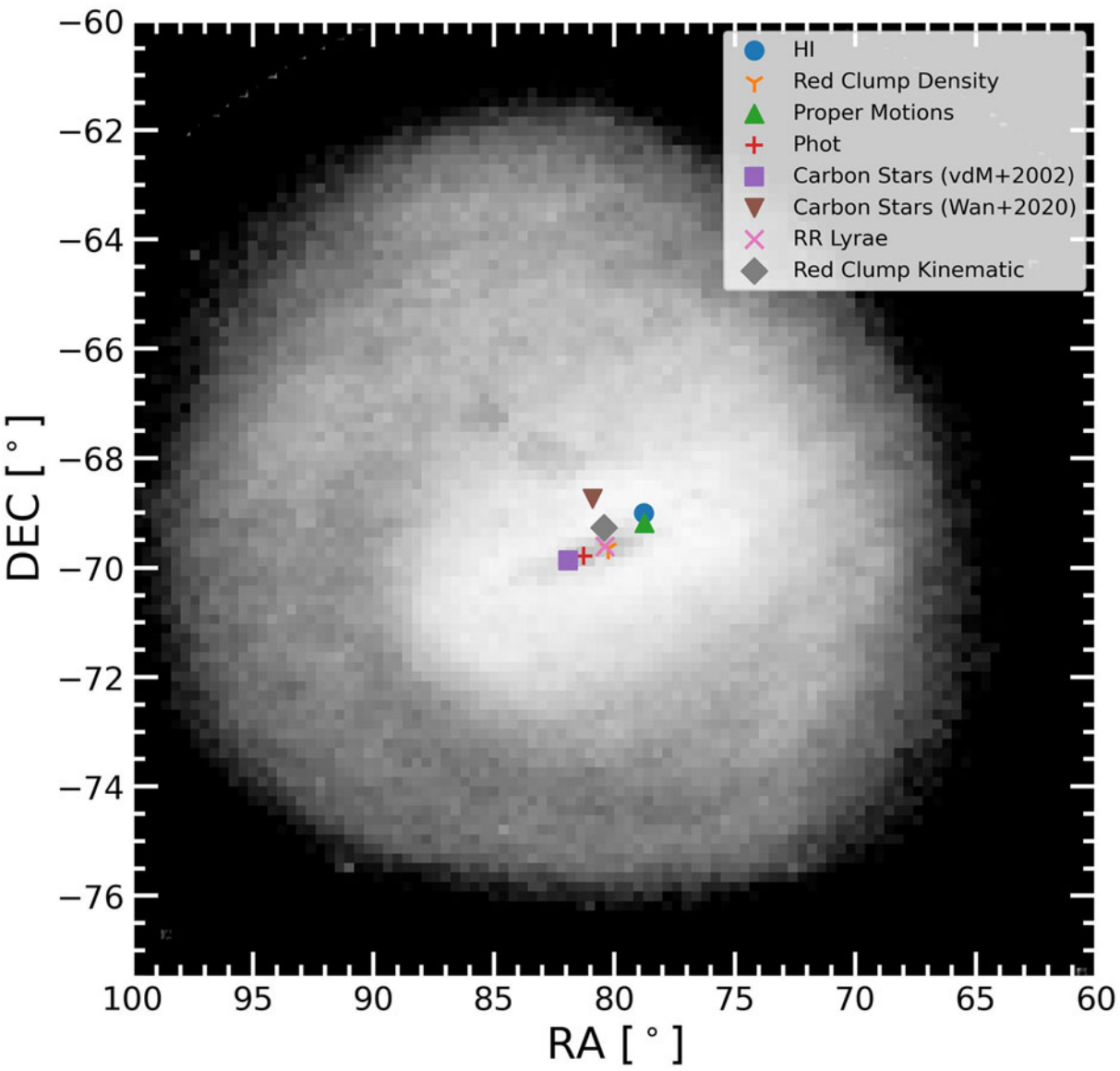

**Fig. 1** Compilation of the centers derived for the LMC in various works against a backdrop of *Gaia* DR3 stellar density. From top to bottom in the inset: HI is an average of the centers from Kim et al. (1998) and Luks and Rohlfs (1992); Red Clump Density is from Rathore et al. (2025b); Proper Motions is from van der Marel and Kallivayalil (2014); Phot is from van der Marel (2001); Carbon Stars is from van der Marel et al. (2002) and Wan et al. (2020), respectively; RR Lyrae is from Jacyszyn-Dobrzeniecka et al. (2017); Red Clump Kinematic is from Choi et al. (2022). The two representative centers used in this review are labeled Proper Motions and Phot. *Image credit: H. Rathore*

Looking forward, to make progress it is necessary for observational data to be analyzed in concert with simulations of the Clouds, which is the intent of simulation suites, such as KRATOS (Jiménez-Arranz et al. 2024). The holy grail would be to use a forward modeling approach, in which simulation-based realizations of the Clouds' orbital evolution are projected into survey data space with realistic errors.

**Summary:** Even in the *Gaia* era of millions of sources with high per-source precision, the center-of-mass PM uncertainties are dominated by limitations in our understanding of the internal kinematics and geometry of the Clouds. Specifically, what is the relationship between the stellar, HI and dark matter centers? The conversion from a per-field PM to a center-of-mass PM requires knowledge of the internal motions for each Cloud, and these motions depend on the adopted center and vice versa. The center is the biggest unknown that affects PM-based orbital histories. Future practitioners should pay attention to the choice of center, as well as the choice of solar motion, as potentially hidden differences in these input parameters have lead to apparently inconsistent orbital histories in the literature. Given the amount of disequilibrium in the Clouds, to make progress it would be beneficial to analyze the data hand-in-hand with simulations.

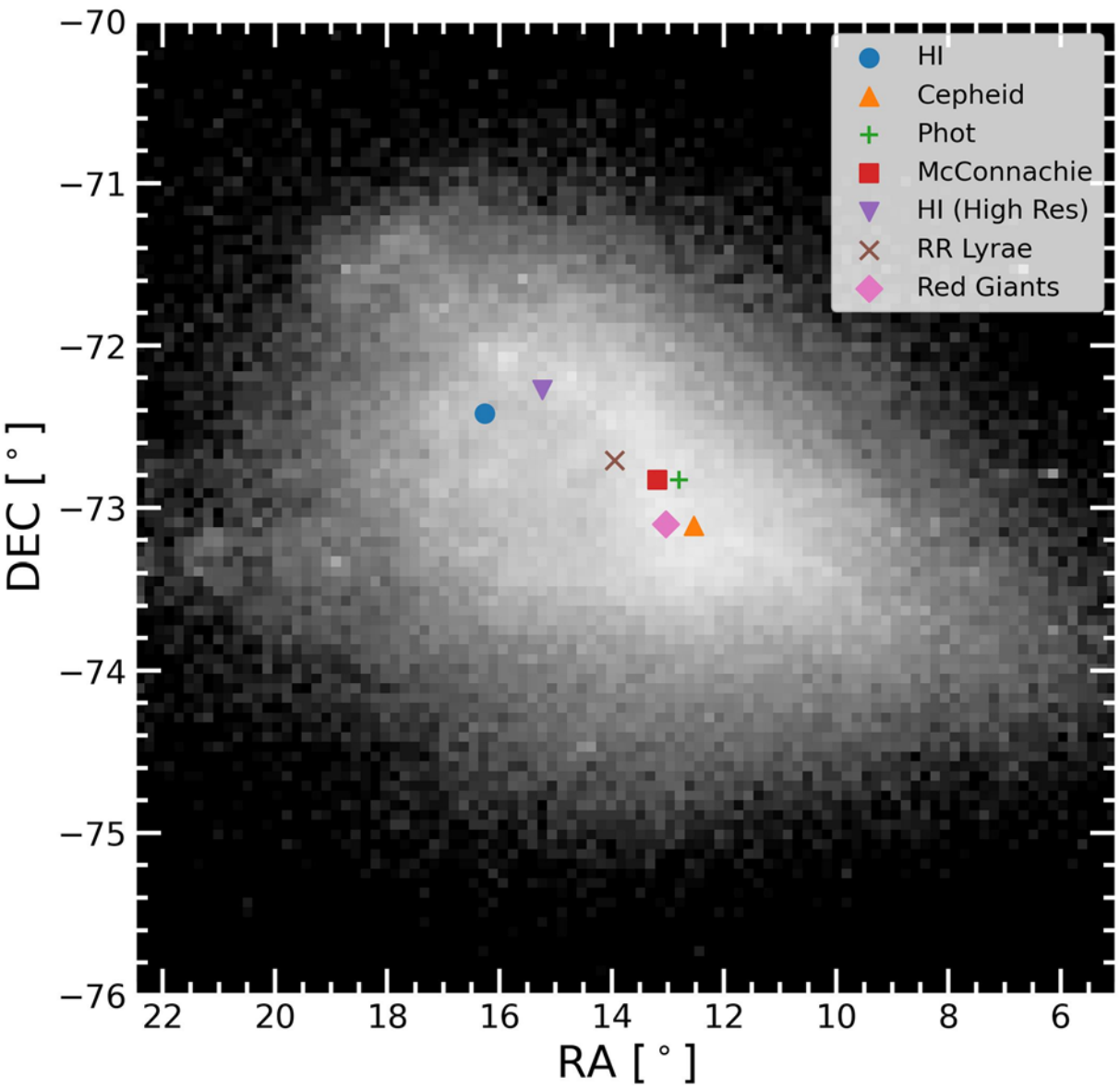

**Fig. 2** Compilation of the centers derived for the SMC in various works against a backdrop of *Gaia* DR3 stellar density. From top to bottom in the inset: HI is from Stanimirović et al. (2004); Cepheid is from Ripepi et al. (2017); Phot is from Gonidakis et al. (2009); McConnachie is listed without attribution in Mcconnachie (2012); HI (High Res) is from Di Teodoro et al. (2019); RR Lyrae is from Jacyszyn-Dobrzeniecka et al. (2017); Red Giants is from Zivick et al. (2021). The two centers used in this review are labeled HI and Phot. *Image credit: H. Rathore*

**Table 2** Galactocentric velocities used in this review

| LMC | $v_{tot}$ | $v_{rad}$ | $v_{tan}$ | $v_X$ | $v_Y$ | $v_Z$ |
|---|---|---|---|---|---|---|
| HI center orbits (K13) | $321 \pm 24$ | $64 \pm 7$ | $314 \pm 24$ | $-57 \pm 13$ | $-226 \pm 15$ | $221 \pm 19$ |
| Phot center orbits (Gaia DR3) | $312 \pm 21$ | $65 \pm 3$ | $305 \pm 22$ | $-72 \pm 6$ | $-216 \pm 12$ | $213 \pm 17$ |
| **SMC** | $v_{tot}$ | $v_{rad}$ | $v_{tan}$ | $v_X$ | $v_Y$ | $v_Z$ |
| HI center orbits (Z18) | $250 \pm 20$ | $-10 \pm 1$ | $250 \pm 20$ | $18 \pm 6$ | $-179 \pm 16$ | $174 \pm 13$ |
| Phot center orbits (Gaia DR3) | $238 \pm 20$ | $-12 \pm 1$ | $238 \pm 20$ | $35 \pm 3$ | $-162 \pm 15$ | $171 \pm 12$ |

The Galactocentric velocities (km/s) used in this review using the measurements and conventions listed in Table 1. As adopted in Kallivayalil et al. (2006b) and subsequent work in the literature, these velocities are in a Cartesian coordinate system $(X, Y, Z)$ with the origin at the Galactic center, the $Z$-axis pointing toward the Galactic north pole, the $X$-axis pointing in the direction from the Sun to the Galactic center, and the $Y$-axis pointing in the direction of the Sun's Galactic rotation. For the LMC, $V_{LOS} = 262 \pm 3$ km/s (van der Marel et al. 2002), and for the SMC, it is $146 \pm 1$ km/s (Harris and Zaritsky 2006). The errors in the table are obtained from a monte carlo of the measurement errors in PM, $V_{LOS}$ and 0.1 for the distance modulus. We do not include the RA/DEC errors, since they are often not listed, and as discussed in the text, the implied center defers are a function of stellar population, far dwarfing any reported measurement errors. Therefore, we choose to present values for two representative centers instead

**Table 3** LMC galactocentric velocities

| Work | $v_{tot}$ | $v_{rad}$ | $v_{tan}$ | $v_X$ | $v_Y$ | $v_Z$ |
|---|---|---|---|---|---|---|
| Kroupa94 | 296 ± 139 | 95 ± 27 | 280 ± 139 | −244± 163 | −154 ± 82 | 66 ± 120 |
| Jones94 | 202 ± 63 | 46 ± 11 | 197 ± 66 | 49 ± 62 | −146 ± 39 | 131 ± 57 |
| KB97 | 332 ± 70 | 50 ± 14 | 328 ± 71 | 28 ± 84 | −225 ± 41 | 242 ± 60 |
| Drake01 | 211 ± 85 | 68 ± 10 | 200 ± 90 | −79 ± 59 | −158 ± 57 | 116 ± 82 |
| Pedreros02 | 328 ± 51 | 81 ± 9 | 318 ± 52 | −156 ± 48 | −220 ± 30 | 187 ± 42 |
| K06 (IAU + phot) | 378 ± 31 | 89 ± 5 | 367 ± 31 | −86 ± 14 | −268 ± 18 | 252 ± 25 |
| K06 | 356 ± 29 | 73 ± 4 | 349 ± 29 | −110 ± 14 | −245 ± 17 | 235 ± 23 |
| P08 (IAU + phot) | 361 ± 25 | 88 ± 5 | 350 ± 26 | −83 ± 11 | −258 ± 15 | 238 ± 20 |
| P08 | 339 ± 24 | 72 ± 4 | 331 ± 24 | −107 ± 11 | −234 ± 14 | 221 ± 18 |
| Costa09 | 338 ± 37 | 72 ± 8 | 330 ± 38 | −105 ± 42 | −234 ± 22 | 220 ± 30 |
| Vieira10 | 317 ± 92 | 66 ± 11 | 310 ± 94 | −66 ± 64 | −223 ± 55 | 215 ± 79 |
| K13 | 321 ± 24 | 64 ± 7 | 314 ± 24 | −57 ± 13 | −226 ± 15 | 221 ± 19 |
| K13 (IAU) | 340 ± 23 | 86 ± 5 | 329 ± 24 | −59 ± 12 | −252 ± 15 | 221 ± 19 |
| K13 (phot) | 328 ± 23 | 65 ± 5 | 322 ± 24 | −77 ± 8 | −224 ± 14 | 227 ± 18 |
| K13 (IAU + phot) | 347 ± 23 | 87 ± 5 | 336 ± 24 | −78 ± 8 | −250 ± 14 | 227 ± 18 |
| Cioni14 | 542 ± 77 | 123 ± 12 | 528 ± 77 | −409 ± 74 | −283 ± 44 | 215 ± 61 |
| vdMS16 | 312 ± 23 | 64 ± 4 | 305 ± 24 | −55 ± 14 | −220 ± 14 | 214 ± 19 |
| Gaia DR2 (HI) | 307 ± 22 | 64 ± 4 | 300 ± 22 | −57 ± 9 | −218 ± 13 | 209 ± 17 |
| Gaia DR3 (phot) | 312 ± 21 | 65 ± 3 | 305 ± 22 | −72 ± 6 | −216 ± 12 | 213 ± 17 |
| vdM02 | 293 ± 39 | 84 ± 7 | 281 ± 41 | −56 ± 36 | −219 ± 23 | 186 ± 35 |
| MF80 | 344 | 92 | 340 | -13 | −234 | 252 |
| HR94 | 368 | 107 | 352 | -10 | −287 | 230 |
| GSF94/GN96 | 297 | 82 | 287 | −5 | −226 | 194 |
| M05 | 249 | 74 | 238 | -4 | −182 | 170 |

LMC Galactocentric velocities (km/s) for literature values of PM, with the conventions/distance/$V_{LOS}$ velocities as in previous tables. Unless otherwise listed, the HI center is adopted, so the reader can see the relative spread in values. Listed also in the bottom half of the table are the values adopted in theoretical works, before measurements for the LMC PM were available. Note, in GN96, the $Y$ component of the velocity is −225 km/s rather than −226 km/s, but otherwise identical to GSF94

## 3 New tools to understand the dynamics of the clouds: dark matter halo models

This section identifies four distinct classes (A–D) of theoretical frameworks used in the literature to characterize the orbit of the LMC about the MW. These four classes are

**Table 4** SMC galactocentric velocities

| Work | $v_{tot}$ | $v_{rad}$ | $v_{tan}$ | $v_X$ | $v_Y$ | $v_Z$ |
|---|---|---|---|---|---|---|
| K06 | 292 ± 57 | 2 ± 7 | 292 ± 57 | −70 ± 51 | −227 ± 46 | 170 ± 40 |
| K06 (IAU) | 302 ± 57 | 23 ± 7 | 301 ± 57 | −86 ± 49 | −248 ± 46 | 150 ± 39 |
| P08 | 253 ± 26 | −14 ± 2 | 253 ± 26 | 40 ± 17 | −172 ± 21 | 181 ± 18 |
| P08 (IAU) | 259 ± 26 | 7 ± 4 | 259 ± 26 | 23 ± 16 | −197 ± 22 | 166 ± 17 |
| Vieira10 | 242 ± 86 | −3 ± 11 | 242 ± 86 | −34 ± 82 | −185 ± 72 | 152 ± 63 |
| K13 | 217 ± 26 | −11 ± 5 | 217 ± 26 | 19 ± 18 | −153 ± 21 | 153 ± 17 |
| K13 (IAU) | 236 ± 26 | 6 ± 4 | 236 ± 26 | 18 ± 17 | −179 ± 21 | 153 ± 17 |
| Cioni16 | 233 ± 28 | −10 ± 3 | 233 ± 28 | 15 ± 20 | −167 ± 22 | 162 ± 19 |
| vdMS16 | 259 ± 26 | −9 ± 3 | 259 ± 26 | 9 ± 18 | −188 ± 22 | 178 ± 17 |
| Zivick18 | 250 ± 20 | −10 ± 1 | 250 ± 20 | 18 ± 6 | −179 ± 16 | 174 ± 13 |
| Gaia DR2 (HI) | 248 ± 22 | −12 ± 1 | 248 ± 22 | 26 ± 8 | −174 ± 17 | 175 ± 14 |
| Gaia DR3 (phot) | 238 ± 20 | −12 ± 1 | 238 ± 20 | 35 ± 3 | −162 ± 15 | 171 ± 12 |
| MF80 | 297 | 70 | 289 | 30 | −220 | 198 |
| GN96 | 255 | 56 | 249 | 40 | −185 | 171 |

SMC Galactocentric velocities (km/s) for literature values of PM, with the conventions/distance/$V_{LOS}$ velocities as in previous tables. Unless otherwise listed, the HI center is adopted, so the reader can see the relative spread in values. Listed also in the bottom half of the table are the values adopted in theoretical works, before measurements for the SMC PM were available

**Table 5** SMC–LMC relative galactocentric velocities

| Work | $v_{tot}$ | $v_{rad}$ | $v_{tan}$ | $v_X$ | $v_Y$ | $v_Z$ |
|---|---|---|---|---|---|---|
| HI center orbits (K13_Z18) | 103 ± 26 | 92 ± 29 | 43 ± 11 | 75 ± 17 | 47 ± 22 | −47 ± 23 |
| Phot center orbits (Gaia DR3) | 127 ± 29 | 112 ± 4 | 60 ± 29 | 108 ± 7 | 53 ± 19 | −42 ± 21 |
| K13 | 128 ± 32 | 112 ± 32 | 61 ± 16 | 76 ± 22 | 73 ± 26 | −68 ± 25 |
| K06IAU | 127 ± 46 | 77 ± 50 | 88 ± 45 | 0 ± 51 | 20 ± 49 | −103 ± 46 |
| Vieira10 | 80 ± 170 | 71 ± 16 | 36 ± 126 | 32 ± 105 | 38 ± 88 | −63 ± 100 |
| Gaia DR2 | 99 ± 33 | 87 ± 3 | 48 ± 31 | 83 ± 12 | 43 ± 21 | −34 ± 22 |
| K13_Cioni16 | 110 ± 44 | 99 ± 5 | 48 ± 36 | 72 ± 23 | 59 ± 26 | −58 ± 26 |
| P08IAU | 145 ± 30 | 136 ± 33 | 46 ± 18 | 106 ± 20 | 61 ± 26 | −72 ± 26 |
| vdMS16 | 80 ± 43 | 74 ± 5 | 31 ± 35 | 64 ± 23 | 32 ± 26 | −35 ± 25 |
| MF80 | 70 | 70 | 9 | 43 | 14 | −54 |
| GN96 | 65 | 59 | 27 | 45 | 41 | −23 |

Relative velocities (km/s) for the HI center and Phot center orbits used in this review, along with literature compilations so the reader can see the spread in values. Note that since these are relative velocities, the adopted center drops out, so these values reflect differences in the PM values (when center is held constant) defined by different assumptions about the total mass and distribution of dark matter in the MW, as outlined below.

**Key Definition:** In this review, **"first infall"** is defined as an orbit in which the LMC does not complete two pericentric approaches to the MW within 5 Gyr.

This definition is motivated by studies that find non-cosmological backward integration schemes to be unreliable on timescales longer than ∼5 Gyr as they do not account for structure formation (satellite mass loss, anisotropic accretion) and the growth of the MW halo over time (Lux et al. 2010; D'Souza and Bell 2022; Santistevan et al. 2024) (see Sect. 10).

Furthermore, "rigid" halo models refer to models, where the dark matter mass distribution of the MW remains static and spherically symmetric as a function of time.

- *Model A:* Low MW mass ($M_{tot} \approx 1 - 5 \times 10^{11}$ $M_{\odot}$), point mass, rigid MW halo models. Assuming HI center velocities, orbit solutions for the LMC are hyperbolic, where the LMC's first inward crossing of the MW's outskirts (∼ 300 kpc) occurred ∼1 Gyr ago.

- *Model B:* High MW mass ($M(< 300\,\mathrm{kpc}) \approx 3-4.4 \times 10^{12}$ $M_{\odot}$), rigid, logarithmic MW halo models. Assuming HI center velocities, orbit solutions for the LMC are bound to the MW for more than 5 Gyr with orbital periods of ∼ 2 Gyr. Using the new SMC velocities (HI or Phot center), the SMC is not bound to the LMC in the Model B Framework.

- *Model C:* Cosmological MW mass profiles, rigid halo models. Most Model C studies assume a "Fixed Center of Mass (COM)" for the MW. More recently, Model C studies have adopted a "Moving COM", wherein MW halo models are rigid, but the center of mass of the MW is allowed to move in response to the gravitational force of the LMC.

- *Model D:* Cosmological MW mass profiles, non-rigid halos. Here, the MW dark matter halo is modeled as a "live" system that deforms in response to the passage of the LMC.

The above classification scheme of orbit model frameworks effectively captures the vast majority of published studies of the orbital history of the LMC about the MW, as summarized for Models A-D in Table 6. This classification also roughly serves as a chronological history of orbital modeling efforts in the literature (A to D), where changes between model classes denote leaps in our understanding of the structure of the dark matter halos of galaxies in Λ Cold Dark Matter (ΛCDM) cosmology.

In Table 6, the Model A-D frameworks are further subdivided based on the assumed mass ratio of the MW to the LMC: High mass ratios (LMC:MW > 1:15) or Low mass ratios (LMC:MW < 1:169). The changes in the assumed mass ratio of the LMC:MW over time is key to understanding the various orbit solutions in the literature.

In the earliest studies of the LMC's orbit, the LMC:MW mass ratio was generally assumed to be high (1:10, Model A; e.g., Avner, 1965), as the MW mass was thought to be of order $10^{11}$ $M_\odot$ and the LMC mass of order $10^{10}$ $M_\odot$. Later studies adopted very massive MW models (Logarithmic Potentials; ($M$ <300 kpc) ≈3–4.4 $\times 10^{12}$ $M_\odot$), while the LMC mass remained low ($\sim 1-3\times 10^{10}$ $M_\odot$), causing the LMC:MW mass ratio to decrease substantially (< 1:169 Model B; e.g., Murai and Fujimoto, 1980; Gardiner and Noguchi, 1996).

More recently, cosmologically motivated dark matter distributions of halos (Hernquist 1990; Navarro et al. 1996) result in estimates for total MW mass within 300 kpc to be $\sim 1-2\times 10^{12}$ $M_\odot$. At the same time, studies of the halo mass-stellar mass connection (Wang et al. 2006; Behroozi et al. 2010; Moster et al. 2010) motivate increasing the infall mass of the LMC by a factor of ∼10 to $\sim 10^{11}$ $M_\odot$. As such, recent studies have returned to the high LMC:MW mass ratio (>1:10) scenarios of the mid 1960s (Models C and D, e.g. Besla et al. 2010).

Significant to understanding more recent orbit models (Model D) is the change from rigid to time-evolving potentials for the MW. This change is a direct consequence of the high mass ratio between the LMC and the MW, which can be as high as 1:4 in some models (Garavito-Camargo et al. 2019). In such a scenario, the LMC cannot be thought of as a negligible perturber to the MW. Instead, the LMC must be a significant contributor of mass to, and distorter of, the MW's dark matter distribution (Besla 2015).

In Table 6, "MW Mass" refers to the mass enclosed within 300 kpc for Models A and B. For Models C and D it is the halo virial mass, defined as the mass enclosed within the radius, where the dark matter halo density is 360 times the average dark matter density of the universe.

For the column marked "Halo Models", B02 refers to the cored halo model outlined in Yang et al. (2014) and Barnes (2002). H90 is Hernquist (1990), NFW is Navarro et al. (1996), K66 is King (1966), L95 is the spherical model adopted in Lin et al. (1995), Z96 is double law profile used in Zhao (1996). For Model A, "Point Mass" refers to models, where the bulk of the MW mass is within 40 kpc. Some authors in Model A use extended potentials. For example, Fujimoto and Sofue (1976) use the oblate potential of Innanen (1966), and both Fujimoto and Sofue (1977) and Tanaka (1981) use the Miyamoto-Nagai potential (Miyamoto and Nagai 1975). However, even in these cases, the total mass of the MW is largely contained within 40 kpc, making the MW effectively a point mass given the LMC's current separation of 50 kpc.

The column marked "Velocity" refers to the LMC 3D Galactocentric velocity vector assumed by the listed studies. "Pre PM" refers to studies completed before accurate proper motions were defined. MF80 refers to Murai and Fujimoto (1980) and GN96 refers to Gardiner and Noguchi (1996), where 3D velocities are derived from theoretical models of the Stream. K06 refers to Kallivayalil et al. (2006a, b), P08 denotes Piatek et al. (2008), V10 denotes Vieira et al. (2010). HI and Phot denote the velocities used in this review, referenced in Table 2.

H15 refers to works that follow the methodology of Hammer et al. (2015), where the authors model the north component of the center of mass motion of the LMC to be $\mu_N \sim 0$ (e.g., Wang et al. 2019, 2022). This value of $\mu_N$ deviates from published observational studies by $\sim 3\sigma$ for HST and $\sim 5\sigma$ for Gaia. The authors illustrate that $\mu_N \sim 0$ is possible if the observed proper motion data is corrected using a model for the internal kinematics of the LMC that includes significant non-circular motions. However, although the proper motion residuals in HST and Gaia data reveal some deviations from circularity (Choi et al. 2022; Gaia Collaboration et al. 2021), the proper motion data are reasonably modeled using an ordered stellar disk that is not in a state of disruption (van der Marel and Kallivayalil 2014; Gaia Collaboration et al. 2021).

**Table 6** Classes of MW and LMC halo models used in LMC orbit studies

| LMC:MW Mass ratio | MW Mass $10^{11}(M_\odot)$ | MW model | Velocity | Work |
|---|---|---|---|---|
| **A: Low mass MW, rigid potential** | | | | |
| Hyperbolic LMC orbit; SMC unbound to LMC | | | | |
| High 1:10 | 1–2.75 | Point mass | Pre PM | [1–6] |
| Low 1:25–50 | 1 | Point mass | Pre PM | [7–8] |
| Low >1:156 | 4.3–4.7 | Point mass | Pre PM | [9–10] |
| Low >1:156 | 4.0–6.6 | B02 | H15 | [11–13] |
| **B: High mass MW, rigid potential** | | | | |
| Bound LMC orbit (Torb ∼ 2 Gyr); SMC unbound to LMC | | | | |
| Low >1:169 | 30–44 | Logarithmic | MF80, GN96 | [13–28] |
| Low >1:169 | 34–40 | Logarithmic | K06 | [29–31] |
| Low 1:17 | 34 | Logarithmic | K06 | [32] |
| **C: Cosmological, rigid potential** | | | | |
| Bound LMC orbit (first infall ortTorb ∼ 5 Gyr); SMC bound or unbound to LMC | | | | |
| *Fixed center of mass* | | | | |
| Low >1:45 | 9–20 | NFW | K06,P08,V10 | [33–41] |
| Low 1:67–167 | 10–13 | L95 | MF80 | [42] |
| High 1:4–10 | 5–20 | NFW, H90 | HI, K06 | [43–51] |
| *Moving center of mass* | | | | |
| High 1:4–10 | 8–20 | NFW, Z96, H90 | HI/Phot | [52–61] |
| **D: Cosmological, non-rigid potential** | | | | |
| Bound, first infall LMC orbit; SMC bound or unbound to LMC | | | | |
| *N-body* | | | | |
| High 1:5–10 | 0.5–18 | NFW, Z96, H90 | HI/Phot | [49], [52], [61–69] |

Some of the studies listed in Table 6 also consider a range of LMC:MW mass ratios. For example, Gómez et al. (2015); Patel et al. (2017); Laporte et al. (2018a); Patel et al. (2020); Garavito-Camargo et al. (2019) all considered a wide range of LMC masses ($\sim$3–25 $\times 10^{10}$ $M_\odot$) and MW models ($\sim$1–2 $\times 10^{10}$ $M_\odot$). Table 6 highlights the models favored by these authors, which tend toward a high mass ratio.

**Table 6** continued

| LMC:MW Mass ratio | MW Mass $10^{11}(M_\odot)$ | MW model | Velocity | Work |
|---|---|---|---|---|
| *Expansions* | | | | |
| High 1:5–15 | 7.9–16 | NFW, H90 | HI/Phot | [70–74] |
| *Perturbation theory* | | | | |
| Low 1:67–167 | 10–13 | K66 | MF80 | [75,76] |
| High 1:6 | 12 | H90 | HI | [77] |

For each class of model, the typical orbital solution for the LMC is listed, assuming an NFW halo for the MW and velocities from Kallivayalil et al. (2013) and Zivick et al. (2018). The MW mass refers to either the virial mass, or the mass enclosed within 300 kpc if a non-extended halo model is adopted

Model A: [1] Avner (1965), [2] Avner and King (1967), [3] Hunter and Toomre (1969), [4] Fujimoto and Sofue (1976), [5] Fujimoto and Sofue (1977), [6]Tanaka1981, [7] Idlis (1959), [8] Elwert and Hablick (1965), [9] Lin and Lynden-Bell (1977), [10] Meurer et al. (1985), [11] Hammer et al. (2015), [12] Wang et al. (2019), [13] Wang et al. (2022)

Model B: [13] Murai and Fujimoto (1980), [14] Lin and Lynden-Bell (1982), [15] Wayte (1990), [16] Gingold (1984), [17]Gardiner et al. (1994), [18] Heller and Rohlfs (1994), [19] Moore and Davis (1994), [20] Gardiner and Noguchi (1996), [21] Yoshizawa and Noguchi (2003), [22] Connors et al. (2004), [23] Bekki et al. (2004), [24] Bekki and Chiba (2005), [25] Connors et al. (2006), [26] Bekki and Chiba (2007), [27] Růžička et al. (2007), [28] Muller and Bekki (2007), [29] Růžička et al. (2009), [30] Růžička et al. (2010), [31] Diaz and Bekki (2011), [32] Nichols et al. (2011)

Model C: [33] Mastropietro et al. (2005), [34] Weinberg and Blitz (2006), [35] Besla et al. (2007), [36] Shattow and Loeb (2009), [37] Yang and Hammer (2010), [38] Diaz and Bekki (2012), [39] Zhang et al. (2012), [40] Yozin and Bekki (2014a), [41] Guglielmo et al. (2014), [42]Lin et al. (1995) (standard model), [43] Besla et al. (2010), [44] Besla et al. (2012), [45] Kallivayalil et al. (2013), [46] Jethwa et al. (2016), [47] Pardy et al. (2018), [48] Zivick et al. (2018), [49]Tepper-García et al. (2019), [50] Lucchini et al. (2020), [51] Craig et al. (2022), [52] Gómez et al. (2015), [53] Patel et al. (2017), [54] Erkal et al. (2019), [55] Patel et al. (2020), [56] Shipp et al. (2021), [57] Cullinane et al. (2022a), [58] Cullinane et al. (2022b), [59] Correa Magnus and Vasiliev (2022), [60] Battaglia et al. (2022), [61] Vasiliev (2024)

Model D: [49], [52], [61], [62] Mackey et al. (2016), [63] Laporte et al. (2018a), [64] Laporte et al. (2018b), [65] Garavito-Camargo et al. (2019), [66] Jiménez-Arranz et al. (2024), [67] Sheng et al. (2024), [68] Stelea et al. (2024), [69] Garver et al. (2026), [70] Petersen and Peñarrubia (2020), [71] Petersen and Peñarrubia (2021), [72] Garavito-Camargo et al. (2021), [73] Vasiliev et al. (2021), [74] Lilleengen et al. (2023), [75] Weinberg (1998), [76] Weinberg (2000), [77] Rozier et al. (2022)

**Summary:** Four classes of theoretical frameworks (Model A, B, C, D) broadly encompass the range of orbital studies of the Clouds and are roughly chronological. The Model A to B transition owed to an order of magnitude increase in the assumed mass of the MW ($\sim 10^{11}$ to $\sim 10^{12}$ M$\odot$). Model B to C was driven by the advent of cosmologically motivated dark matter profiles (from rigid Isothermal spheres to rigid NFW halos). Model C is divided into two subclasses, where the impact of a massive LMC ($\gtrsim 10^{11}$ $M_\odot$; LMC:MW $\gtrsim$ 1:10) on the orbital barycenter of the MW–LMC system is accounted for (Model C: Moving COM) or not (Model C: Fixed COM). The Model C to D transition is driven by the understanding that a high LMC:MW mass ratio encounter ($\gtrsim 1:10$) will drive distortions in the MW's dark matter halo that mandate time-evolving potentials.

### 3.1 What is the mass of the LMC?

Studies classified in this review as Models A-D adopt differing assumptions for both the peak dark matter halo mass of the MW and the peak dark matter halo mass of the LMC. Peak dark matter halo mass refers to the maximal dark matter mass of either galaxy's dark matter halo, throughout its history. Since the LMC is now a satellite of the MW, the peak halo mass refers to the LMC's halo mass at infall to the MW. Studies in the literature assume either a fixed LMC halo mass or allow the LMC halo mass to evolve as a function of time. In this review, the MW's peak halo mass is synonymous with the virial mass. The LMC's peak halo mass is also assumed to be the virial mass of the LMC as well as the halo mass of the LMC at the time of infall ( "infall mass").

The distribution of dark matter about the LMC is a significant uncertainty in the orbital history of the Clouds. The mass of the LMC controls the orbit of the SMC, ultimately determining the eccentricity of the SMC's orbit and how long the two galaxies have interacted with each other as a binary pair (Kallivayalil et al. 2013; Bekki and Chiba 2005). Furthermore, dynamical friction causes the orbit of the LMC about the MW to decay at a rate proportional to the inverse of the LMC's infall mass (Binney and Tremaine 1987). More recently, it has been demonstrated that the location of the MW–LMC orbital barycenter depends on the assumed halo mass of both the MW and the LMC (Gómez et al. 2015; Garavito-Camargo et al. 2019; Erkal et al. 2019; Petersen and Peñarrubia 2020; Yaaqib et al. 2025; Brooks et al. 2026).

Studies classified as Model A and B typically assume an infall mass for the LMC, ranging from $M_{tot} = 4 \times 10^9 - 3 \times 10^{10}$ M$_\odot$, which is assumed to be fixed in time. The low mass is motivated by the assumption that the LMC's dark matter distribution is quickly truncated by MW tides during pericentric passages about the MW. The LMC tidal radius is traditionally assumed to be of order 8 kpc. However, with deeper photometry, the LMC's stellar disk is now known to extend to a radius of $\sim$19 kpc (Saha et al. 2010; Mackey et al. 2016, 2018; Nidever et al. 2019; Grady et al. 2021). The extent of the LMC's stellar disk immediately implies that the total mass of the LMC has been underestimated in the past. But by how much?

The rotation curve of the LMC stellar disk is observed to remain flat to a radius of 8.7 kpc and peak at 91.7 $\pm$18.8 km/s. This implies a dynamical mass of 1.7 $\pm$ 0.7 $\times 10^{10}$ M$_\odot$ within 8.7 kpc (van der Marel and Kallivayalil 2014). If the rotation curve remains flat to $\sim$ 19 kpc the enclosed dynamical mass is $\sim$ 3.7$\times$ $10^{10}$M$_\odot$, providing a lower bound on the infall mass of the LMC. This lower bound is higher than the infall mass assumed by all studies of the LMC prior to 2010.

This review considers infall LMC halo masses that range from $3 - 25 \times 10^{10}$ M$_\odot$. Many Model C studies adopt infall LMC halo masses across this entire mass range, whereas most Model D studies favor high LMC infall mass ($8 - 25 \times 10^{10}$ M$_\odot$). This shift to high LMC mass is motivated by several arguments, outlined below, following Besla (2015) and Garavito-Camargo et al. (2019).

- *Cosmological Expectations:* Within the $\Lambda$CDM paradigm, it is expected that isolated low mass galaxies reside within massive dark matter halos, and have much larger mass-to-light ratios than galaxies like the MW. This means that, upon capture by the MW, the LMC was likely significantly more massive than traditionally modeled by Model A and B studies. If the LMC was relatively isolated in the past and is now on first infall, then, in a ΛCDM framework, the relevant infall mass for orbit calculations is approximately the cosmologically expected halo mass for an isolated central galaxy with the same stellar mass at the time of infall.
The total stellar mass of the LMC is $\sim 2.7 \times 10^9$ M$_\odot$ (van der Marel et al. 2009). Using abundance matching, a statistical technique used to assign a dark matter halo mass to an isolated central galaxy of a given stellar mass, the mean infall mass of the LMC, assuming infall at z=0, should be $\sim 1.6 \times 10^{11}$ M$_\odot$ (Moster et al. 2013).
The existence of the SMC as a satellite of the LMC further suggests that the LMC must have been relatively massive at infall. In cosmological simulations, galaxies with stellar masses similar to the LMC that also have an SMC companion usually reside in dark matter halos with a virial mass of $M_{\rm vir} = 3.4$ (+1.8/–1.2) $\times 10^{11}$ M$_\odot$ (Shao et al. 2018). This high mass is consistent with studies of cosmological dwarf galaxy pairs in the field (Besla et al. 2018). At such halo masses, cosmological models predict that a massive LMC should have several satellite companions (Sales et al. 2011; Jahn et al. 2022). Indeed, there are several candidates for such satellites (Jethwa et al. 2016; Kallivayalil et al. 2018; Patel et al. 2020; Battaglia et al. 2022; Correa Magnus and Vasiliev 2022; Martínez-García et al. 2026), supporting the cosmological expectation of an LMC-group infall (D'Onghia and Lake 2008).

- *The Timing argument:* The timing argument is a method that compares the galaxies' currently observed positions and velocities to the solution of their equations of motion in an expanding universe (Kahn and Woltjer 1959). Peñarrubia et al. (2016) applied a Bayesian inference method to constrain the total mass of the LMC using the timing argument, finding an infall mass of the LMC of $2.5^{+0.9}_{-0.8} \times 10^{11}$ M$_\odot$. In this review, $2.5 \times 10^{11}$ M$_\odot$ is adopted as the upper range for the LMC's infall mass, but this does not rule out the possibility of it being higher. Because the LMC mass profile is constrained by the observed rotation curve, increasing the mass of the LMC requires adding mass at large distances ($>$ 50 kpc, i.e., larger than the LMC's distance from the MW). As such, LMC models with infall mass as high as $2.5 \times 10^{11}$ M$_\odot$ do not cause perturbations in the MW disk that violate observational constraints on the kinematics of stars in the Solar neighborhood (Laporte et al. 2018a).

- *Perturbations to Stellar Streams:* Vera-Ciro and Helmi (2013) posited that the properties of the Sagittarius stellar stream can best be explained by the combined gravitational potential of the MW and a massive LMC ($\sim 10^{11}$ M$_\odot$). There is increasing observational evidence for kinematic and spatial perturbations in recently discovered MW stellar streams, where the motions of stars in the stream are misaligned with the stream track on the sky (Koposov et al. 2019; Shipp et al. 2019, 2021). The corresponding properties of these streams (particularly the Orphan-Chenab stream) are best explained by gravitational perturbations from an LMC with infall mass of order $\sim 1.3 \times 10^{11}$ M$_\odot$ (Erkal et al. 2019; Vasiliev et al. 2021; Lilleengen et al. 2023), and a total mass within ∼30 kpc of order $4.7 - 7.92 \times 10^{10}$ M$_\odot$ (Koposov et al. 2023; Warren et al. 2025).

- *The Formation of the Stream:* Several authors have demonstrated that a massive LMC (infall mass ∼ $10^{11}$ M$_\odot$) and SMC (∼ $10^{10}$ M$_\odot$) is not at odds with the existence of the gaseous Stream that trails behind the Clouds or the Bridge of gas that connects them (Besla et al. 2010, 2012, 2013; Pardy et al. 2018; Tepper-García et al. 2019; Lucchini et al. 2020, 2021; Craig et al. 2022).

- *The Binarity of the LMC and SMC:* A massive LMC (∼ $10^{11}$ M$_\odot$) was first introduced in the hydrodynamical studies of the Clouds of Besla et al. (2010). These authors argued that if the LMC were not massive, the Clouds could not maintain a binary state given the high relative speed of the SMC to the LMC measured by Kallivayalil et al. (2006a). In these models, the SMC is also relatively massive (∼ $2 \times 10^{10}$ M$_\odot$), yielding a SMC:LMC mass ratio of 1:10.

With this background knowledge concerning the field's changing understanding of the LMC's infall mass, the orbital solutions adopted by authors over time are examined in detail; Models A through D.

**Summary:** The assumed mass of the LMC at infall has increased by an order of magnitude over time, from ∼ $1 - 3 \times 10^{10}$ M$_\odot$ (Models A and B) to cosmologically motivated infall masses of ∼ $1 - 3 \times 10^{11}$ M$_\odot$ (Models C and D).

## 4 Model A: Low mass MW, rigid potential

The earliest efforts to model the orbital history of the LMC involved rigid, low mass MW models ($M_{tot} = 1 - 7.7 \times 10^{11}$ M$_\odot$). Model A studies typically used either a point mass approximation or moderately extended density profile to describe the MW's dark matter content, where the majority of the MW's mass is within ∼40 kpc. These studies are outlined under Model A in Table 6.

Model A studies were largely completed before accurate proper motions were measured, or they adopted center of mass proper motions that deviate significantly from recent measurements. Instead, Model A studies constrain the LMC's orbit using specific assumptions, such as whether the LMC is on a bound, unbound, or first infall orbit, whether the LMC is in a state of disruption and kinematically disordered internally, or whether the LMC is currently at the apocenter or pericenter of its orbit.

Model A studies are classified by the assumed mass ratio between the LMC and MW.

- *High Mass Ratio (LMC:MW = 1:10)*: The first published study to analyze the consequences of a high mass ratio (1:10) between the LMC and MW is the PhD thesis of Dr. Elaine Avner (Avner 1965), who was the first graduate student of Dr. Ivan King. In her thesis, and subsequent publication (Avner and King 1967), Dr. Avner made several astute realizations.

First, Dr. Avner recognized that if the LMC:MW mass ratio is 1:10, the LMC should have a non-negligible impact on the structure of the MW disk. The impact of a 1:10 encounter on the MW disk would be further developed by several authors (e.g., Hunter and Toomre 1969; Laporte et al. 2018a; Stelea et al. 2024), with recent studies expanding these efforts to study the impact of the LMC on the structure of the MW dark matter halo (e.g., Garavito-Camargo et al. 2021; Vasiliev 2023). As stated in the conclusion of Avner and King (1967), these early studies adopting a 1:10 mass ratio realized that "contrary to a widely held belief, the gravitational effect of the... Clouds on the MW is indeed serious".
Second, Dr. Avner recognized that it would be difficult for the LMC and SMC to remain a binary in the presence of the MW, given their relative separation and proximity to the MW. Avner and King (1967) suggest instead that "the Clouds are bound to each other, but passing our Galaxy in a hyperbolic orbit". Indeed, later studies utilizing newer halo models and recent proper motions have concluded the Clouds are most likely on their first passage about the MW, but on a bound orbit (Besla et al. 2007).
Subsequent studies (Fujimoto and Sofue 1976, 1977; Tanaka 1981), also assumed a high LMC:MW ratio, but identified bound, multiple-orbit configurations for the LMC's orbit around the MW. This was done by keeping the LMC's speed a free parameter. In particular, the work of Fujimoto and Sofue (1976) considered orbital scenarios, where the LMC is currently at apocenter, rather than pericenter (see Sect. 1.3).

- *Low Mass Ratio (LMC:MW $<$ 1:25)*: The first published study to analyze the orbital history of the LMC about the MW using a MW mass model of order $10^{11}$ $M_\odot$ is Idlis (1959). In this study, the LMC is assumed to be very low in mass, of order $2 - 4 \times 10^9$ $M_\odot$ (see also, Elwert and Hablick 1965). Later studies would ignore the LMC's mass entirely (Lin and Lynden-Bell 1982), or increase the mass of the MW to make the mass ratio even smaller ($<$ 1:156; Meurer et al. 1985). Such a small mass ratio is similar to that assumed in Model B. There are recent studies that have returned to low mass MW dark matter halo models ($4 - 6.6 \times 10^{11}$ $M_\odot$) and very low mass LMC models ($3 \times 10^9$ $M_\odot$) (Hammer et al. 2015; Wang et al. 2019, 2022). These studies generally find the LMC to be on first infall.

Using a low mass MW dark matter halo ($M_{\rm vir} = 1 - 6.6 \times 10^{11}$ $M_\odot$), a 1:10 or smaller LMC:MW mass ratio, and the HI/Phot Center Velocities, Model A studies result in orbital solutions, where the LMC is on first infall. If the MW is modeled as an NFW halo (ignoring the disk and bulge mass), the escape speed at 50 kpc for such low mass MW models would be less than the current speed of the LMC. Thus, using the latest measurements, the Model A framework (rigid, low mass MW halo) results in an LMC on a hyperbolic orbit about the MW. Such hyperbolic orbits are similar to orbits illustrated in Fig. 3 of Besla et al. (2007) with $\mu_W$ -1 to -4 $\sigma$ from the mean value of K06.

**Summary:** Model A studies adopt low mass MW models that necessitate high mass ratio LMC:MW encounters. Avner (1965) were the first to recognize that: (1) the impact of a 1:10 (LMC:MW) mass ratio encounter would be significant for the MW; (2) the existence of the Clouds as a binary today, despite the tidal field of the MW, suggests a recent infall. Model A orbital solutions for the Clouds using the HI/Phot Center velocities are hyperbolic orbits.

## 5 Model B: High mass MW, rigid potential

In this section, the Model B framework is outlined and the results of canonical studies are described. In Sect. 5.1, Model B orbit solutions are updated using our new understanding of the 3D velocities of the Clouds.

The Model B framework refers to very massive, rigid, MW models, where the dark matter distribution is described using a static Logarithmic potential. The most general form of this potential is the Singular Isothermal Sphere (Appendix A), which has a density distribution parameterized by $V_C$, the circular speed at the distance of the Sun (see Eq. A1). The corresponding mass profile increases linearly with radius (see Eq. A2, which yields very massive MW models.

The canonical work of Murai and Fujimoto (1980) adopts a circular speed for the MW of $V_C = 250$ km/s, resulting in $M_{\rm Iso}(< 300$ kpc$) = 4.4 \times 10^{12}$ M$_\odot$. Lin and Lynden-Bell (1982) adopt $V_C = 244$ km/s, where $M_{\rm Iso}(< 300$ kpc$) = 4.15 \times 10^{12}$ M$_\odot$. Most other studies follow the work of Gardiner et al. (1994), who adopt the IAU value of $V_C = 220$ km/s. This yields a mass enclosed of $M_{\rm Iso}(< 300$ kpc$) = 3.38 \times 10^{12}$ M$_\odot$ (Heller and Rohlfs 1994; Moore and Davis 1994; Gardiner and Noguchi 1996; Yoshizawa and Noguchi 2003; Connors et al. 2004; Bekki et al. 2004; Bekki and Chiba 2005; Connors et al. 2006; Bekki and Chiba 2007; Muller and Bekki 2007; Besla et al. 2007; Diaz and Bekki 2011; Nichols et al. 2011).

Given the high MW dark matter mass enclosed within 300 kpc, the LMC is necessarily bound to the MW, completing an orbit within roughly 2 Gyr in most instances.

Starting with Murai and Fujimoto (1980), Model B studies typically assume that the Clouds are subject to dynamical friction owing to their motion through the MW's dark matter halo (Chandrasekhar 1943), see Appendix F.

The majority of Model B studies do not include dynamical friction acting on the SMC owing to its motion through the dark matter halo of the LMC. The exceptions are the studies of Bekki and Chiba (2005) and Besla et al. (2007). Bekki and Chiba (2005) were the first to adopt a dynamical friction term that acts on the SMC when it enters the current tidal radius of the LMC. See equations listed in Appendix F.

Orbits for the LMC and SMC are computed using a backward integration scheme to numerically solve the differential equation of motion (Murai and Fujimoto 1980). For the LMC, the equation of motion includes the acceleration from the MW Isothermal halo ($\mathbf{a}_{\rm Iso}$; Eq. A4), the SMC halo ($\mathbf{a}_{\rm S_h}$) and the deceleration owing to dynamical friction from the MW halo ($\mathbf{a}_{\mathrm{DF,Iso,MW}}$; Eq. F27):

$$\mathbf{a}_{\mathrm{L}} = \mathbf{a}_{\mathrm{Iso}}(\mathbf{r}_{\mathrm{L-MW}}) + \mathbf{a}_{\mathrm{S_h}}(\mathbf{r}_{\mathrm{L-S}}) + \mathbf{a}_{\mathrm{DF,Iso,MW}}(\mathbf{r}_{\mathrm{L-MW}}, \mathbf{v}_{\mathrm{L-MW}}) \tag{3}$$

where $\mathbf{r}_{\mathrm{L-MW}}$ and $\mathbf{v}_{\mathrm{L-MW}}$ is the relative position and velocity vector of the LMC COM with respect to the MW COM, which is assumed to be fixed at [0,0,0]. $\mathbf{r}_{\mathrm{L-S}}$ is the distance vector of the LMC COM relative to the SMC COM.

For the SMC, the equation of motion includes the acceleration from the MW's Isothermal halo, the LMC halo ($\mathbf{a}_{\mathrm{L_h}}$), and the deceleration owing to dynamical friction from both the MW halo, and from the LMC halo ($\mathbf{a}_{\mathrm{DF,Iso,L}}$, see F.3 and Eq. F33):

$$\mathbf{a}_{\mathrm{S}} = \mathbf{a}_{\mathrm{Iso}}(\mathbf{r}_{\mathrm{S-MW}}) + \mathbf{a}_{\mathrm{L_h}}(\mathbf{r}_{\mathrm{S-L}}) + \mathbf{a}_{\mathrm{DF,Iso,MW}}(\mathbf{r}_{\mathrm{S-MW}}, \mathbf{v}_{\mathrm{S-MW}}) + \mathbf{a}_{\mathrm{DF,Iso,L}}(\mathbf{r}_{\mathrm{S-L}}, \mathbf{v}_{\mathrm{S-L}}) \tag{4}$$

where $\mathbf{r}_{\mathrm{S-L}}$ is the distance vector of the LMC COM relative to the SMC COM and $\mathbf{r}_{\mathrm{S-MW}}$ and $\mathbf{v}_{\mathrm{S-MW}}$ is the relative position and velocity vector of the SMC COM with respect to the MW COM.

In the Model B framework, the acceleration from the LMC and SMC halos ($\mathbf{a}_{\mathrm{L_h}}$ and $\mathbf{a}_{\mathrm{S_h}}$, respectively) are generally modeled using Plummer Potentials (Plummer 1911); see Appendix B and Eq. (B10). The LMC and SMC baryonic components are typically ignored in these calculations.

The majority of authors use a Leap Frog Integrator to compute the orbit of the Clouds. Following Springel et al. (2001), the first step of this process is to predict the position of the satellite at the next half time step ($dt/2.0$) using the current position $r_i$ and velocity $v_i$ of the satellite. To integrate backward in time, $dt$ is negative. In this review we typically adopt $dt = -0.01$ Gyr unless noted otherwise. The three steps in the Leap Frog Integration scheme are

$$\mathbf{r}_{\mathrm{half}} = \mathbf{r_i} + \mathbf{v_i}\frac{dt}{2} \tag{5}$$

Next, the velocity of the satellite is computed at the next full time step using the equations for the acceleration for the satellite in question ($\mathbf{a}_{\mathrm{sat}}$, from Eq. 3 or 4), computed at $\mathbf{r}_{\mathrm{half}}$ (Eq. 5):

$$\mathbf{v} = \mathbf{v_i} + \mathbf{a}_{\mathrm{sat}}(\mathbf{r}_{\mathrm{half}})dt \tag{6}$$

Finally, the position of the satellite is computed at the next timestep, using the predicted velocity from Eq. (6):

$$\mathbf{r} = \mathbf{r_i} + \frac{1}{2}(\mathbf{v} + \mathbf{v_i})dt \tag{7}$$

Most Model B studies were conducted prior to accurate measurements for the proper motion of the LMC. Instead, the present-day 3D velocity vector of the LMC is constrained using numerical models that match the properties of the Stream (see Sect. 1.3). The SMC velocity vector is derived assuming that the SMC maintains a roughly constant separation from the LMC in the presence of the MW. The exception to the above are studies computed after 2006 (Besla et al. 2007; Diaz and Bekki 2011; Nichols et al. 2011).

In the majority of Model B studies, the LMC is assumed to have a total mass of order 1–3 $\times 10^{10}$ $M_\odot$ and the SMC of order 1–5 $\times 10^9$ $M_\odot$. Given the high mass of Isothermal Sphere models of the MW ($3-4.4\times 10^{12}$ $M_\odot$), orbit solutions for Model B are necessarily low mass ratio encounters, LMC:MW $< 1:169$. As such, the impact of the LMC on the MW itself is typically neglected.

Figures 3 and 4 illustrate the orbital solution for the LMC and SMC determined by Murai and Fujimoto (1980) and Gardiner et al. (1994), respectively. These are the canonical examples of orbit solutions using the Model B framework. The specific properties assumed by the authors are listed in the figure caption.

Model B studies result in LMC orbits about the MW with orbital periods of ∼1.5–2 Gyr, where the LMC is bound to the MW and orbits on a decaying orbit for a Hubble time (eccentricity ∼0.3–0.4). The SMC is assumed to be bound to the LMC and the relative separation between the Clouds is roughly constant, except for a close encounter with the LMC in the past 0.2 Gyr.

In Murai and Fujimoto (1980), the present-day SMC–LMC relative velocity is assumed to be 71 km/s, which is lower than the escape speed from the LMC of $\sqrt{-2\Phi_{\rm Plummer}}$= 84 km/s at the relative distance of the SMC (23 kpc), but above the circular speed of $\sqrt{GM_{\rm Plummer}(<r)/r}$ = 60 km/s at that distance. In Gardiner et al. (1994) the SMC is currently moving at the circular speed of 64 km/s at a distance of 21 kpc from the LMC.

The N-body studies (live LMC/SMC, but rigid MW) of Gardiner and Noguchi (1996) (hereafter GN96), Connors et al. (2004), and Connors et al. (2006) are similar to that of Gardiner et al. (1994) except that the mass of the SMC is assumed to be $3\times 10^9$ $M_\odot$ and there is a slight difference in the present-day Y component of the LMC velocity vector, which is taken as Vy = −226 km/s instead of the Vy = −225 km/s assumed in Gardiner et al. (1994).

Figure 5 illustrates the orbital solutions for the LMC and SMC determined by Bekki and Chiba (2005). These authors adopt the same assumptions as GN96 for the masses, velocities and positions of the LMC, SMC, and MW. The primary difference between this study and others in Model B is the inclusion of dynamical friction acting on the SMC as the LMC orbits about the SMC. Bekki and Chiba (2005) conclude that dynamical friction from the LMC halo makes it unlikely for the SMC to remain bound to the LMC for more than 6 Gyr.

The described LMC and SMC orbit solutions are consistent across all Model B studies that do not use HI/Phot Center velocities.

**Summary:** Model B studies adopt rigid, massive Isothermal Sphere MW halos (M($<$ 300 kpc) $= 3-4.4\times 10^{12}$ $M_\odot$), and low mass ratio encounters, LMC:MW $< 1:169$. Canonical orbit solutions yield decaying orbits, where the Clouds complete multiple orbits about the MW with an orbital period of ∼ 1.5–2 Gyr and eccentricity of ∼ 0.3–0.4. The SMC is assumed to be bound to the LMC for at least ∼ 6 Gyr, on a roughly circular orbit, with a recent close encounter ∼ 0.2 Gyr ago.

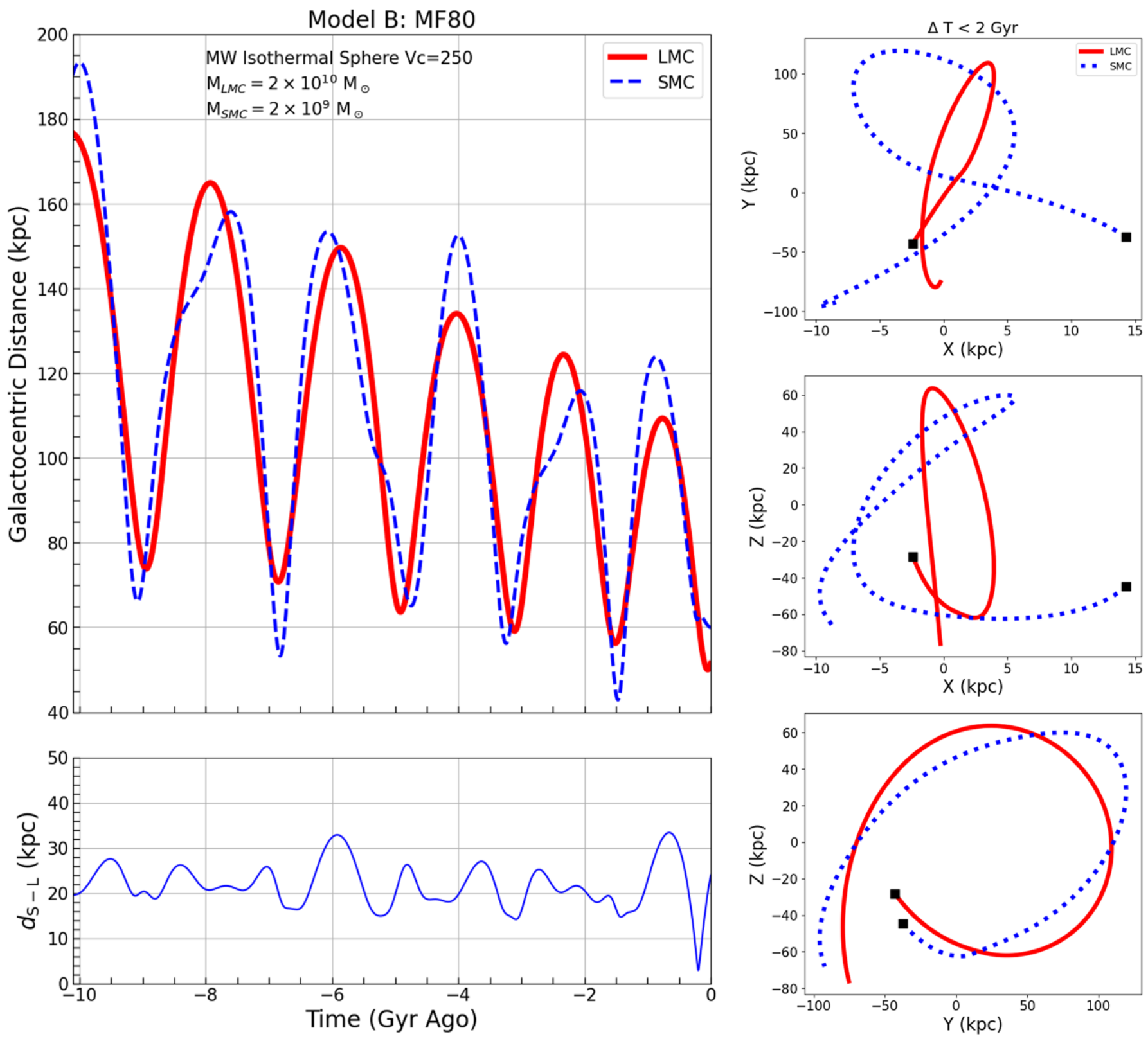

**Fig. 3** Orbit of the SMC and LMC around the MW in the Model B Framework using velocities assumed by Murai and Fujimoto (1980) [MF80]. The LMC is modeled as a Plummer potential (Eqs. B7–B10), with a scale length of $k$=3 and a total mass of $M_{\rm sat} = 2 \times 10^{10}$ M$_\odot$. The SMC is modeled as a Plummer potential with scale length of $k$=2 and total mass of $M_{\rm sat} = 2 \times 10^{9}$ M$_\odot$. The MW is modeled as an Isothermal Sphere (Eq. A3) with $V_C = 250$ km/s. Dynamical friction owing to the MW is modeled using Eqs. (F27) and (F28), where $b_{\rm min} = k$, the scale length of the LMC's Plummer profile. The forces felt by the LMC and SMC are described in Eqs. 3 and 4, except that there is no dynamical friction from the LMC acting on the SMC ($a_{\rm DF,LMC} = 0$). To accurately reproduce the results from Murai and Fujimoto (1980), the time step is dt = −0.002 Gyr in the Leap Frog integration scheme. The LMC's current position is adopted by these authors as L[x,y,z] = [−2.4, −42.9, −28.3] kpc, and the SMC is S[x,y,z] = [14.3,−37.4,−44.6] kpc, in Galactocentric Coordinates, where [0,0,0] is the center of the MW. The velocity vectors for the LMC and SMC are listed in Tables 3 and 4 (values for MF80). *Left Top:* Galactocentric distance of the LMC (red, solid) and SMC (blue, dashed) as a function of look-back time. Time today is 0. The orbital period of the Clouds about the MW is ∼ 2 Gyr. *Left Bottom:* The relative separation between the Clouds as a function of time. The SMC is bound to the LMC on a circular orbit. There is a close encounter between the Clouds ∼0.2 Gyr ago. *Right Panels:* The orbit of the LMC (red, solid) and SMC (blue dashed) is computed over the past 2 Gyr and projected in Galactocentric planes

## 5.1 Understanding the model B framework

The majority of Model B studies were completed prior to the first accurate proper motion measurements for the Clouds by Kallivayalil et al. (2006b) and Kallivayalil et al. (2006b). The following describes the impact our new understanding of the Galac-tocentric 3D velocity vector and mass of the LMC on the orbit solutions for the LMC (Sects. 5.1.1 and 5.1.2) and SMC (Sect. 5.1.3) using the Model B framework, wherein the MW is modeled as an Isothermal Sphere.

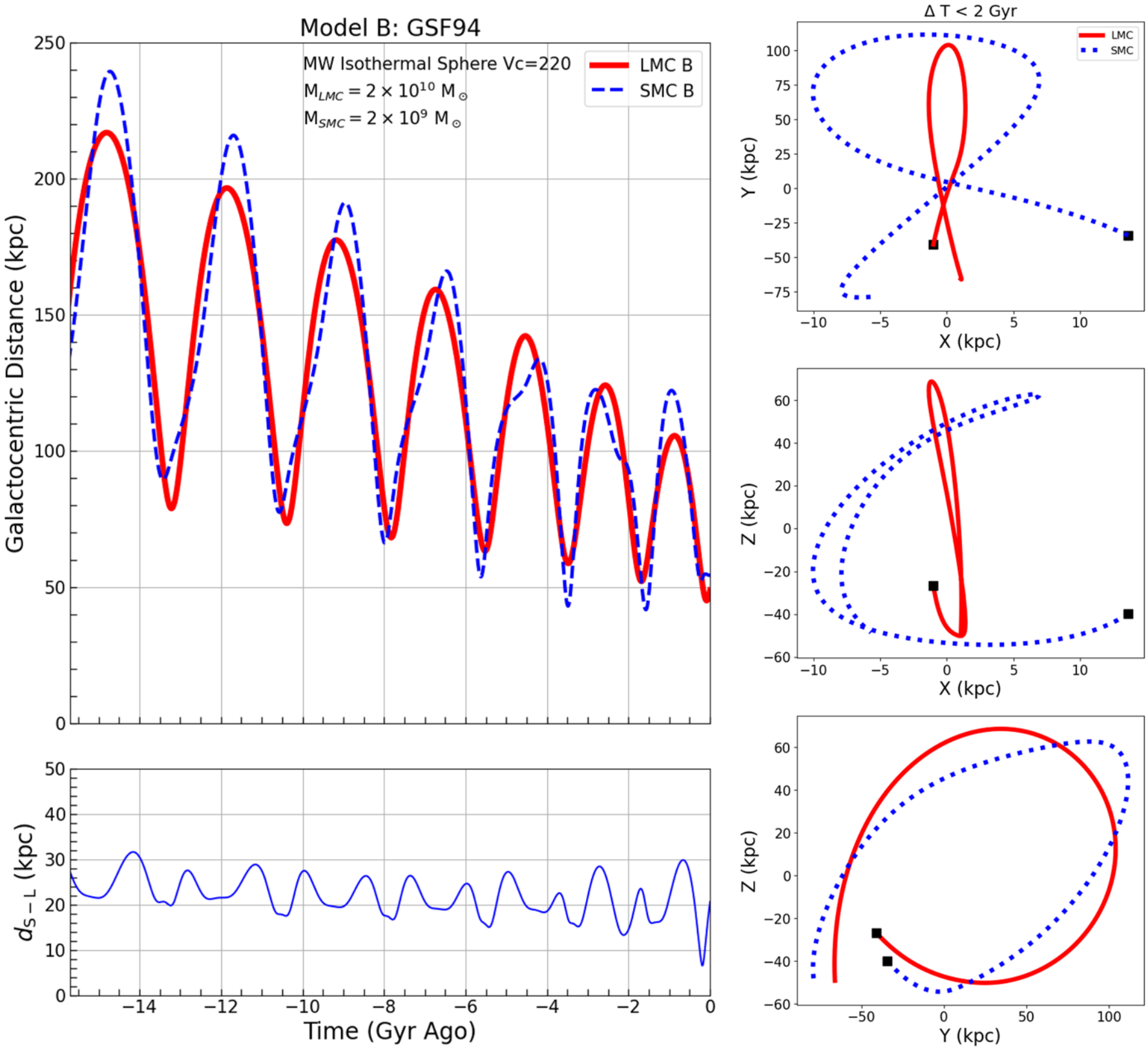

**Fig. 4** Orbit of the SMC and LMC around the MW in the Model B Framework using velocities assumed by Gardiner et al. (1994) [GSF94]. The LMC and SMC orbit computation methodology and figure panels are described in Fig. 3. However, the MW Isothermal Sphere model adopts $V_C$ = 220 km/s. In addition, the LMC and SMC position and velocity vectors are assumed to be: L[x,y,z] = [−1,−40.8, −26.8] kpc, VL[x,y,z] = [−5, −225, 194] km/s, S[x,y,z] = [13.6,−34.3, −39.8] kpc, VS[x,y,z] = [40, −185, 171] km/s. Dynamical Friction is computed as described in Eq. (F27), with fixed $\Delta_C = 3$ for each galaxy as it orbits the MW. Dynamical friction between the Clouds is not included

#### 5.1.1 New proper motions, $V_C$, and LMC center: model B LMC orbit

As described in Shattow and Loeb (2009) and Růžička et al. (2010), the 3D velocity vector of the LMC depends on the assumed value of the MW's Local Standard of Rest ($V_C$). Furthermore, as described in Sect. 2, the 3D velocity vector also depends on the assumed location of the LMC's kinematic center (HI or Phot). Figure 6 illustrates how these assumptions change the LMC orbit solutions in the Model B framework. Dynamical friction is ignored in these orbit calculations to highlight changes driven by the velocity vector, LMC center, and $V_C$.

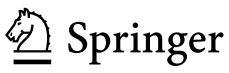

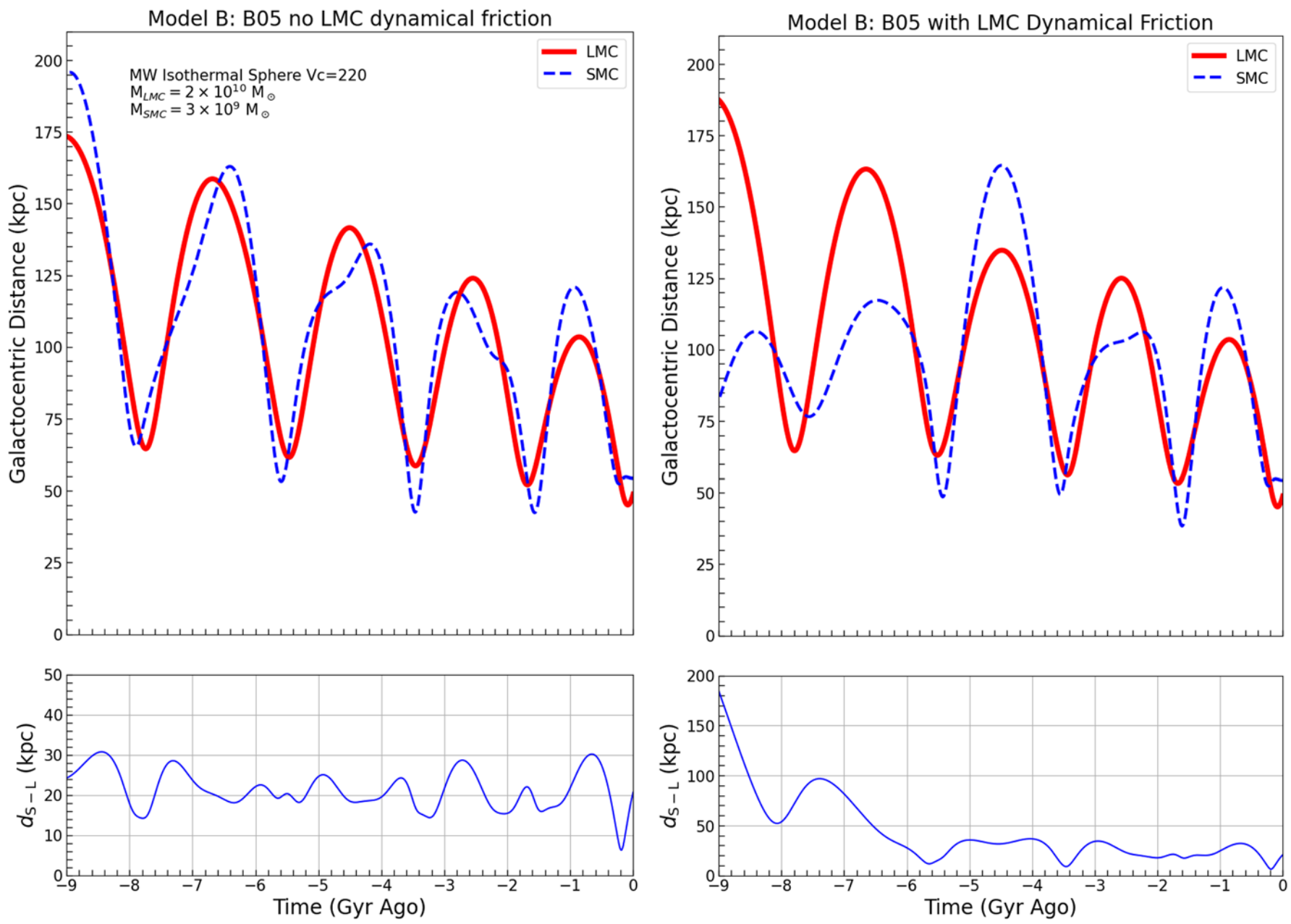

**Fig. 5** Orbit of the SMC and LMC around the MW in the Model B framework using velocities adopted by Bekki and Chiba (2005) [B05]. The LMC/SMC orbit computation methodology and figure panels are as described in Fig. 4. However, the SMC is modeled as a Plummer potential with scale length of $k$=2 and total mass of $M_{\rm sat} = 3 \times 10^9$ M$_\odot$, and the LMC velocity vector is assumed to be VLxyz = [−5, −226, 194] km/s, as in Gardiner and Noguchi (1996). *Left:* The LMC and SMC orbits about the MW with no dynamical friction acting on the SMC from the LMC. *Right:* The same LMC and SMC orbit, but with dynamical friction acting on the SMC from the LMC, according to Eq. (F33). The SMC does not maintain a binary state with the LMC for more than 6 Gyr. These panels correspond to the BO5 models C and D, illustrated in their Fig. 4

The canonical Model B LMC orbit solution is marked as "GN96," which uses the LMC's 3D velocity vector adopted by (Gardiner and Noguchi 1996; Bekki and Chiba 2005, see Fig. 5). Orbits are also computed using the mean K13 velocities from Table 3 for different choices of LMC center (HI center or Phot center) and $V_C$ (IAU 220 km/s, or the McMillan 2011 value of 239 km/s). The different values of $V_C$ change both the mass of the MW halo and the velocity vector of the LMC.

In the left panel of Fig. 6 ($V_C$ = 220 km/s; IAU), the choice of LMC center changes the LMC's orbital period from 1.5 Gyr (GN96) to 2 Gyr, and increases its orbital eccentricity ($e$ =0.33 to $e$ ∼0.5). Using the "Phot" center yields the highest current speed for the LMC, and thus the most eccentric orbit.

In the right panel, a higher $V_C$ = 239 km/s (McMillan 2011) is assumed, meaning the MW is more massive (M(<300 kpc) $\sim 4 \times 10^{12}$ M$_\odot$ if $V_C$ = 239 km/s, vs. $\sim 3.4 \times 10^{12}$ M$_\odot$ if $V_C$ = 220 km/s). In all cases, the orbit solutions using the new velocities are somewhat more eccentric than the canonical Model B orbits ($e$ = 0.26 increases to $e$ ∼0.31). Interestingly, the new orbit solutions in this more massive halo

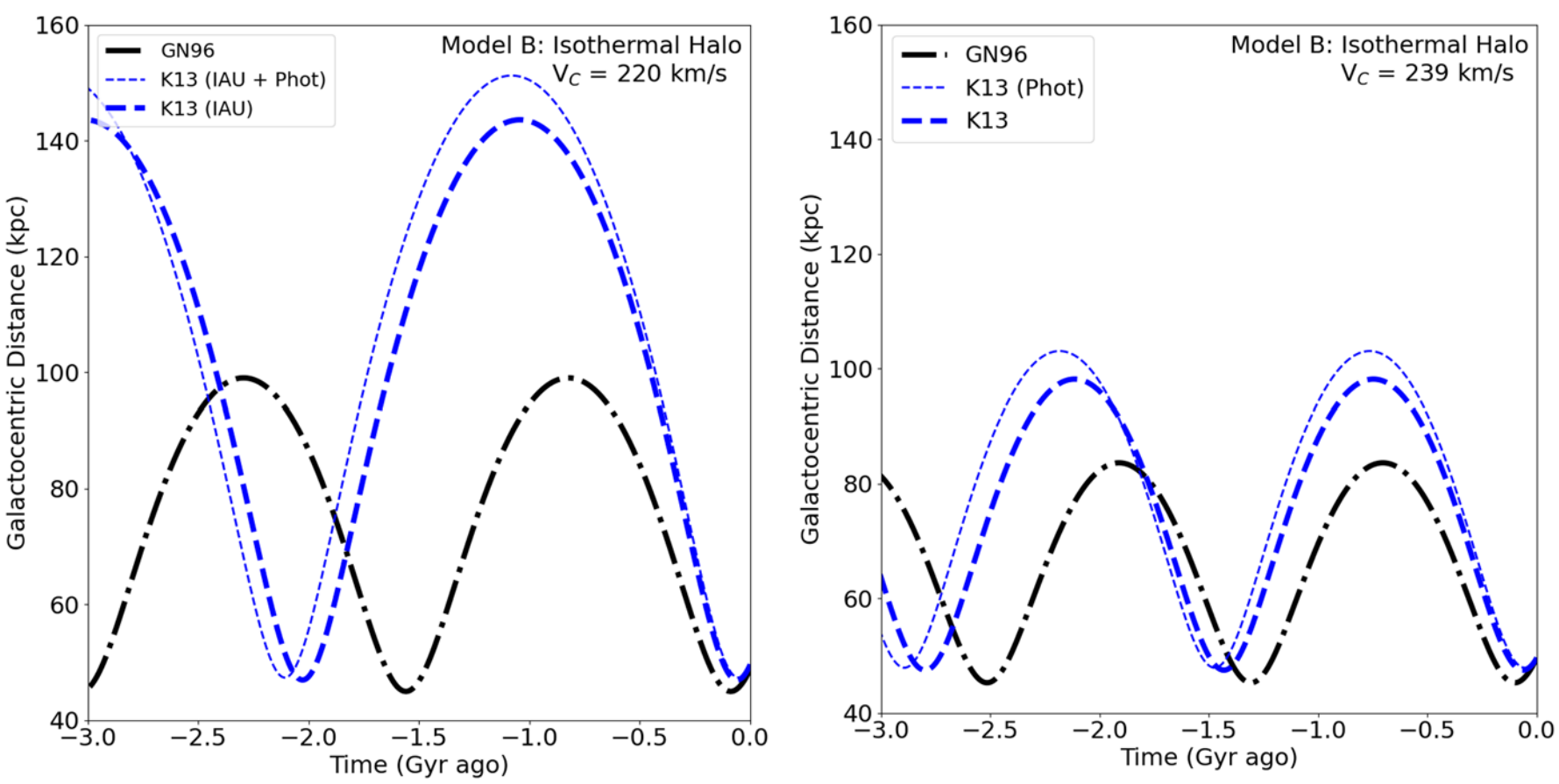

**Fig. 6** Impact of $V_C$ and the assumed LMC center on the orbital history of the LMC in the Model B Framework. The Galactocentric distance of the LMC is plotted as a function of time in the past, where the present day is at t=0. The LMC's orbital history is computed without dynamical friction assuming an Isothermal Sphere model for the MW. $V_C$ is assumed to be 220 km/s (IAU standard; left panel) or 239 km/s (McMillan 2011 right panel). The MW is more massive in the right panel than in the left. The present-day galactocentric position vector of the LMC is that of the HI center orbits in all cases (see Table 1). The canonical Model B velocity vector (GN96, black dashed line; Fig. 5 Gardiner and Noguchi 1996) is the same in both panels; changes in the orbit reflect the higher mass of the MW model. Orbits are also computed using velocities derived from the K13 proper motions, depending on the choice of LMC center (Phot center: thin, dashed, blue line; HI center: thick, dashed, blue line) and $V_C$ (220 km/s or 239 km/s); see Table 3. This figure illustrates how the choice of $V_C$ and LMC Center impact the LMC's orbit. The K13 LMC orbits in the Model B framework with $V_C = 239$ km/s (higher MW mass) are similar to the canonical Model B solution (GN96/B05, $V_C = 220$ km/s), regardless of the choice of LMC center

are similar to the GN96 orbit with $V_C = 220$ km/s, regardless of the choice of center. The choice of $V_C$ has a significant impact on the derived orbit.

The larger velocity of the LMC derived by recent studies (Gaia/HST: K13, DR2, DR3) relative to the lower velocity adopted by early theoretical models (B05, GN96, GSF94, MF80) is mainly due to an increase in the west component of the proper motion ($\mu_W$). However, the LMC velocity did not change only in magnitude. The direction of motion of the LMC on the sky also changed.

As discussed in Besla et al. (2007), because the LMC orbit is primarily in the YZ Galactoceneric plane (i.e., polar relative to the MW's disk), the north component of the proper motion controls the location of the LMC's past orbit when projected on the sky ($\mu_N$). This is illustrated in Fig. 7, where the LMC's past orbital trajectory is projected on the sky in the Model B framework. In most Model B studies, the LMC's 3D velocity vector was chosen, such that the location of the LMC's past orbit matched that of the Stream on the sky (indicated by the orange dashed line marked MF80). The projected LMC Model B orbits derived using the HI Center velocities (HST K13 HI, Gaia DR2HI) and the Phot Center velocities (Gaia DR3 Phot) *all* deviate from the canonical Model B orbits (i.e., location of the Stream; MF80) by ∼10 degrees. The Phot Center velocities are more offset than the HI Center velocities, due to differences in $\mu_N$. Other proper motion studies result in larger deviations between the the past

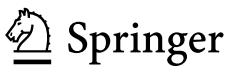

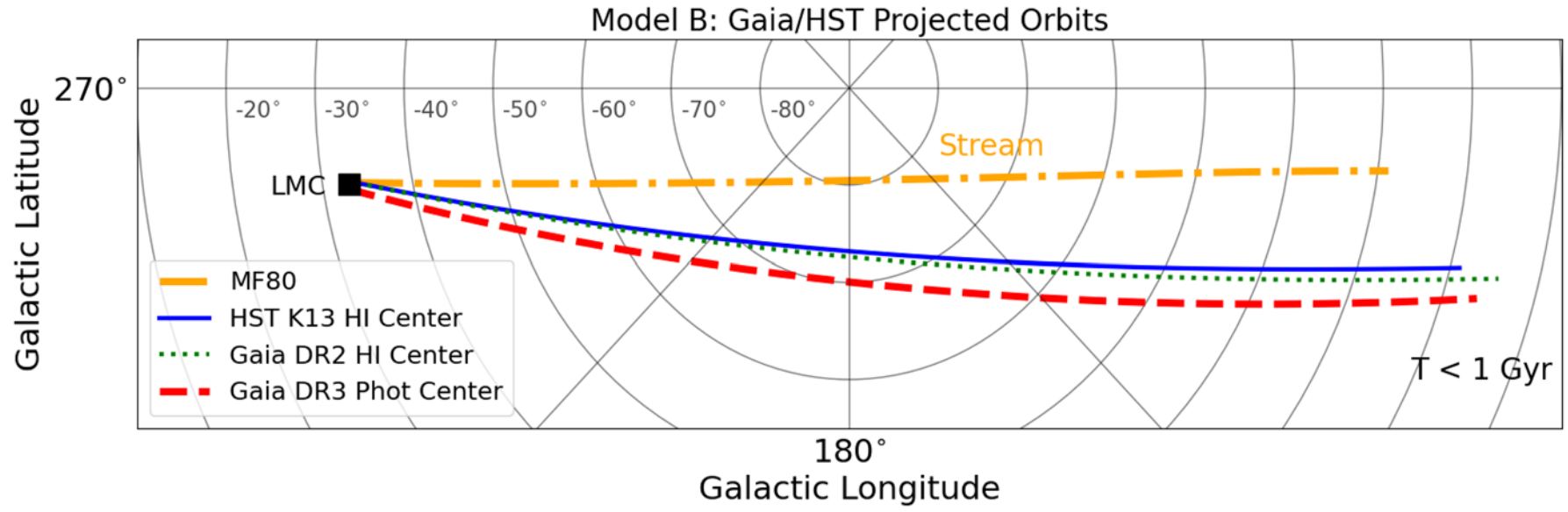

**Fig. 7** LMC projected orbits in Galactic coordinates in the Model B framework. The MW is modeled as an Isothermal Sphere with $V_C$=239 km/s. No dynamical friction is assumed. The projected location of the LMC's orbit on the sky is controlled by the north component of the LMC's proper motion vector ($\mu_N$). Canonical Model B orbit solutions, like MF80 or GN96, adopted $\mu_N \sim 0$, so that the LMC's past orbit trace the Stream location by design. Recent velocity measurements using the HI center (HST K13 HI, Gaia DR2 HI) or the Phot center (Gaia DR3 Phot) find $\mu_N > 0$, regardless of choice of LMC center or $V_C$. The resulting projected orbits all broadly agree in the Model B framework and are all offset from the location of the Stream by ~10 degrees. However, the LMC Phot Center velocity orbits are more offset than the HI Center velocities owing to differences in $\mu_N$

LMC orbit and Stream (e.g., Kroupa et al. 1994; Jones et al. 1994; Drake et al. 2001; Pedreros et al. 2002; Kallivayalil et al. 2006b; Piatek et al. 2008; Cioni et al. 2014). The choice of $V_C$ or LMC center does not change the result that the location of the LMC's projected orbit on the sky is offset from the Stream (e.g., DR2 HI vs. DR3 Phot). This result is true for any spherically symmetric halo model (e.g., Fig. 18). Non-spherical halos are discussed in Sect. 6.3.3 (see also Růžička et al. 2007, 2009).

In summary, we have discussed how Model B canonical orbital solutions are impacted by three ways in which the 3D velocities of the Clouds have changed: (1) adopting a higher $V_C$ (239 km/s vs. 220 km/s) causes the orbits to be less eccentric owing to the correspondingly larger MW mass; (2) changing the center of the LMC from the HI to the Phot center causes the orbits to be somewhat more eccentric; (3) adopting the new HST/Gaia proper motions causes the LMC's orbit to not track the Stream on the sky vs. canonical orbits (MF80).

**Summary:** The net impact on the LMC Model B orbits of the new proper motions, $V_C$, and LMC center is that the LMC's orbital eccentricity (~ 0.3) and period (~ 1.5 Gyr) remain similar to that of canonical Model B studies (MF80, GSF94, GN96, B05; see K13 orbits in the right panel of Fig. 6). However, the past orbit of the LMC deviates markedly from the location of the Stream when projected on the sky, regardless of the choice of LMC center (Fig. 7).

### 5.1.2 New velocities and a higher LMC mass: model B LMC orbit

As discussed in Sect. 3.1, the total mass of the LMC at infall is cosmologically expected to be of order $10^{11}$ $M_\odot$, a factor of 10 higher than traditionally modeled. The highest LMC mass considered by Model B studies is $2 \times 10^{11} M_\odot$ (Nichols et al. 2011). In the following, the impact of higher LMC mass on the orbital evolution of the LMC

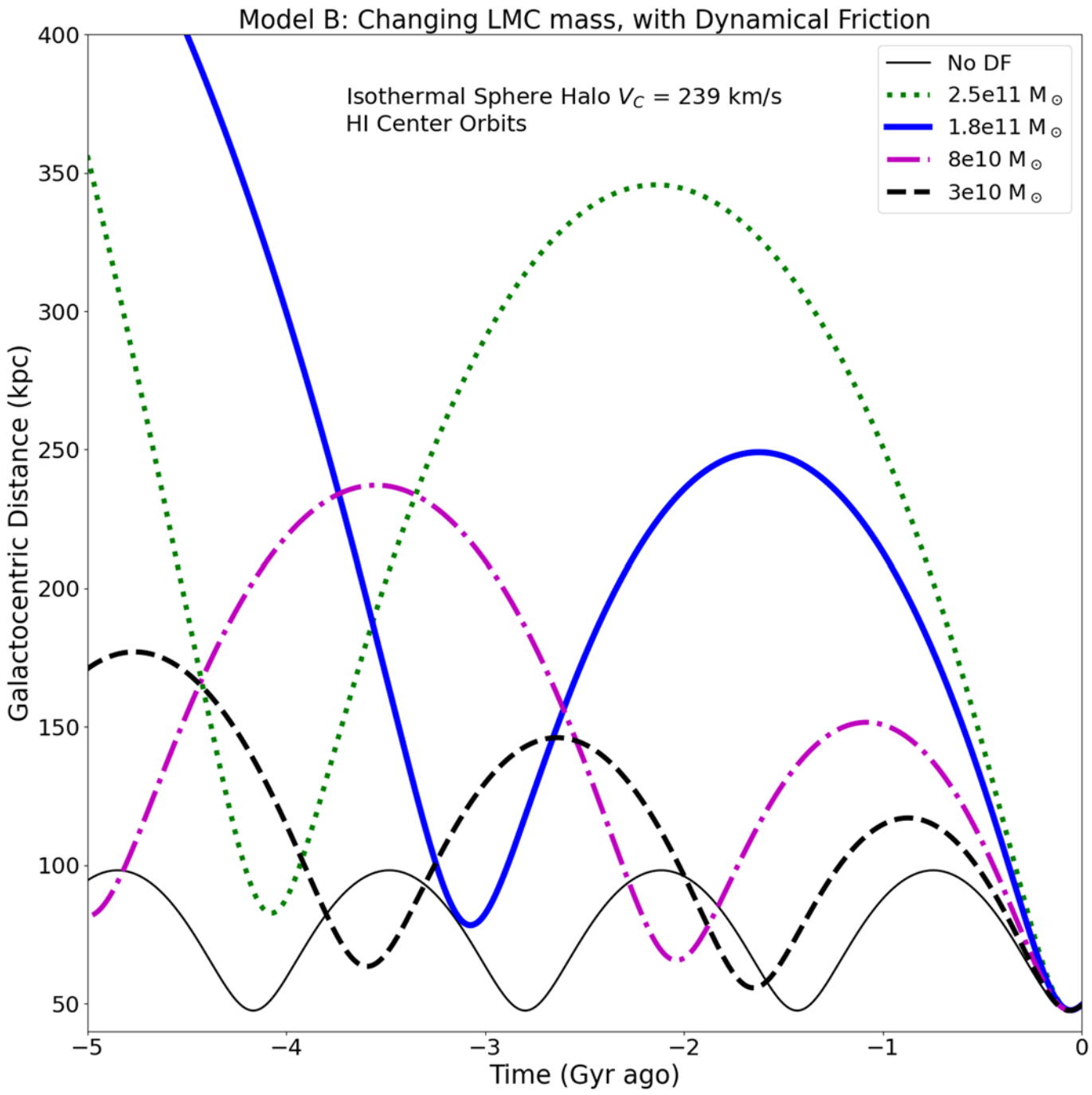

**Fig. 8** Galactocentric distance of the LMC as a function of lookback time, plotted using the LMC HI Center velocities (Table 2) in the Model B framework, where the MW is modeled as an Isothermal Sphere with $V_C = 239$ km/s. Results are similar if the LMC Phot center velocities are used instead. The gravitational force from the SMC is not included. The mass of the LMC is varied from 3–25 $\times 10^{10}$ $M_{\odot}$ using a Plummer profile (see Table 7). Dynamical friction (Eq. F27) owing to the LMC's motion through the MW's dark matter halo is included in all cases, except for the solid black line. As the LMC mass is increased, the orbit becomes increasingly eccentric owing to dynamical friction (0.3 to 0.75)

in the Model B framework is examined, in addition to the new present-day velocity vector (HI/Phot Center Velocities). For the following calculations, dynamical friction is included.

In Fig. 8, the LMC's Galactocentric distance is plotted backward in time in an Isothermal Sphere model with $V_C = 239$ km/s. The LMC is modeled as a Plummer sphere and its total mass is varied as: 3, 8, 18, 25 $\times 10^{10}$ $M_{\odot}$. Following Patel et al. (2020), the softening length for each model is determined, such that the total mass enclosed within 10 kpc is roughly the dynamical mass inferred from the LMC's rotation curve of M($<$ 10kpc) = 1.2 $\times$ $10^{10}$ $M_{\odot}$ (van der Marel and Kallivayalil 2014). The resulting parameters for the LMC Plummer potentials are listed in Table 7. The orbits become more eccentric with increasing LMC mass. These trends are the same for orbits computed with the HI or Phot center, since the orbit changes owing to dynamical friction dominate the velocity uncertainties.

**Table 7** LMC plummer potential parameters

| Mass ($10^{10}$ $M_\odot$) | $k$ | $V_{esc}$ (23 kpc) km/s |
|---|---|---|
| 2 | 3 | 86 |
| 3 | 9.18 | 102 |
| 5 | 12.61 | 128 |
| 8 | 15.94 | 156 |
| 10 | 17.64 | 172 |
| 11 | 18.38 | 179 |
| 18 | 22.54 | 219 |
| 25 | 25.63 | 234 |

The SMC is modeled as Plummer potential with total mass of $3 \times 10^9$ $M_\odot$ and $k = 2$ (Bekki and Chiba 2005)

**Summary:** As the LMC mass is increased from 3–25 $\times 10^{10}$ $M_\odot$, dynamical friction owing to the LMC's motion through the MW halo causes the LMC orbit to become significantly more eccentric (0.3 to 0.75), reaching larger apocenter distances ($\sim$ 100 to 350 kpc).

### 5.1.3 New velocities and LMC mass: model B SMC orbit

Next, the SMC's orbital solution is explored within the Model B framework using the HI/Phot Center Velocities and a change of $V_C = 239$ km/s.

In Fig. 9, the orbits of the Clouds about the MW are computed, where the LMC and SMC are modeled as Plummer spheres, as in (Bekki and Chiba, 2005, Fig. 5). The LMC mass is the canonical low mass ($2 \times 10^{10} M_\odot$) value used in Model B studies like (Bekki and Chiba 2005), as described in Sect. 5.1.1.

Dynamical friction due to the MW acts on both the LMC and SMC, but dynamical friction acting on the SMC from the LMC is omitted. The inclusion of dynamical friction between the Clouds would not improve their chances for binarity (Bekki and Chiba 2005). The orbits of the Clouds are computed following the equations of motion in Eq. (4), with $\mathbf{a}_{\rm DF,IsoL}$=0.

The HI center velocity ($V_C$=239 km/s; Table 2) is adopted for the LMC, but the orbits are similar if the Phot center is adopted. The resulting LMC orbit is similar to that found by Murai and Fujimoto (1980); Gardiner et al. (1994); Gardiner and Noguchi (1996); Bekki and Chiba (2005), as expected from Fig. 6.

The HI center velocity is also adopted for the SMC (Table 2). The relative speed of the SMC with respect to the LMC determined by HST and Gaia ($v_{rel} > 100$ km/s; see Table 5) is higher than the LMC's escape speed of 86 km/s for this Plummer potential (at the SMC's current separation of 23 kpc). This means the SMC is not bound to the LMC. Using the Phot center velocities, the relative speed is higher than with the HI velocities (see Table 5), which will only exacerbate the problem of keeping the SMC bound. As such, both Gaia and HST measurements imply that the SMC is currently *not bound* to the LMC in the Model B framework, which requires a low mass LMC model.

Because the Clouds are not bound in this scenario, the SMC only coincidentally follows a similar orbit as the LMC for the past 3 Gyr. At earlier times, the SMC orbits around the MW within 100 kpc, while the LMC orbits in roughly the same orbital plane, but at larger distances because of dynamical friction.

The higher relative speed of the SMC determined by Gaia and HST (Table 5) means that the Model B assumption that the SMC is in a circular orbit about the LMC is no longer valid (see Sect. 1.3). Moreover, the Clouds are not a binary pair in models that assume a massive Isothermal Sphere MW model and low mass LMC.

If the mass of the LMC is increased, the escape speed at the location of the SMC increases (see Table 7), suggesting that the LMC and SMC binarity could be maintained for higher LMC mass models. However, the escape speed calculation does not account for the tidal force of the MW, which is substantial for such massive Isothermal Sphere halos.

This issue is illustrated in Fig. 10, where the HI Center orbits of the SMC and a massive LMC are plotted as a function of lookback time, in the Model B framework. The Clouds are again modeled as Plummer spheres, where now the LMC is the most massive model considered in this review ($2.5 \times 10^{11}$ $M_{\odot}$, $k = 25.63$; Table 7). Even in this scenario, the LMC is unable to hold on to the SMC owing to the strength of the tidal field of the massive MW ($V_C = 239$ km/s, M(<300 kpc) $\sim 4 \times 10^{12}$ $M_{\odot}$). Furthermore, although the LMC is bound to the MW, its orbit decays rapidly, making it unlikely to have survived multiple orbits about the MW. Again, the relative SMC–LMC speed is higher if the Phot centers are used, making it even harder to maintain the binary.

**Summary:** Long-lived LMC–SMC binary orbits are not possible under Model B if the Clouds move with 3D velocities as measured by HST or Gaia, regardless of LMC mass (3–25 $\times 10^{10}$ $M_{\odot}$). Instead, in Model B the LMC and SMC *coincidently* orbit in similar planes about the MW for a Hubble time and only happen to approach one another at the present day.

## 6 Model C: Advances in cosmological halo profiles

In the $\Lambda$CDM paradigm, the dark matter halos of massive galaxies are now understood to be poorly represented by Isothermal Spheres at large Galactocentric distances. Instead, it is expected that the dark matter density profile falls off more sharply, making it easier for satellites to travel to larger distances.

In the Model B framework, Isothermal Spheres are adopted to describe the dark matter distribution of the MW. Such potentials do a reasonable job at reproducing the observed rotation curve of the MW disk. However, given that the mass profile of an Isothermal Sphere increases linearly with radius, the total mass at large distances is much larger than observational estimates or cosmological expectations for the total dark matter content of the MW (Shen et al. 2022). As a result, trajectories of halo

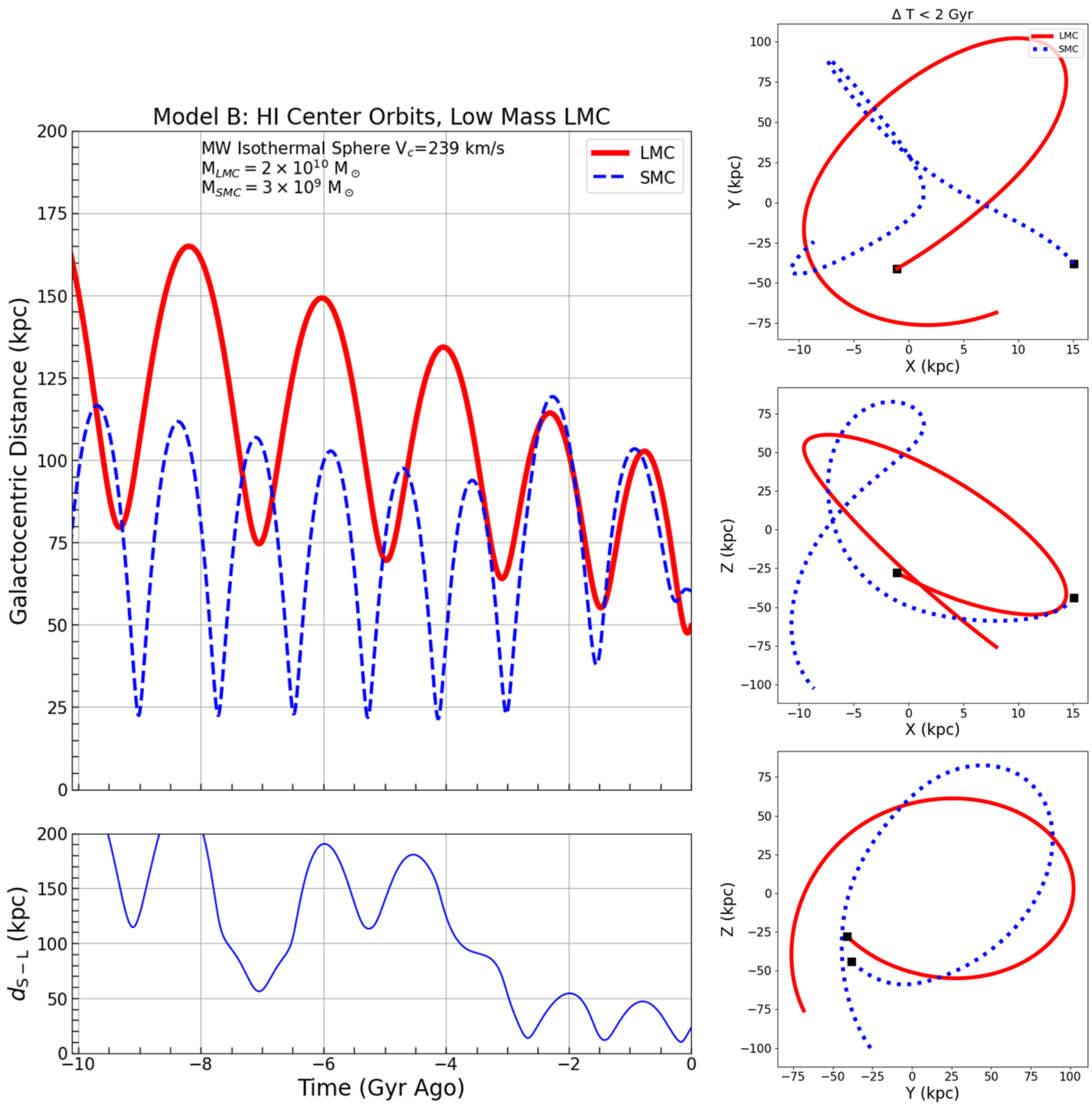

**Fig. 9** Orbit of the Clouds as a function of look back time, computed using the Model B framework (low mass LMC) with the Galactocentric positions and HI Center velocities for the LMC and SMC from Tables 2 and 1. The MW is modeled as a static Isothermal Sphere with $V_C = 239$ km/s. The LMC is modeled as a Plummer Sphere with a low halo mass of $2 \times 10^{10}$ $M_\odot$ and $k = 3$, and the SMC as a Plummer Sphere of halo mass of $3 \times 10^9 M_\odot$ and $k = 2$, as in Bekki and Chiba (2005). The orbital history of the Clouds is computed as in Figs. 4 and 3, with: Galactocentric distance (left upper panel), relative LMC–SMC separation (left lower panel), and orbits for the past 2 Gyr plotted in the Galactocentric XY, XZ, YZ planes (right). Dynamical friction owing to the motion of the Clouds within the MW's halo is included. However, dynamical friction between the Clouds is not, since this would only serve to further disrupt the binary. The SMC is not bound to the LMC owing to the high relative speed of the Clouds as determined by HST and Gaia ($v_{rel} > 100$ km/s). Instead, the SMC is *coincidentally* orbiting in the same orbital plane of the LMC and only approaches the LMC within the past 3 Gyr. If the LMC Phot Center velocities were used, the relative speed between the Clouds would be higher (Table 5)

tracers at large Galactocentric distances ($> 50$ kpc) are poorly estimated in such potentials.

With the advent of large volume simulations of structure formation, based on ΛCDM cosmology, different models have been introduced to describe the density profile of dark matter surrounding galaxies. Model C studies adopt cosmologically

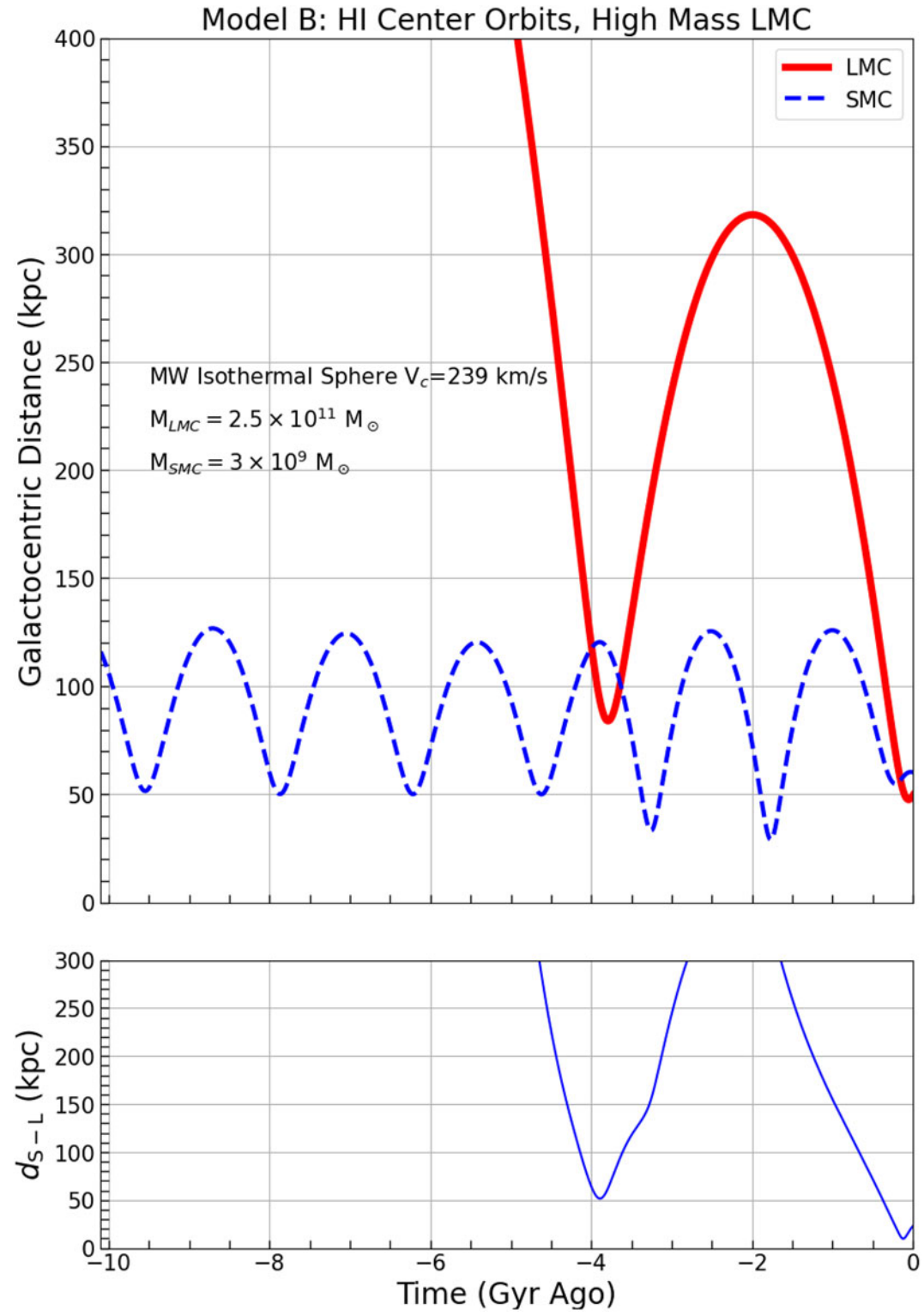

**Fig. 10** Orbital history of the Clouds in the Model B Framework with a massive LMC, using the HI Center velocities (Table 2). The orbital history is computed as in Fig. 4. The SMC is modeled as a Plummer potential with a mass of $3 \times 10^9$ M$_\odot$ and $k = 2$, following Bekki and Chiba (2005). The LMC is modeled using the most massive model considered in this review ($2.5 \times 10^{11}$ M$_\odot$, $k = 25.63$, see Table 7). *Top Panel:* Galactocentric distance of a high mass LMC (solid, red) and SMC (dashed, blue) as a function of lookback time. *Bottom Panel:* The relative separation between the Clouds. Dynamical friction from the MW is acting on the LMC and SMC but not between the Clouds (this would only serve to further disrupt the binary). The LMC is unable to hold on to the SMC owing to the tidal field of the MW, even though the escape speed at the distance of the SMC is well above their relative speed (see Table7). Using the Phot Center velocities would not improve the results as the relative velocity between the Clouds would be even higher. Binary LMC–SMC orbits are not sustainable in Model B owing to the high relative speeds of the Clouds determined by HST and Gaia and the high mass of Isothermal Sphere MW models

motivated dark matter density profiles for the MW. These studies were typically completed after accurate proper motions were measured.

The NFW density profile (Navarro et al. 1996) is commonly adopted by Model C studies and is used in this review. Another commonly used density profile is the Hernquist profile (Hernquist 1990, hereafter H90). Both these profiles are defined in Appendix C and Appendix E. Studies classified under Model C are listed in Table 6. Note that the study of Lin et al. (1995) adopt a different halo profile, but they show that it is similar to the Hernquist profile.

As described in Sect. 3, Model C studies further require the MW potential to be rigid, meaning that the MW's dark matter distribution maintains a static, spherically symmetric density profile as a function of time.

Most Model C studies assume that the location of the center of mass of the MW is fixed in time, which means that the MW does not respond to the gravitational forces imparted by the LMC (Lin et al. 1995; Mastropietro et al. 2005; Weinberg and Blitz 2006; Besla et al. 2007; Shattow and Loeb 2009; Besla et al. 2010; Yang and Hammer 2010; Diaz and Bekki 2012; Zhang et al. 2012; Kallivayalil et al. 2013; Yozin and Bekki 2014a; Guglielmo et al. 2014; Jethwa et al. 2016; Pardy et al. 2018; Zivick et al. 2018; Tepper-García et al. 2019; Lucchini et al. 2020; Craig et al. 2022). In this review, these studies are referred to as Model C: Fixed COM.

However, starting with the study of Gómez et al. (2015), there is a subset of Model C studies of the LMC's orbit that account for the LMC's gravitational force acting on the MW itself (Patel et al. 2017; Erkal et al. 2019; Patel et al. 2020; Shipp et al. 2021; Cullinane et al. 2022a, b; Correa Magnus and Vasiliev 2022; Battaglia et al. 2022). This effect is significant for studies that consider LMC models with virial masses of order $10^{11}$ $M_{\odot}$. Note that the MW is still modeled as a rigid spherical mass distribution. In this review, these studies are referred to as Model C: Moving Center of Mass (COM).

Most Model C studies describe the LMC and SMC using Plummer potentials (Eqs. B7–B10), but some Model C studies also adopt cosmological potentials to describe the LMC and SMC and also include a disk potential. In these cases, the LMC and SMC are either modeled as live N-body systems or rigid potentials. A further subset of these studies adopt a wide range of LMC and MW masses to explore the range of orbital solutions that are possible (Besla et al. 2007; Shattow and Loeb 2009; Kallivayalil et al. 2013; Gómez et al. 2015; Guglielmo et al. 2014; Patel et al. 2020; Cullinane et al. 2022a, b).

The outline for the following sections is as follows. To broadly understand the orbit solutions enabled by the Model C framework, a standard set of NFW halo profiles are defined to describe the MW, LMC, and SMC (Sects. 6.1, 6.2 and 6.4.1, respectively). This enables the gravitational forces to be computed across the orbital parameter space in a standardized fashion.

After defining the MW models, Sect. 6.1 describes the LMC orbit solution in the Model C: Fixed COM framework. The orbits are explored without dynamical friction to illustrate the impact of adopting cosmological halo density profiles on the LMC orbit solution.

After defining the LMC models, Sect. 6.3 explores the impact of the LMC mass on LMC orbit solutions in the context of: (1) dynamical friction (Fixed COM framework); and (2) the motion of the LMC–MW orbital barycenter. The intuition from both effects are combined to understand the LMC orbit solutions in the Model C: Moving COM framework in Sects. 6.3.3 and 6.4, respectively.

**Summary:** Model C studies adopt cosmologically motivated dark matter density profiles for the MW, such as NFW or Hernquist profiles. These models are less massive than Model B Isothermal Spheres, enabling the LMC/SMC to travel to larger Galactocentric distances in the past. All Model C studies adopt rigid MW halos and account for dynamical friction, but more recent Model C studies include the gravitational force of the LMC acting on the MW (Moving COM), whereas older studies did not (Fixed COM).

### 6.1 MW mass profile

In this review, a fiducial set of three MW models are adopted with virial masses of $M_{\rm vir} = [1, 1.5, 2] \times 10^{12}$ M$_\odot$. The virial mass is defined as the mass contained within a radius at which the average dark matter density of the halo is $\Delta_{\rm vir}$ ~360 times the average dark matter density of the universe (assuming $\Omega_m = 0.3$, $\Omega_\Lambda = 0.7$, $h = 0.7$; see Appendix Eq. E20). The MW is modeled as an NFW halo with parameters given in Table 8. The relevant equations are described in Appendix E. The NFW profile is truncated at the virial radius, such that, at larger radii, the MW is treated as a point mass.

For each model, the disk of the MW is modeled as a Miyamoto-Nagai disk (Miyamoto and Nagai 1975) and the bulge is modeled as an H90 profile. Model parameters are given in Table 8 and the relevant equations are described in Appendix D and Appendix C.

The corresponding mass profiles and rotation curves for the three MW models are illustrated in Fig. 11. The total mass of each model within 50 kpc is given in Table 8 and agrees with observational constraints by design.

The escape speed at the location of the LMC is listed in the right panels for each model. Note that for model MW1, the escape speed is roughly the same as the LMC velocity of 380 km/s reported by Kallivayalil et al. (2006b). This fact was the original motivation for Besla et al. (2007) to revisit the orbital history of the Clouds. The recent Gaia/HST LMC velocities are lower (~320 km/s), which means that the LMC is bound to the MW in all models considered in this review (with the exception of hyperbolic orbit solutions in Model A; Sect. 4). However, the LMC's speed is still sufficiently high that the LMC is moving supersonically (Mach ~ 2) through the MW's CGM and should generate a large scale bow shock (Setton et al. 2023; Carr et al. 2025; Lucchini et al. 2026).

The three MW models in this review are chosen to be representative of the range of MW models adopted by Model C studies, and should not be taken as a definitive statement on the mass profile for the real MW.

In particular, adiabatic contraction of the halo owing to the growth of the disk is not included in this analysis. Instead, the concentration parameters ($C_{\rm vir}$) are selected, such that each model reaches a peak rotation speed of ~ 240 km/s at 8.5 kpc. The required concentration parameters are higher than the average values for cosmological halos at these masses (~9.3–9.9; Klypin et al. 2011). However, several authors have successfully reproduced the rotation curve of the MW using halo models with adiabatic

**Table 8** Milky Way models

| Property | MW1 | MW2 | MW3 |
|---|---|---|---|
| $M_{vir}$ ($10^{12}M_\odot$) | 1.0 | 1.5 | 2.0 |
| $C_{vir}$ | 22 | 18 | 16 |
| $R_{vir}$ (kpc) | 261 | 299 | 329 |
| $M_{disk}$ ($10^{10}M_\odot$) | 7 | 7 | 7 |
| $R_{disk}$ (kpc) | 4 | 4 | 4 |
| $z_{disk}$ (kpc) | 0.53 | 0.53 | 0.53 |
| $M_{bulge}$ ($10^{10}M_\odot$) | 1 | 1 | 1 |
| $R_{bulge}$ (kpc) | 0.7 | 0.7 | 0.7 |
| $M_{tot}$ (< 50 kpc) ($10^{10}M_\odot$) | 47 | 57 | 65 |

The halo is modeled following a Navarro et al. (1996) profile. The virial concentration parameter, $C_{vir}$, is computed, such that the rotation curve reaches ∼240 km/s at 8.5 kpc. Adiabatic contraction is not included in these models; as such a higher concentration is required than expected from average halos in cosmological simulations. The disk is modeled following a Miyamoto and Nagai (1975) profile with disk scale length, $R_{disk}$, and scale height, $z_{disk}$. The bulge is modeled following an Hernquist (1990) profile with scale length, $R_{bulge}$. See the Appendix sections Appendix D, Appendix C and Appendix E for relevant equations. $M_{tot}$ refers to the total mass enclosed within 50 kpc (disk + bulge + halo). Observationally, this value ranges from 2–8 $\times 10^{11}$ $M_\odot$ (Eadie and Jurić 2019) contraction for the appropriate range of concentrations (Besla et al. 2007; Shattow and Loeb 2009; Kallivayalil et al. 2013; Patel et al. 2017, 2020). As such, it is not claimed here that the MW rotation curve is inconsistent with typical parameters for cosmological NFW halos. The LMC orbits presented in this analysis are broadly consistent with the findings of studies that use adiabatic contraction, validating the use of a simpler framework to describe the MW halo for this class of models.

In Fig. 12, the orbit of the LMC is computed using the three fiducial NFW MW models (Table 8) in the Model C: Fixed COM framework, where the halo is rigid and the center of mass is fixed in time. The LMC orbit is also plotted in the Isothermal Sphere MW model of Model B ($V_C$ = 239 km/s). The SMC is ignored in these calculations.

All orbits are computed using the HI center velocities (Table 2), where results are similar if the Phot center velocity is used instead. The exception is the orbit marked "B12", which is computed using the LMC velocity vector from Kallivayalil et al. (2006b) and represents the LMC orbit in the N-body simulation of Besla et al. (2012).

The goal of Fig. 12 is to illustrate the change in the LMC orbit that results from different choices of halo profile (Model B Isothermal Sphere halo vs. Model C cosmological NFW halo). Dynamical friction is not included in this orbit calculation, and the MW does not respond to the LMC's gravitational force. As such, the LMC's mass is irrelevant to this calculation. The LMC's equation of motion is thus:

$$\mathbf{a}_{\rm L} = \mathbf{a}_{{\rm MW}_{\rm h+d+b}}(\mathbf{r}_{\rm L-MW}) \quad (8)$$

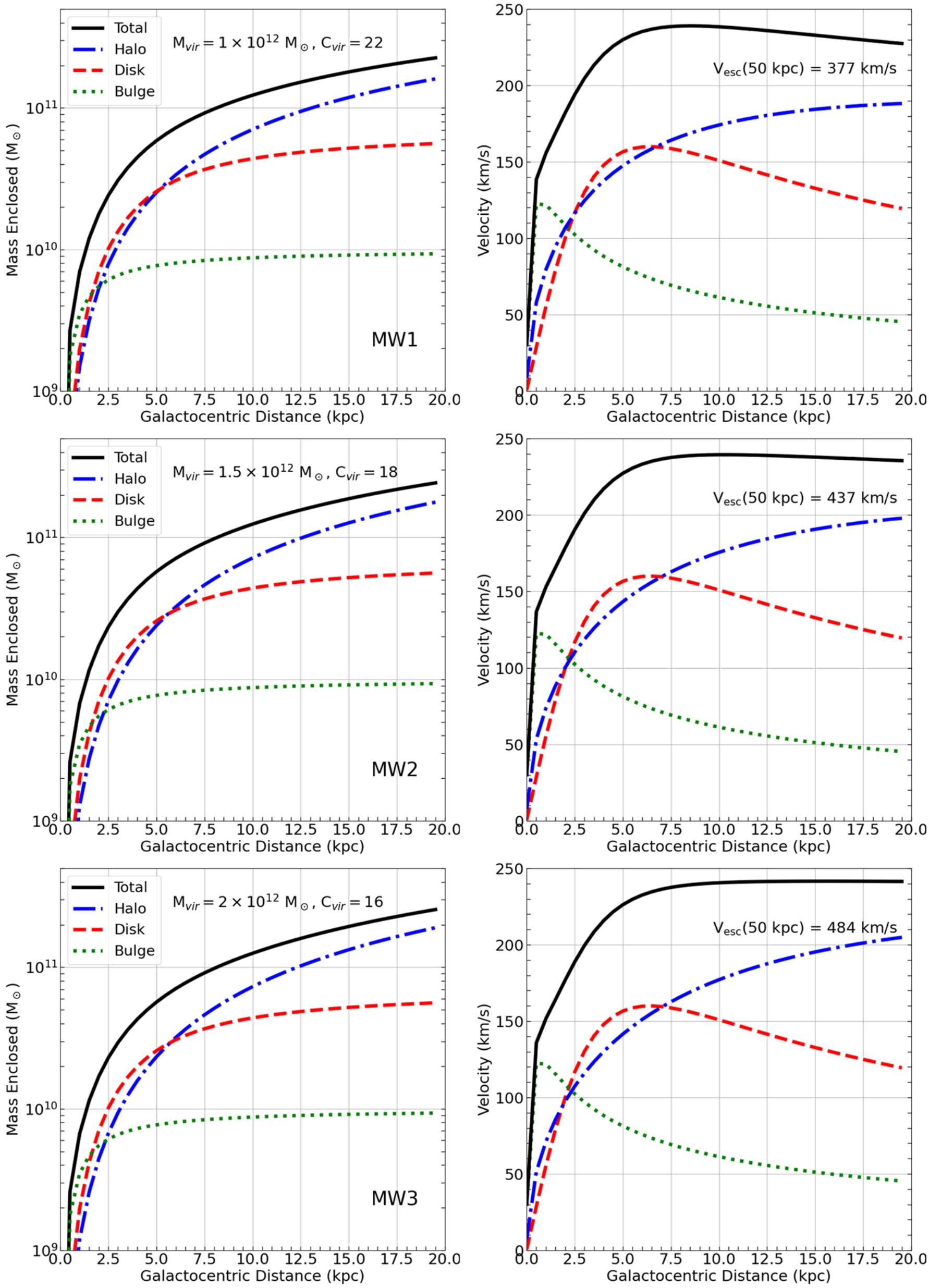

**Fig. 11** Mass profile (left) and rotation curve (right) for the three MW models outlined in Table 8 (one per row). Halos are modeled as NFW profiles with different virial masses, as marked. Different lines denote: halo profiles (blue, dash-dotted lines), Miyamoto and Nagai (1975) disk profiles (red, dashed lines), and H90 bulge profiles (green, dotted lines). Disk and bulge profiles are constant across all three models. The concentration parameters ($C_{\rm vir}$) are adjusted to reach a peak rotation speed of $\sim 240$ km/s at 8.7 kpc. The escape speed at the location of the LMC (50 kpc) is marked for each model

where $\mathbf{a}_{\mathrm{MW_{h+d+b}}}$ is either the acceleration from an Isothermal Sphere (Eq. A4) or the combined acceleration for an NFW halo, Miyamoto and Nagai (1975) disk, and H90 bulge ($\mathbf{a}_{\mathrm{MW_{h+d+b}}} = \mathbf{a}_{\mathrm{NFWhalo}} + \mathbf{a}_{\mathrm{MNdisk}} + \mathbf{a}_{\mathrm{H90bulge}}$; see relevant equations in the Appendix).

The LMC orbits are computed backward in time for the past 5 Gyr. It has been shown using cosmological simulations that analytic orbit reconstruction in static potentials are generally unreliable past 5 Gyr as they do not account for the growth of the primary halo and perturbations from substructure (e.g. Lux et al. 2010; D'Souza and Bell 2022).

Figure 12 illustrates the dramatic orbit difference that results from changing the halo model (Model B to Model C). Recent advances in our understanding of how dark matter is distributed about galaxies result in an LMC orbit that is significantly more eccentric than solutions using the Model B framework. This was first illustrated in Besla et al. (2007).

As the MW halo mass increases, the LMC orbit approaches the Model B, Isothermal Sphere solution, which is the most massive model considered by any study. However, even in the most massive NFW model (MW2 = $2\times10^{12}$ $\mathrm{M}_\odot$), the LMC orbital period and apocenter distance are still twice as large as those of the Isothermal Sphere solution. In the lowest mass NFW MW model (MW1 = $10^{12}$ $\mathrm{M}_\odot$) the LMC is on first infall, crossing the virial radius ($R_{\mathrm{vir,MW1}} = 260$ kpc) almost 1.5 Gyr ago.

Plotted for reference is the LMC orbit from the simulations of (Besla et al. 2012, B12) and (Besla et al. 2010, B10), where the MW model is similar to the MW2, NFW model. However, the LMC velocity vector is that of K06, which is substantially higher than the HI Center and Phot Center velocities (Table 3). As such, in the B12/B10 canonical studies, the LMC is on first infall even in fairly massive MW models, owing to both a change in the halo model and a substantially faster LMC.

The next sections examine the impact of LMC mass in the orbit calculations for Model C - both in terms of dynamical friction and the motion of the center of mass of the MW in response to the LMC's gravitational force (Gómez et al. 2015).

**Summary:** Three fiducial MW models are defined using NFW halos with virial masses of $M_{\mathrm{vir}} = 1, 1.5$ and $2\times10^{12}$ $\mathrm{M}_\odot$ (MW1, MW2, MW3, respectively). Ignoring dynamical friction and the COM motion of the MW, LMC orbit solutions in the Model C: Fixed COM framework are significantly more eccentric than Model B solutions (Isothermal Sphere MW models). In MW1, the LMC is on first infall. Advances in modeling the structure of dark matter halos have significantly changed our understanding of the orbital history of the Clouds.

### 6.2 LMC mass profile

In this review, a fiducial set of LMC mass profiles is defined in order to study the orbit solutions for Model C in a standardized manner.

Observations of the LMC's rotation curve and total mass at different radii are used to constrain the parameters for a suite of analytic NFW halo profiles that span the range of LMC halo masses adopted by Model C studies ($M_{\mathrm{vir}} = 3, 5, 8, 10, 18, 25 \times 10^{10}$

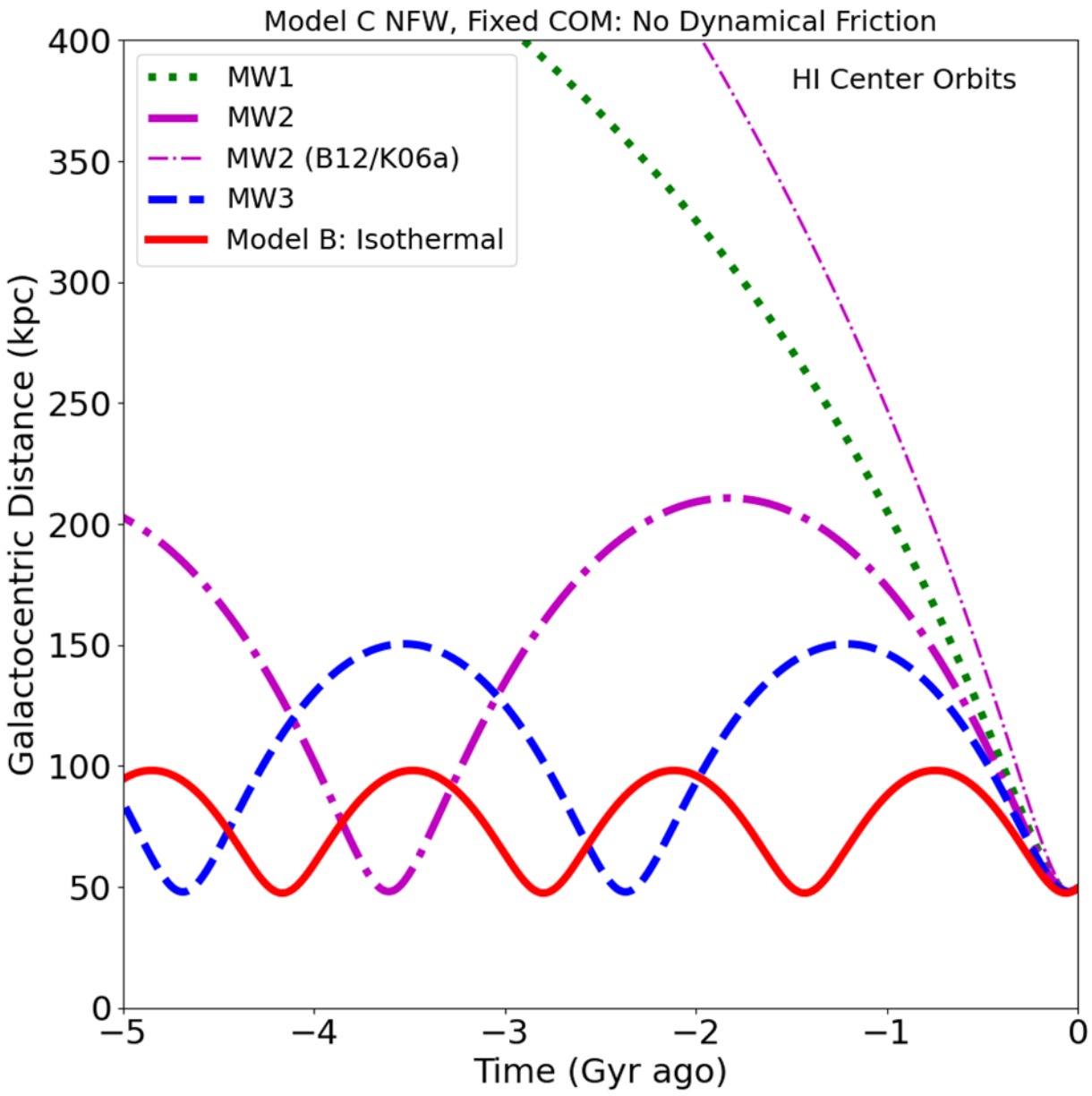

**Fig. 12** Impact of the MW model on the orbit of the LMC: the Galactocentric distance of the LMC is plotted as a function of time in the past, where the present day is at t=0. The orbital history is computed using the HI Center velocity in the Model C: Fixed COM framework, without dynamical friction, using three cosmologically motivated MW models (MW1, MW2, MW3; Table 8; NFW halos+disk+bulge). Results are similar if the LMC Phot Center velocity is used instead. The orbit solution for the Model B Isothermal Sphere model ($V_C$ = 239 km/s) is also plotted (red solid line); the total mass enclosed within 100(260) kpc is 1.3(3.4) $\times$ $10^{12}$ M$_\odot$ for this MW model. The (Besla et al. 2012, B12) LMC orbit, using Kallivayalil et al. (2006b) velocities and the MW2 model, is also plotted for reference. The NFW dark matter profile is truncated at the virial radius, but the Isothermal Sphere profile is not truncated, in keeping with the literature. This figure illustrates that changing the model for the dark matter distribution of the MW, specifically moving away from Isothermal Spheres (Model B) to cosmologically motivated halos (Model C), is one of the main factors that has changed our understanding of the orbital history of the LMC about the MW

M$_\odot$). Parameters for all six LMC models are summarized in Table 9. In this review, the virial mass is assumed to be the infall mass of the LMC.

As in the case of the MW, the LMC dark matter halo density profile is truncated at the LMC's virial radius and the LMC is treated as a point mass outside this radius (total mass = $M_{\rm halo} + M_{\rm disk}$). The concentration parameter, $C_{\rm vir}$, is computed following equation 10 in Klypin et al. (2011) for the mean concentration of isolated cosmological halos at redshift 0:

$$C_{\rm vir} = 9.60\left(\frac{M_{\rm vir}}{10^{12}h^{-1}M_\odot}\right)^{-0.075} \tag{9}$$

The LMC halos are not adiabatically contracted in response to the growth of the LMC's disk.

The LMC disk is modeled using a (Miyamoto and Nagai 1975) profile with total mass 4.2 $\times$ $10^9$ M$_\odot$. The disk scale length is 2.0 kpc and the disk scale height is 0.2 times the scale length. The corresponding Miyamoto and Nagai (1975) disk density

**Table 9** LMC NFW halo models

| $M_{vir}$ ($10^{10}$M$_\odot$) | $C_{vir}$ | $R_{vir}$ (kpc) | $M_{tot}(<8.7$ kpc) ($10^{10}$M$_\odot$) | $M_{tot}(<13$ kpc) ($10^{11}$M$_\odot$) | $M_{tot}(<30$ kpc) ($10^{11}$M$_\odot$) | $V_{peak}$ (km/s) |
|---|---|---|---|---|---|---|
| 3 | 12.8 | 81 | 0.8 | 1.2 | 2.0 | 69 |
| 5 | 12.3 | 96 | 1.0 | 1.5 | 2.8 | 73 |
| 8 | 11.9 | 113 | 1.2 | 1.8 | 3.7 | 76 |
| 10 | 11.7 | 121 | 1.3 | 2.0 | 4.2 | 78 |
| 18 | 11.2 | 148 | 1.6 | 2.6 | 6.0 | 82 |
| 25 | 10.9 | 165 | 1.8 | 3.0 | 7.3 | 85 |
| Observed | | >18.5 [1] | 1.7 ±0.7 [2] | 2.3–3.08 [3] | 4.7–7.92 [4,5] | 91.7 ±18.8 [2] |

M(<8.7), M(<13), and M(< 30) refer to the total mass enclosed within 8.7, 13, and 30 kpc, respectively (disk + halo). In all models, the LMC disk is modeled as a Miyamoto and Nagai (1975) profile with total mass of $M_{disk} = 4.2 \times 10^9$ M$_\odot$, $R_{disk} = 2.0$ kpc, and $z_{disk} = 0.4$ kpc. References: [1] Mackey et al. (2016); [2] van der Marel and Kallivayalil (2014); [3] Watkins et al. (2024), but note that error range may be larger Roychowdhury et al. (2026); [4] Koposov et al. (2023); [5] Warren et al. (2025)

profile matches that of an exponential disk profile with a disk scale length of 1.67 kpc, as found in Choi et al. (2018). This model yields a total disk mass of $\sim 3.27 \times 10^9$ M$_\odot$ within 8.7 kpc, consistent with the observed stellar mass ($\sim$2.7 $\times 10^9$ M$_\odot$; van der Marel et al. 2009) and gas mas (CO+HI mass $\sim 5.2 \times 10^8$ M$_\odot$; Meixner et al. 2013 and references in their Table 1) of the LMC disk.

The mass profiles and rotation curves for the six LMC models (NFW halo + Miyamoto-Nagai Disk) explored in this review are illustrated in Fig. 13. Selected observational constraints on the LMC's mass profile and rotation curve are listed in Table 9 and marked in the figures.

Examining Table 9, the mass profiles, and the rotation curves, it is apparent that a high mass LMC model ($M_{vir} > 10^{11}$ M$_\odot$) is needed to satisfy observational constraints, such as: the LMC's rotation curve (van der Marel and Kallivayalil 2014); the properties of the Orphan-Chenab Stream, which is perturbed by the LMC (Warren et al. 2025; Koposov et al. 2023; Erkal et al. 2019); the 3D kinematics of LMC globular clusters (Watkins et al. 2024, but see caveats noted in Roychowdhury et al. 2026); and estimates from precision timing of millisecond pulsars in the solar neighborhood in comparison with simulations of the LMC–Sgr–MW interaction (Donlon et al. 2026 $4.1 \times 10^{10}$M$_\odot$ within 16.6 kpc). The observational constraints require the LMC to be a dark matter dominated galaxy (1:3–7 mass ratio of baryons:dark matter within 8.7 kpc) with a high halo concentration ($C_{vir} \sim 11$–12); the LMC is well-described using a cuspy, NFW dark matter profile.

Several studies in Model C have used Plummer or H90 models to model the LMC. The choice of an NFW halo for the LMC in this review does not change the broad conclusions for the Model C orbit solutions. Moreover, by choosing one consistent set of halo profiles, other parameters that impact the orbit solutions (such as the LMC:MW mass ratio, the LMC's velocity vector, etc.) can be isolated.

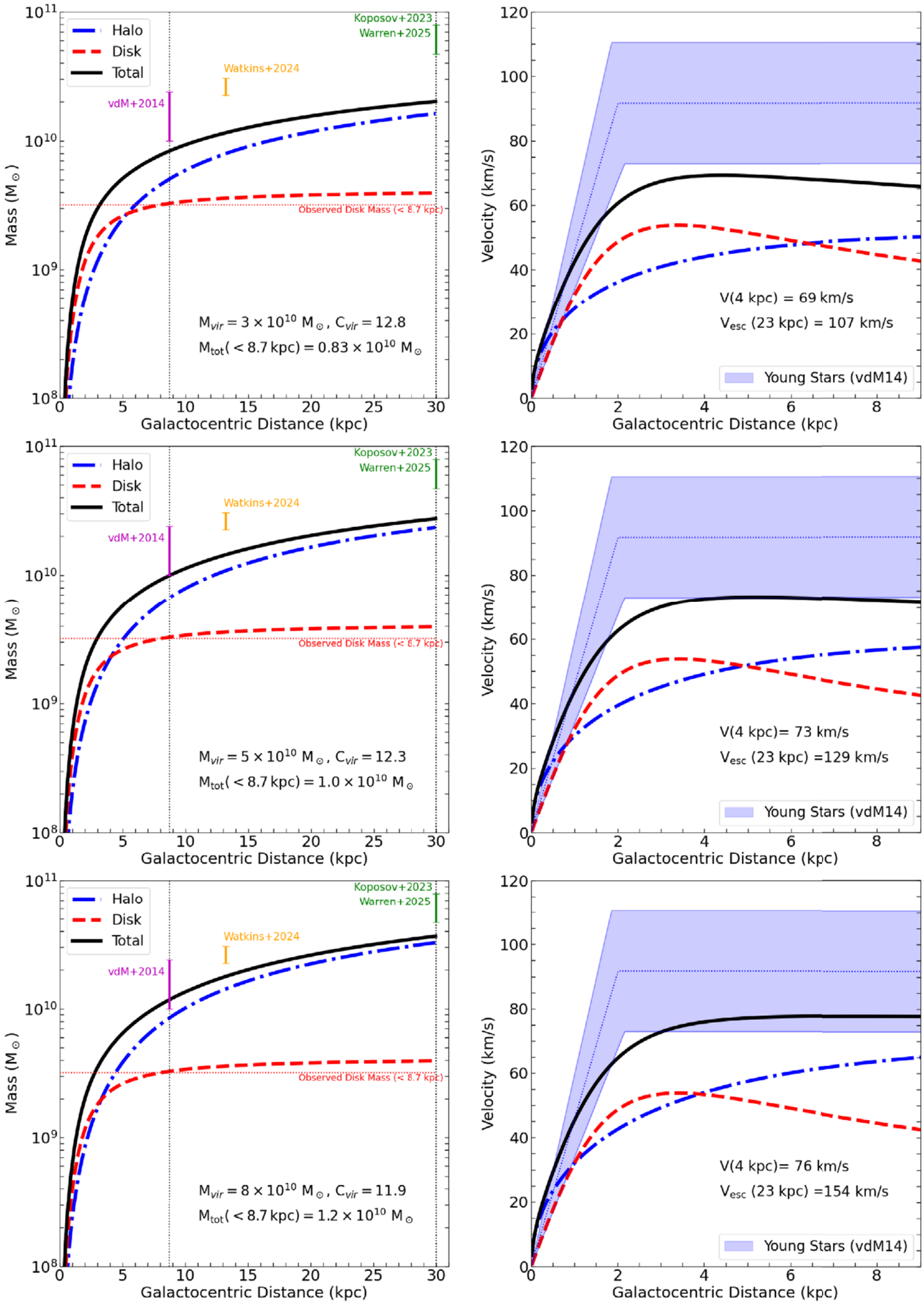

**Fig. 13** Mass profile (left) and rotation curve (right) for the LMC, assuming an NFW mass profile for the dark matter halo with six different virial masses, ($M_{\rm vir}$ = 3,5,8,10,18,25 $\times 10^{10}$ $M_{\odot}$, one model per row). Model parameters and observational constraints are summarized in Table 9. The LMC halo is modeled as an NFW profile (blue, dash-dotted lines). The disk profile (red, dashed line) follows a Miyamoto and Nagai (1975) profile that agrees with the observed disk mass (gas+stars) of $3.2\times10^9$ $M_{\odot}$ (red dotted line) at 8.7 kpc (black, vertical dotted line). The observational constraint on the peak rotation speed using young stars (van der Marel and Kallivayalil 2014) is indicated by the blue shaded regions in all panels, where the mean value is denoted by the dotted blue line ($V_{\rm peak}$ = 91.7 ± 18.8 km/s). High mass LMC models ($M_{\rm vir} > 10^{11}$ $M_{\odot}$) are needed to satisfy all observational constraints

**Fig. 13** continued

**Summary:** Six LMC models are defined using NFW halos with virial masses of $M_{\rm vir} = 3, 5, 8, 10, 18, 25 \times 10^{10}$ M$_\odot$. These models are used throughout this review to compute the LMC orbits. LMC models with $M_{\rm vir} > 10^{11}$ M$_\odot$ and concentrations of $C_{\rm vir}$ ~11 are required to match observational constraints on the mass profile and rotation curve of the LMC. The LMC is dark matter dominated and well-described using a cuspy dark matter profile.

### 6.3 The LMC's orbital history: impact of LMC halo mass

The LMC's halo mass at infall is critical to two key processes that impact the orbital solutions: (1) dynamical friction (Sect. 6.3.1); and (2) the motion of the center of mass of the MW about the MW–LMC orbital barycenter (Sect. 6.3.2).

The Model C: Fixed COM framework accounts for dynamical friction only. The Model C: Moving COM framework accounts for both effects. In the following, we illustrate the impact of each effect on the LMC's orbit, before combining the effects to understand the full orbit solutions in the Model C: Moving COM framework in Sect. 6.3.3.

#### 6.3.1 LMC mass and dynamical friction - Model C: fixed COM framework

In this section, the impact of dynamical friction on the orbit of the LMC is explored in the Model C: Fixed COM framework, where the MW and LMC are modeled using NFW halos (Sects. 6.2 and 6.1) and the center of mass of the MW does not move in response to the infall of the LMC. The LMC HI center velocity is adopted at present day (Table 2).

The Chandrasekhar formula (Chandrasekhar 1943) is used to describe dynamical friction for a host and satellite system, where both are modeled as NFW halos. The relevant equations are outlined in Appendix F. The equation of motion for the LMC is thus:

$$\mathbf{a}_{\rm L} = \mathbf{a}_{{\rm MW}_{\rm h+d+b}}(\mathbf{r}_{\rm L-MW}) + \mathbf{a}_{\rm DF,NFW,MW}(\mathbf{r}_{\rm L-MW}, \mathbf{v}_{\rm L-MW}) \qquad (10)$$

where $\mathbf{r}_{\rm L-MW}$ is the Galactocentric positon vector of the LMC COM from the MW COM, where the latter is fixed at (0,0,0). $\mathbf{v}_{\rm L-MW}$ is the Galactocentric velocity vector of the LMC with respect to the MW, which is fixed at (0,0,0).

In Fig. 14, the orbital history of the LMC is computed using MW models MW1 and MW2, and 4 LMC models (NFW virial masses: 3, 8, 18, 25 $\times 10^{10}$ M$_\odot$). The orbits are computed with dynamical friction, following the equation above. An orbit without dynamical friction is also shown for comparison.

In the case of MW1 ($M_{\rm vir} = 10^{12}$ M$_\odot$), the LMC orbit is a first infall scenario (as in Fig. 12). For MW1, in all cases the LMC has only one pericentric approach to the MW within 5 Gyr and crosses inward of the virial radius ~1 Gyr ago. As concluded by Besla et al. (2007), dynamical friction does not play a significant role in changing the orbital history of the LMC for low mass MW models.

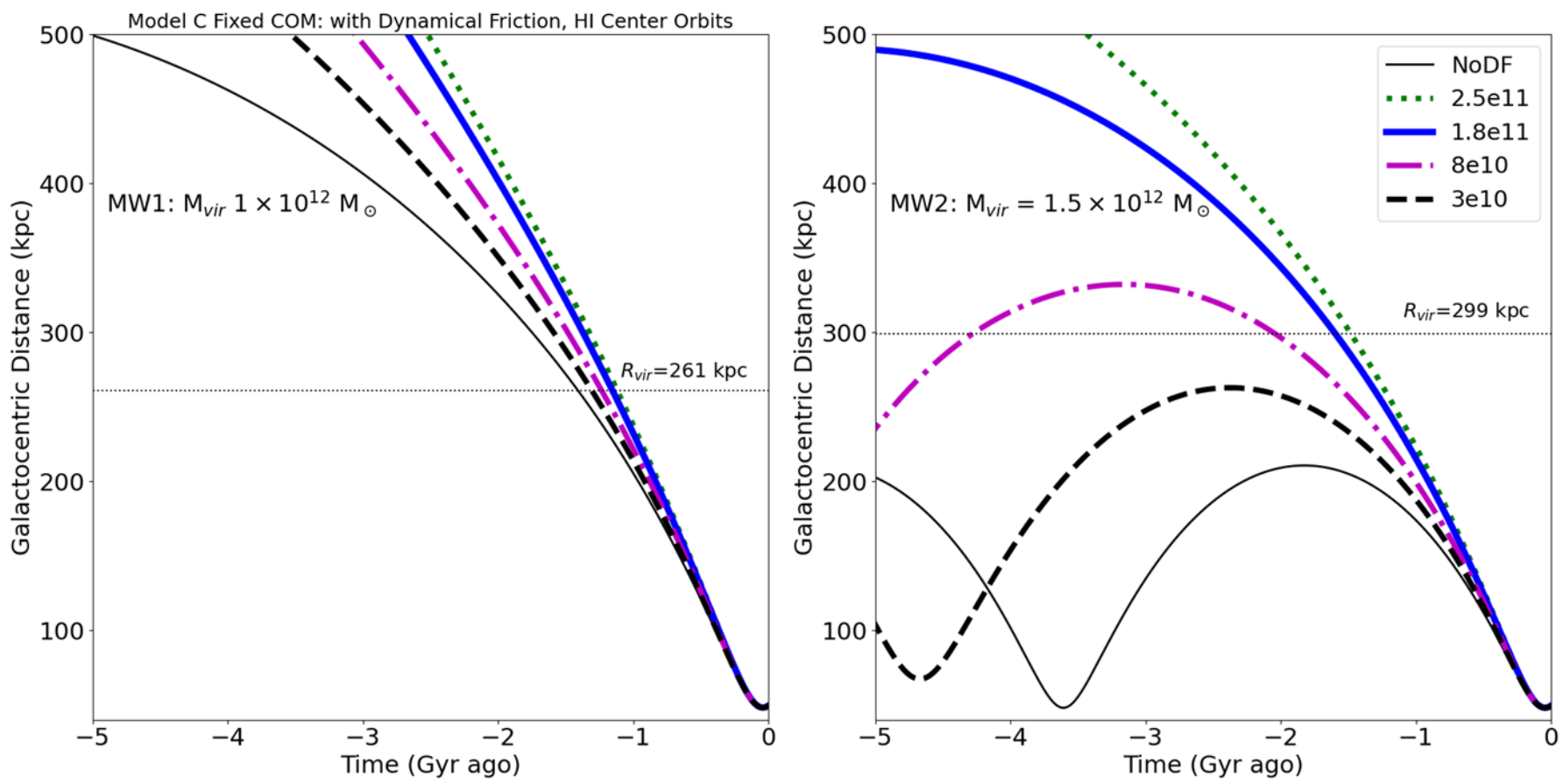

**Fig. 14** Impact of dynamical friction on the orbital history of the LMC using HI Center velocities in the Model C: Fixed COM framework. The Galactocentric distance of the LMC is plotted as a function of time in the past, where the present day is at t=0. The orbits are computed for MW NFW models MW1 (left) and MW2 (right); see Table 8. The virial radius for each MW model is marked by the dotted lines. The LMC is modeled as an NFW halo with virial masses (infall masses): 1, 8, 18, 25 $\times 10^{10}$ $M_\odot$ (Table 9). The gravitational force of the LMC acting on the MW is not included in these calculations (the MW COM is fixed). Dynamical friction is included in all cases except that marked by "NoDF". This figure illustrates that dynamical friction changes the orbital solution for the LMC. This effect is most important for MW models with virial masses larger than $10^{12}$ $M_\odot$ and high mass LMC models. For low mass MW models (MW1), the LMC orbital solutions are consistently on first infall, regardless of LMC mass (only one pericentric approach to the MW within 5 Gyr). For higher mass MW models (MW2), the LMC is on first infall if its initial mass is larger than $8 \times 10^{10}$ $M_\odot$. As the MW mass is increased the frequency of first infall solutions decreases, but the LMC's orbital period and eccentricity are always larger with dynamical friction

For the higher mass MW models, the LMC's orbit is less eccentric than in MW1. Consequently, the LMC spends more time in the MW's halo, and therefore, the LMC infall mass is more important to the orbital solution, owing to dynamical friction. Orbit solutions in MW2 for low mass LMC models (e.g., $M_{\rm vir} = 3\times10^{10}$ $M_\odot$) can complete an orbit about the MW within 5 Gyr. However, higher mass LMC models have longer orbital periods, where LMC models with virial mass larger than $8 \times 10^{10} M_\odot$ are on first infall.

An important caveat to this analysis is that the LMC virial mass (infall mass) is fixed as a function of time in the Model C framework. The Chandrasekhar prescription for dynamical friction does not account for mass loss of the satellite due to the tidal forces of the host. This is true for all orbital solutions in Model C and Model B. N-body simulations of Garavito-Camargo et al. (2021) find that the LMC can lose 40–50% of its infall mass, even in a first infall scenario. This will be discussed in Sect. 7.

**Summary:** In the Model C: Fixed COM framework, dynamical friction causes the LMC orbit to become more eccentric as a function of increasing LMC mass, reaching larger distances over longer periods of time. In the lowest mass MW model, the LMC is on first infall, regardless of infall mass.

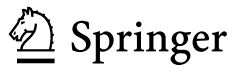

### 6.3.2 LMC mass and the motion of the orbital barycenter of the MW–LMC system

In the Models A, B, and C: Fixed COM frameworks, the gravitational force of the LMC on the MW is ignored. This is appropriate for cases, where the mass ratio between the LMC:MW is small. However, as the mass ratio increases (LMC:MW >1:10), the assumption that the LMC is a negligible perturber of the MW is no longer valid.

Gómez et al. (2015) were the first to illustrate that the acceleration of the MW's center of mass owing to the LMC is non-negligible. They find that the orbital barycenter of the MW–LMC system can be displaced by 5–40 kpc and 15–100 km/s if the LMC is modeled as a Plummer potential with total mass ranging from $3-25\times10^{10}$ $M_\odot$. As such, Gómez et al. (2015) argue that the approximation of an inertial Galactocentric reference frame is not valid in the presence of a massive LMC.

Such studies are categorized as Model C: Moving COM. In this scenario, the orbital history of the LMC is computed in the Model C framework (analytic, non-deforming NFW halos), but changes in the orbital barycenter of the MW–LMC system as the LMC orbits the MW are also accounted for.

The orbit of the MW is computed using the following equation of motion:

$$\mathbf{a}_{\rm MW} = \mathbf{a}_{\rm L_{h+d}}(\mathbf{r}_{\rm MW-L}) \tag{11}$$

where $\mathbf{a}_{\rm L_{h+d}} = \mathbf{a}_{\rm NFW} + \mathbf{a}_{\rm MN}$, following an NFW profile for the halo and Miyamoto and Nagai (1975) profile for the disk, as described in Table 9 and equations in Appendix E and Appendix D. Note that the distance between the LMC and the MW ($\mathbf{r}_{\rm MW-L}$) accounts for the motion of the MW's center of mass.

As an exercise, dynamical friction acting on the LMC is first ignored to isolate the impact of the change in the orbital barycenter. The equation of motion for the LMC is thus:

$$\mathbf{a}_{\rm L} = \mathbf{a}_{\rm MW_{h+d+b}}(\mathbf{r}_{\rm L-MW}) \tag{12}$$

In Fig. 15, the relative separation between the LMC and the MW is plotted as a function of look-back time, using 4 LMC models (3, 8, 18, 25 $\times10^{10}$ $M_\odot$; see Table 9).

Because dynamical friction is not included in the orbit calculation, differences in the orbits for a given MW model are a direct result of the changing MW–LMC orbital barycenter. As the mass of the LMC increases, the larger the deviation of the orbit relative to the case, where the MW's center of mass is fixed. The shift of the orbital barycenter is maximized in the case of a low mass MW and a high mass LMC. This is the opposite trend caused by dynamical friction (compare to Fig. 14).

In Model C studies that consider this effect (marked as "Moving Center of Mass" in Table 6), the MW is treated as a rigid sphere. As such, when the center of mass of the MW moves, the entire MW halo moves.

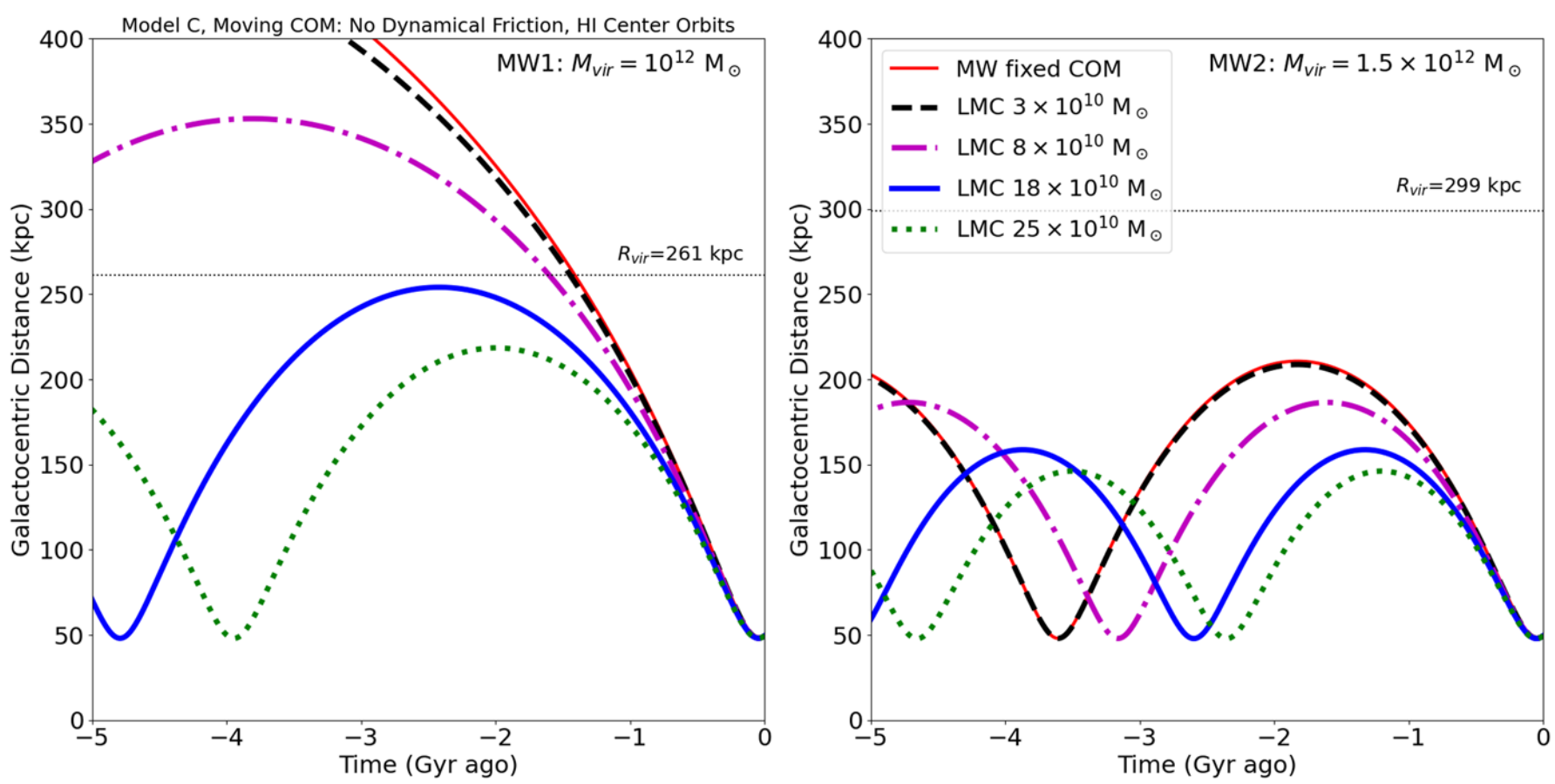

**Fig. 15** Impact of changes in the orbital barycenter of the MW–LMC system on the orbit of the LMC, as a function of LMC and MW mass (Model C: Moving COM). This figure is the same as Fig. 14, except that dynamical friction is not included and the MW is treated as a rigid body that is allowed to move in response to the gravitational force exerted by the LMC. Because there is no dynamical friction, it is only the motion of the MW COM about the new MW–LMC orbital barycenter that causes the orbits to differ. A reference orbit computed in a MW whose COM is fixed at (0,0,0) is plotted as the red solid line (MW fixed COM; same as in Fig. 12). The changes in the orbit are minimized as the mass of the LMC decreases or as the MW mass increases (right panel). Orbits are computed using the HI Center velocities, but results are similar if the Phot Center velocities are used instead

**Summary:** LMC orbits were explored in the Model C framework, ignoring dynamical friction, but allowing the MW to move in response to the LMC's gravitational force. For a given MW mass, as the LMC mass increases, the relative separation between the LMC and MW decreases. Consequently, the orbit becomes less eccentric with increasing LMC mass, reaching smaller distances over a longer period of time - orbits are less likely to be first infall solutions than if this effect were ignored. In MW1, only the lowest mass LMC orbit solutions are first infall. Interestingly, this behavior is the *opposite* effect of dynamical friction, which makes orbits more eccentric.

### 6.3.3 LMC orbits in the model C: moving COM framework

Three key factors have contributed to changes in the modeling of the orbital history of the LMC about the MW: (1) the velocity of the LMC; (2) our understanding of the MW's dark matter distribution; and (3) our understanding of the LMC's dark matter distribution.

The LMC velocity vector is currently measured to be larger than adopted by earlier studies (most Model A, B studies). This means that the LMC can orbit at larger Galactocentric distances than previously believed. However, the LMC is unlikely to be unbound, even in the lightest MW dark matter halo models considered by Model C studies.

In Model C, dark matter halos are recognized to be poorly-described as Isothermal Spheres at large distances. Cosmologically motivated halos (e.g., H90, NFW) are adopted instead. Such MW models enclose less mass at larger radii, allowing the LMC to reach larger Galactocentric distances backward in time.

High LMC masses are adopted by many Model C studies (specifically the "High" mass ratio studies in Table 6). The LMC halo mass at infall is cosmologically expected to be at least a factor of 10 larger than the LMC masses considered in Model A or B. The rate of decay of the LMC's orbit about the MW is strongly dependent on the mass of the LMC, where a higher mass results in more rapid decay owing to dynamical friction. In a backward orbital integration scheme, stronger dynamical friction implies that the satellite must be at larger distances in the past to be at a given distance today. Consequently, the LMC orbit become more eccentric as the LMC's mass increases.

In addition, a subset of Model C studies also recognized that the increased mass of the LMC implies that the orbital barycenter of the MW–LMC system no longer coincides with the center of mass (COM) of the MW disk. As such, the MW COM will move in response to the gravitational force of the LMC ("Moving Center of Mass" in Table 6). Consequently, as the LMC mass increases, the relative separation between the LMC and MW decreases.

With this intuition in mind, the historical and on-going discussion surrounding the orbital history of the LMC–MW system can be understood.

The orbital solutions for the Model C: Moving COM framework account for both dynamical friction on the LMC's orbit owing to its motion through the MW's halo, and the change in the COM position and velocity of the MW owing to the gravitational influence of the LMC. The equation of motion of the LMC becomes:

$$\mathbf{a}_{\rm L} = \mathbf{a}_{\rm MW_{h+d+b}}(\mathbf{r}_{\rm L-MW}) + \mathbf{a}_{\rm DF,NFW,MW}(\mathbf{r}_{\rm L-MW}, \mathbf{v}_{\rm L-MW}) \tag{13}$$

where $\mathbf{v}_{\rm L-MW}$ in Eq. (F29) refers to the velocity of the LMC relative to the velocity of the moving MW's COM. The equation of motion of the MW is the same as in Eq. (11).

In Fig. 16, the orbital history of the LMC is computed, following the above equation of motion, for the three fiducial MW models and 4 LMC models (3, 8, 18, 25 $\times 10^{10}$ $M_\odot$). In all cases, the present-day LMC velocity is adopted as the HI center velocity vector (Table 2). In Sect. 8, observational uncertainties in the velocity vector of the LMC are accounted for, including the choice of dynamical center.

For the lowest MW model (MW1), there is little difference between the orbital histories. The LMC is on first infall, crossing the MW's virial radius for the first time $\sim$ 1.3 Gyr ago. This result is robust against uncertainties in the dynamical friction formalism, as it holds true over a wide range of LMC masses. If the MW's virial mass is $\sim 10^{12}$ $M_\odot$, the LMC must be on first infall, regardless of infall mass.

As the mass of the MW increases, the orbit becomes less eccentric. For MW2 ($M_{\rm vir} = 1.5 \times 10^{12}$ $M_\odot$), the LMC completes an orbit within the past 5 Gyr (two pericentric approaches to the MW) only if the LMC mass is low ($\sim 3 \times 10^{10}$ $M_\odot$).

For MW3 ($M_{\rm vir} = 2 \times 10^{12}$ $M_\odot$), the LMC completes an orbit about the MW in all scenarios. However, as the mass of the LMC increases, the previous pericentric distance increases substantially. This means that even in this scenario, the LMC has only recently made its closest approach to the MW.

In all orbital solutions, the recent close encounter between the LMC and MW, 0.5 Gyr ago at a distance of ∼49 kpc, is robust to uncertainties in LMC or MW halo mass. Owing to the motion of the MW about the changing orbital barycenter, the recent pericentric approach of the LMC is also the closest distance at which the LMC has ever approached the MW.

For these same orbits, Fig. 17 illustrates the motion of the center of mass of the MW over the past 2 Gyr. The results are shown for the MW1 model, but results are similar for the other two models over this same time scale. The MW center of mass moves by 20–120 kpc and by 15–80 km/s over the past 2 Gyr owing to its motion about the new MW–LMC orbital barycenter (see Gómez et al. 2015; Garavito-Camargo et al. 2021).

In Fig. 18, the orbital path of the LMC is projected in Galactic coordinates using the MW1 and LMC $1.8 \times 10^{11}$ $M_{\odot}$ models. Orbits are computed in the Model C: Moving COM framework over the past 1 Gyr. There is no substantial change in the position of the past orbit on the sky when the LMC mass or MW mass is varied. In the top panel, the orbit using the Phot Center velocity is more offset from the Stream than the HI Center velocity orbit. This was also seen in the Model B solution (Fig. 7). This offset occurs because of the different values of $\mu_N$ and is independent of the LMC and MW mass model, provided the MW model is spherical.

In the bottom panel, the projected orbit in the Model C: Moving COM framework is compared to Model B solutions using the HI Center velocities only. The barycenter motion causes the projected Model C: Moving COM orbit to turn more toward the Stream, relative to the orbit, where the barycenter motion is fixed (Model B or Model C: Fixed COM). However, the Model C: Moving COM orbit can be reproduced by the Model B framework using an Isothermal *prolate* halo (q=1.2; see the Appendix equations A5, A6).

Růžička et al. (2007) explored LMC orbit solutions in flattened Isothermal models (Model B framework), using the older Kallivayalil et al. (2006b) LMC velocities. They concluded that *oblate* MW halos ($q < 1.0$) provided a better match to the location of the Stream, demonstrating the sensitivity of such analyses to the proper motion vector of the Clouds.

We conclude that the motion of the MW in response to the orbit of the LMC complicates the assessment of the shape of the dark matter halo of the MW. The impact of the Clouds on the structure of the halo will be discussed in the context of Model D orbit solutions (Sect. 7).

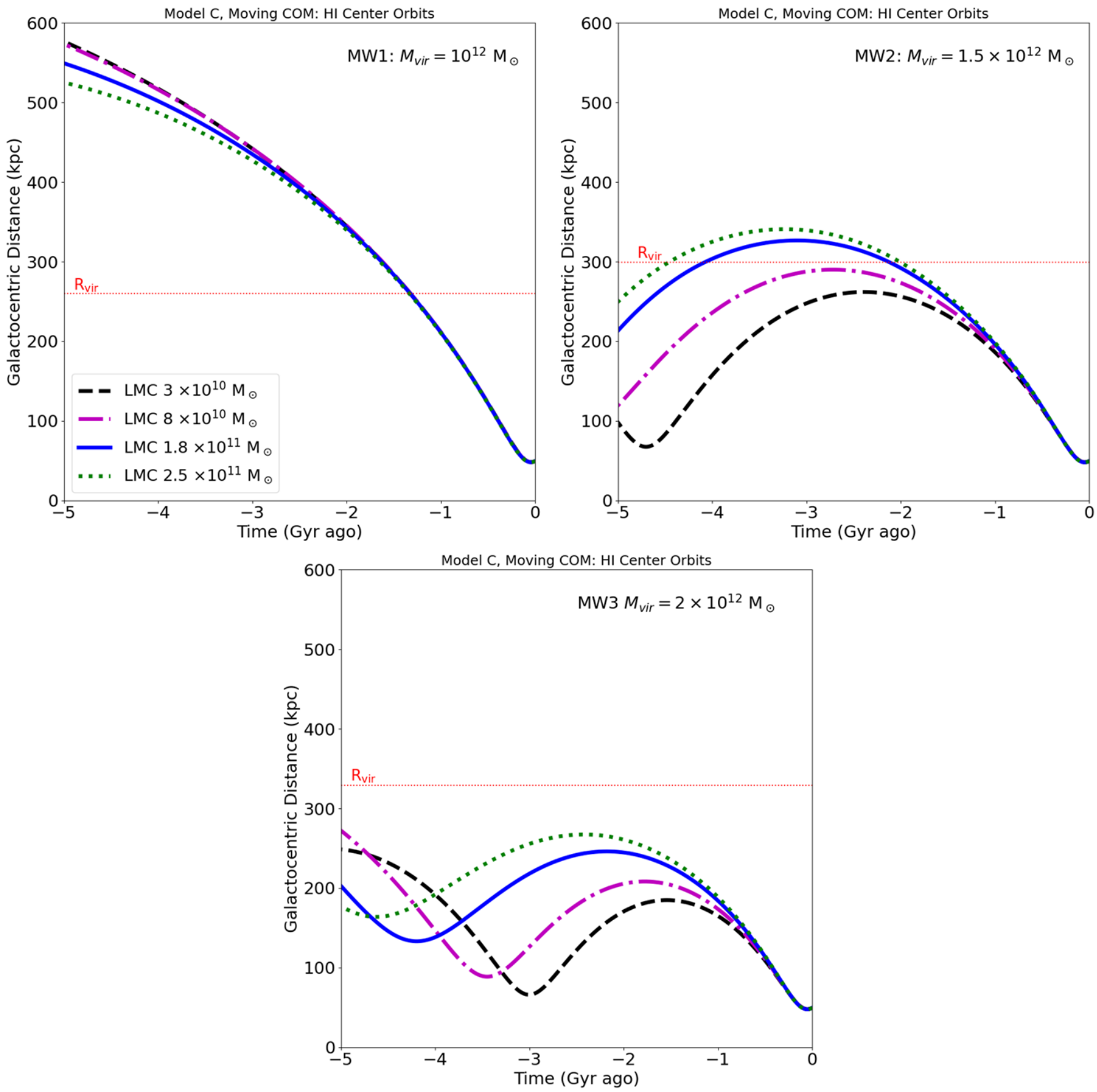

**Fig. 16** LMC orbits in the Model C: Moving Center of Mass (COM) Framework, combining intuition of the impact of dynamical friction and the motion of the orbital barycenter as a function of LMC and MW mass. In each panel the orbit of the LMC is computed using a different NFW MW model (MW1, MW2, MW3; Table 8) and four NFW LMC models (3, 8, 10, 25 $\times 10^{10}$ $M_{\odot}$; Table 9). The orbital barycenter of the MW–LMC system changes as the LMC orbits the MW, which changes the relative separation between the MW–LMC. Consequently the closest approach between the LMC and the MW occurs today in all cases (the pericenter of the LMC orbit is always at a larger Galactocentric distance in the past). In MW1, the differences in the LMC orbits reflect changes in the orbital barycenter—as the LMC mass increases, the orbit is *less* eccentric. For higher mass MW models (MW2, MW3), dynamical friction dominates over the barycenter motion—as the LMC mass increases the orbit becomes *more* eccentric. Orbits are computed using the HI center velocity, but results are similar if the Phot center velocity is used

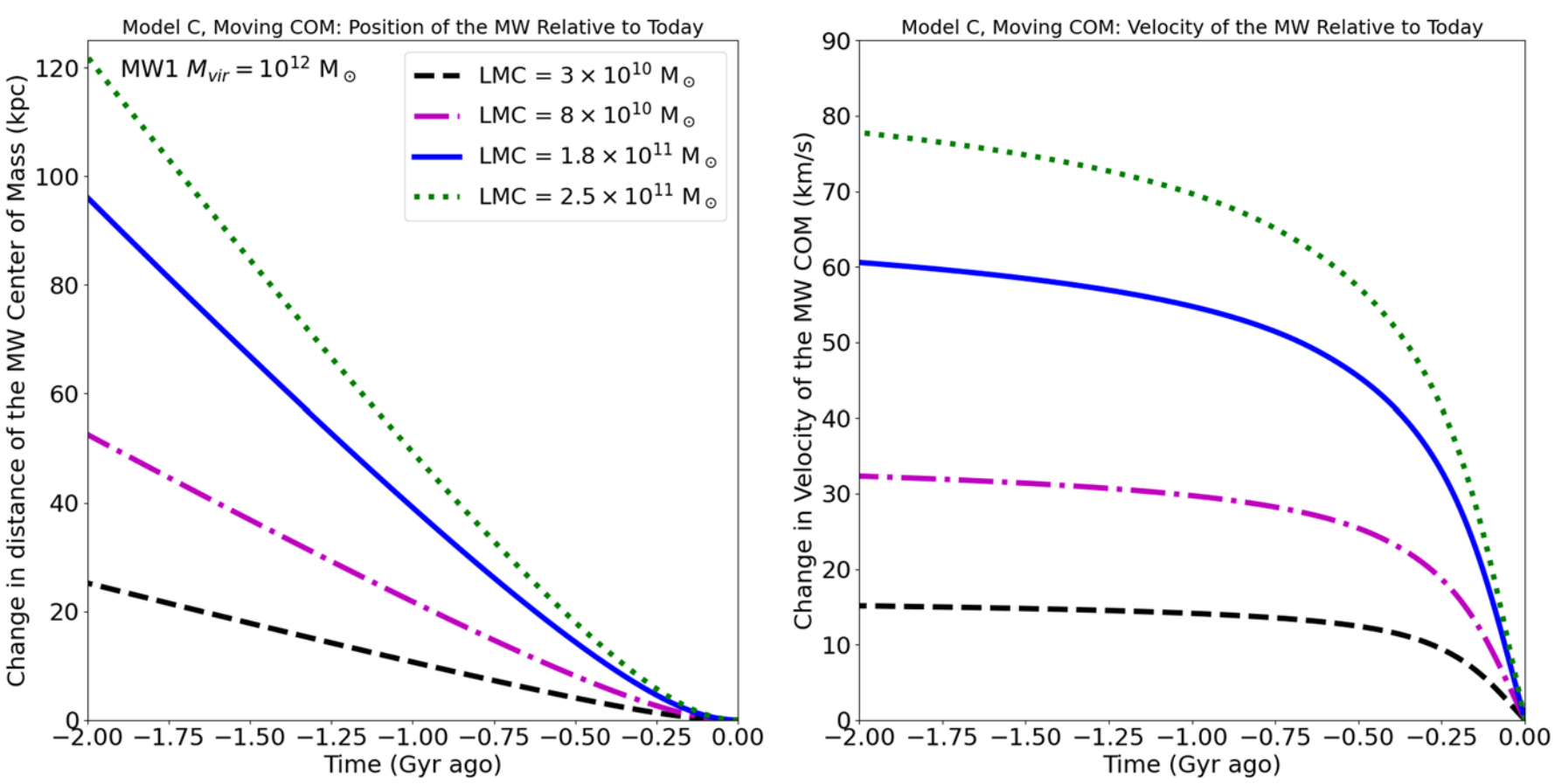

**Fig. 17** Change in the MW's center of mass (COM) position (left) and velocity (right) over the past 2 Gyr, as a function of the LMC mass. Orbits are computed in the Model C Moving COM framework using the NFW MW1 and HI Center Velocity. This figure illustrates the motion of the MW COM corresponding to the top left panel of Fig. 16, where changes in MW position and velocity are measured relative to its present day. The MW is defined to be at rest at Galactocentric coordinates (0,0,0). Results are similar for the more massive MW models over this timescale. Depending on the mass of the LMC, the MW COM moves 20–120 kpc and at speeds of 15–80 km/s over the past 2 Gyr, owing to its motion about the new MW–LMC orbital barycenter. This motion changes the separation between the LMC and MW, causing different orbit solutions than in the Model C: Fixed COM framework

**Summary:** LMC orbits in the Model C: Moving COM framework include the effects of both dynamical friction and the motion of the MW in response to the LMC. Using the HI/Phot Center velocities, the LMC is on first infall for the lowest mass NFW MW model (MW1), regardless of LMC mass. The orbital barycenter motion is most significant for MW1 with high mass LMC models. Previous passage orbit solutions are more prevalent as the MW mass increases, but the orbital period is at least double that of Model B. Any previous pericentric passage occurs at a larger distance than 50 kpc, meaning the most recent pericentric approach (∼50 Myr ago at ∼49 kpc, in all models) is the closest in the LMC's history. Across all models, the LMC's orbital path deviates from the location of the Stream on the sky. However, the COM motion causes the orbit to curve closer to the Stream, which complicates efforts to estimate MW halo shape using orbit trajectories.

## 6.4 The SMC's orbital history

In this section, the SMC is included in the orbit calculations within the Model C: Moving COM framework. As illustrated in Figs. 3, 4, and 5, the SMC can perturb the orbit of the LMC. The derived orbital history of the SMC is strongly dependent on the assumed mass of both the LMC and SMC, as well as the SMC's present-day velocity and assumed dynamical center.

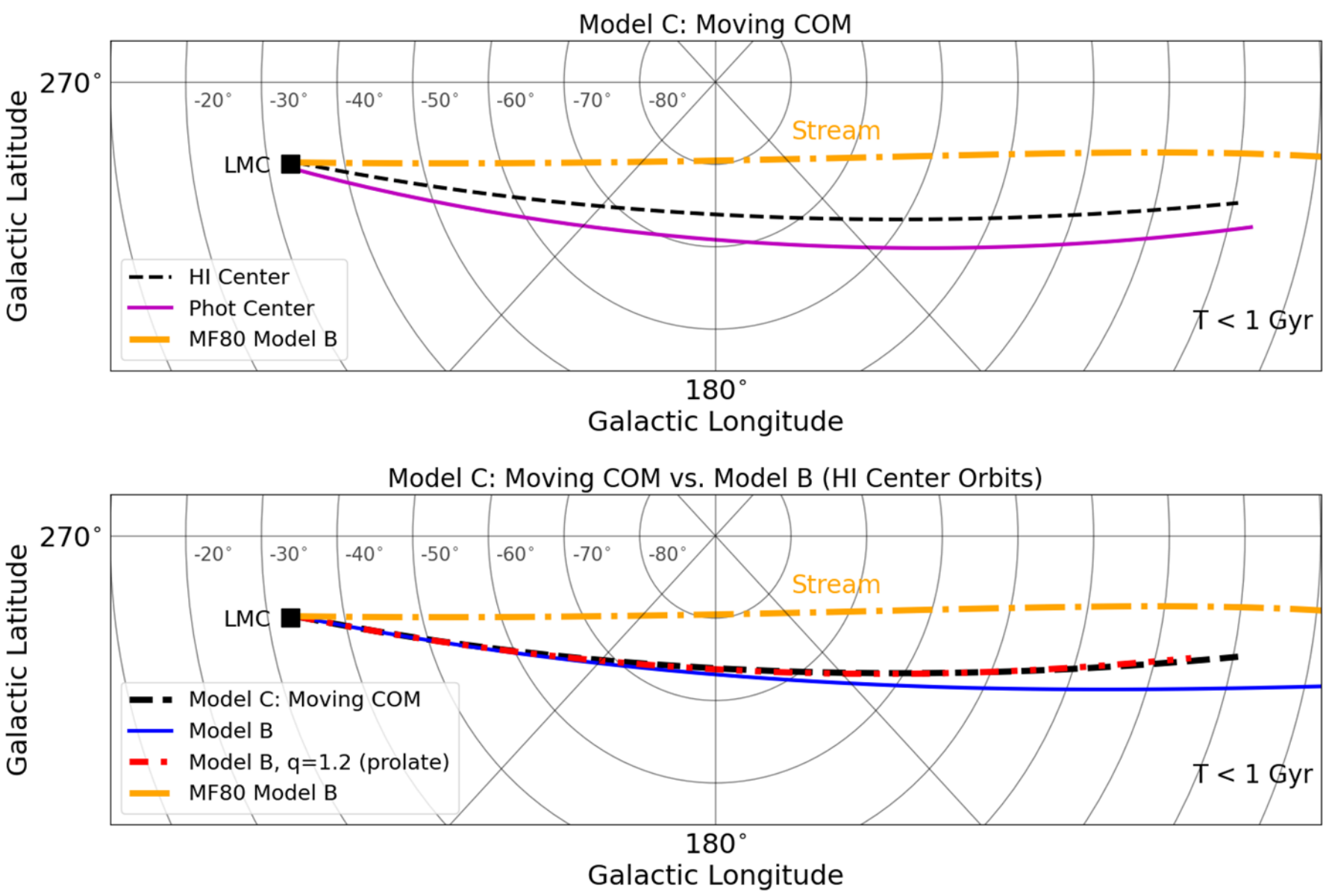

**Fig. 18** LMC orbits projected in Galactic coordinates over the past 1 Gyr computed in the Model C: Moving COM (NFW) or Model B (Isothermal Sphere, fixed COM) framework. Model C orbits are computed using the MW1 model (Table 8) and LMC model $M_{\mathrm{vir}} = 1.8 \times 10^{11}$ $\mathrm{M}_{\odot}$ (Table 9) Results do not change substantially for different LMC or MW masses. The MF80 orbit is the same as in Fig. 7 and traces the location of the Stream in the Model B framework. *Top panel:* LMC orbits are computed using the HI Center or Phot Center velocities, which have different $\mu_N$. The Phot Center orbit is more offset from the Stream than the HI Center orbit. Both orbits curve back toward the Stream as a result of the COM motion of the MW. *Bottom panel:* Model B orbits in comparison with Model C: Moving COM solutions. The Model B orbit (blue) is the same as in Fig. 7 and the Model C: Moving COM orbit is the same as in the top panel. The Model C solution is similar to the Model B orbit in a *prolate* halo with $q = 1.2$ (Eq. A5). This figure demonstrates that the motion of the MW in response to the orbit of the LMC complicates the assessment of the shape of the dark matter halo of the MW (see also Gómez et al. 2015; Garavito-Camargo et al. 2019)

The orbits of both the LMC and SMC are computed using distances given in Table 1 and the HST/Gaia velocities with two different assumptions for the dynamical center: HI or Phot for both the LMC and SMC (Table 2). Uncertainties in the orbital parameters are discussed in Sect. 8.

#### 6.4.1 SMC mass profile

A model for the present-day SMC dark matter distribution is created, following Patel et al. (2020). The SMC is modeled as an NFW dark matter halo with $M_{\mathrm{vir}} = 5 \times 10^9$ $\mathrm{M}_{\odot}$. Model parameters are summarized in Table 10.

The present-day stellar mass of the SMC is $\sim 3 \times 10^8 \mathrm{M}_{\odot}$ (van der Marel et al. 2009; Skibba et al. 2012). If the SMC were an isolated galaxy, cosmological abundance matching prescriptions would assign the SMC a virial mass of $\sim 6 - 8 \times 10^{10}$ $\mathrm{M}_{\odot}$ (Moster et al. 2013; Read and Erkal 2019). However, the SMC has been subjected to tidal forces from the LMC and the MW, complicating estimates of its total mass at infall (Foote et al. 2026). Since the SMC orbit will be integrated backward in time and the reviewed analytic schemes do not include satellite mass loss, the SMC is typically modeled at present day using a halo mass that is an order of magnitude lower than if it were isolated (e.g., Patel et al. 2020). N-body simulations that study the evolution of the SMC forward in time (starting 2–6 Gyr ago) have adopted more massive SMC models (e.g. $2 \times 10^{10}$ $M_{\odot}$; Besla et al. 2010, 2012; Pardy et al. 2018; Lucchini et al. 2020). For the purposes of this review, only low mass SMC models and backward orbital integration schemes are considered.

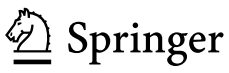

The mass profile corresponding to the SMC model is given in the left panel of Fig. 19. The SMC NFW concentration ($C_{\rm vir} = 37$) is chosen, such that the total mass of the SMC within 2–4 kpc of its center approaches mass estimates in the literature. The chosen concentration is not consistent with cosmological expectations for an isolated central. However, this SMC model does not include a gaseous or stellar component, thus requiring all the mass to be in the form of dark matter with a high concentration.

SMC mass estimates in the literature are summarized in Table 10 and include: 1) the dynamical mass derived from old stars $M(< 3$ kpc) = 2.7–5.1 $\times 10^9$ $M_{\odot}$ Harris and Zaritsky 2006; 2) enclosed mass derived from HI kinematics $M(< 4$ kpc) = 1–1.5 $\times 10^9$ $M_{\odot}$ Di Teodoro et al. 2019; 3) Jeans modeling of inferred bound SMC stars $M(< 3$ kpc) = $2.29 \pm 0.46 \times 10^9$ $M_{\odot}$ (De Leo et al. 2024); and 4) theoretical models of the SMC-induced tilt of the LMC's stellar bar $M(< 2$ kpc) $<$ 0.8–2.4 $\times 10^9$ $M_{\odot}$ (Rathore et al. 2025a). The bar tilt constrains the mass of the SMC *prior* to an LMC–SMC collision; as such, the Rathore et al. (2025a) constraint is an upper limit on the SMC's total mass profile today. The adopted SMC NFW model yields a reasonable analytic approximation of the mass profile of the SMC, given that the present-day SMC must be severely tidally distorted.

The corresponding rotation curve for this NFW SMC model is illustrated in the right panel of Fig. 19. The rotation curve does not match observational constraints from old stars at 2 kpc (10–17 km/s Harris and Zaritsky 2006; Zivick et al. 2021 or from the HI at 3 kpc (55–60 kpc Stanimirović et al. 2004; Di Teodoro et al. 2019). The HI constraints have informed the higher mass SMC models adopted by N-body studies of the SMC (Besla et al. 2010, 2012; Pardy et al. 2018; Lucchini et al. 2020).

As described in Rathore et al. (2026), the very low rotation rate of the old stars in the SMC is well below expectations from the Baryonic Tully Fisher relation (Mcgaugh 2005). Given the SMC's baryonic mass of $\sim 7 \times 10^8$ $M_{\odot}$ ($4.0 \times 10^8$ $M_{\odot}$ gas + $3.1 \times 10^8$ $M_{\odot}$ stars: van der Marel et al. 2009; Stanimirović et al. 2004; Brüns et al. 2005). However, the HI derived rotation curve is consistent with the Baryonic Tully Fisher relation. The discrepancy between the stars and gas is a known puzzle that is likely related to the SMC's recent collision with the LMC, leaving the SMC in a state of disequilibrium, where large scale velocity gradients are induced by the LMC's tidal field that mimic the appearance of rotation (Rathore et al. 2026).

The NFW SMC model rotation curve agrees with observations from old stars at larger radii (Dobbie et al. 2014; Dhanush et al. 2025), but again these observational constraints may reflect velocity gradients induced by the LMC's tidal field (Niederhofer et al. 2021; Vijayasree et al. 2026). The rotation curve mismatch does not invalidate the adopted SMC mass model, since the mass profile reasonably approximates obser-

**Table 10** SMC NFW halo model

| | $\mathbf{M_{vir}}$ ($10^9 M_\odot$) | $\mathbf{C_{vir}}$ | $\mathbf{R_{vir}}$ (kpc) |
|---|---|---|---|
| Model | 5 | 37 | 45 |
| | $\mathbf{M_{tot}}$**(< 2 kpc)** ($10^9 M_\odot$) | $\mathbf{M_{tot}}$**(<3 kpc)** ($10^9 M_\odot$) | $\mathbf{M_{tot}}$**(<4 kpc)** ($10^9 M_\odot$) |
| Model | 0.7 | 1.0 | 1.3 |
| Observed | < 0.8–2.4 [1] | 1.8–5.1 [2,3] | 1–1.5 [4] |
| | $\mathbf{V_{rot}}$**(2 kpc)** (km/s) | $\mathbf{V_{rot}}$**(3 kpc)** (km/s) | |
| Model | 38 | 38 | |
| Observed | 10–17 [2,5] | 20–40 [6,7], 55–60[4,8] | |

M(<2,3,4 kpc) refers to the total mass enclosed within 2,3,4 kpc. Rows marked "Model" refer to the fiducial SMC model in this review. Note the SMC is not modeled with a stellar or gaseous disk, requiring a large concentration to match mass constraints. Observational constraints on the rotation curve from old stars [Refs: 2, 5, 6, 7] yield lower values than from HI [Refs: 4,8]. References: [1] Rathore et al. (2025a): this is an upper limit constraining the enclosed SMC mass prior to an LMC–SMC collision; [2] Harris and Zaritsky (2006); [3] De Leo et al. (2024);[4] Di Teodoro et al. (2019); [5] Zivick et al. (2021); [6] Dobbie et al. (2014); [7] Dhanush et al. (2025); [8] Stanimirović et al. (2004)

vational constraints and the present-day kinematic state of the SMC is not the subject of this review.

The SMC NFW profile is truncated at the virial radius; outside this radius the SMC is treated as a point mass with total mass equal to $M_{\rm vir}$. The acceleration imparted on the LMC from the SMC is then given by Eq. (E25) in the Appendix.

**Summary:** A fiducial SMC model is introduced with an NFW halo profile and virial mass of $M_{\rm vir} = 5 \times 10^9$ M$_\odot$. The corresponding mass profile reasonably matches observational constraints. The rotation curve constraints from observations of gas and stars are in conflict, because the true SMC is in a state of disequilibrium; its kinematics cannot be approximated by an analytic model.

#### 6.4.2 SMC orbits in the model C: moving COM framework

In this section, the orbital history of the LMC in the Model C: Moving COM framework is re-computed, but now also accounting for the orbit and gravitational influence of the SMC. The equation of motion of the LMC becomes:

$$\mathbf{a}_{\rm L} = \mathbf{a}_{\rm MW_{h+d+b}}(\mathbf{r}_{\rm L-MW}) + \mathbf{a}_{\rm DF,NFW,MW}(\mathbf{r}_{\rm L-MW}, \mathbf{v}_{\rm L-MW}) + \mathbf{a}_{\rm S_h}(\mathbf{r}_{\rm L-S}) \quad (14)$$

This is the same as Eq. (13), but now includes the gravitational force from the SMC NFW halo at the relative separation between the Clouds ($\mathbf{r}_{\rm L-S}$).

The equation of motion for the SMC is similar to that of the LMC, but also includes dynamical friction owing to its motion through the LMC's dark matter halo, $a_{\rm DF,Iso,L}$. The dynamical friction term is defined in Eq. (F33) for the SMC moving through the LMC's halo, where $\ln \Lambda_L = 0.3$, as in Patel et al. (2020). Dynamical friction from the LMC is constrained to act on the SMC only, while the SMC is within the virial radius of the LMC. The SMC's equation of motion is thus:

$$\begin{aligned} \mathbf{a}_{\rm S} =& \mathbf{a}_{\rm MW_{h+d+b}}(\mathbf{r}_{\rm S-MW}) + \mathbf{a}_{\rm L_{h+d}}(\mathbf{r}_{\rm S-L}) + \mathbf{a}_{\rm DF,NFW,MW}(\mathbf{r}_{\rm S-MW}, \mathbf{v}_{\rm S-MW}) \\ &+ \mathbf{a}_{\rm DF,Iso,L}(\mathbf{r}_{\rm S-L}, \mathbf{v}_{\rm S-L}) \end{aligned} \tag{15}$$

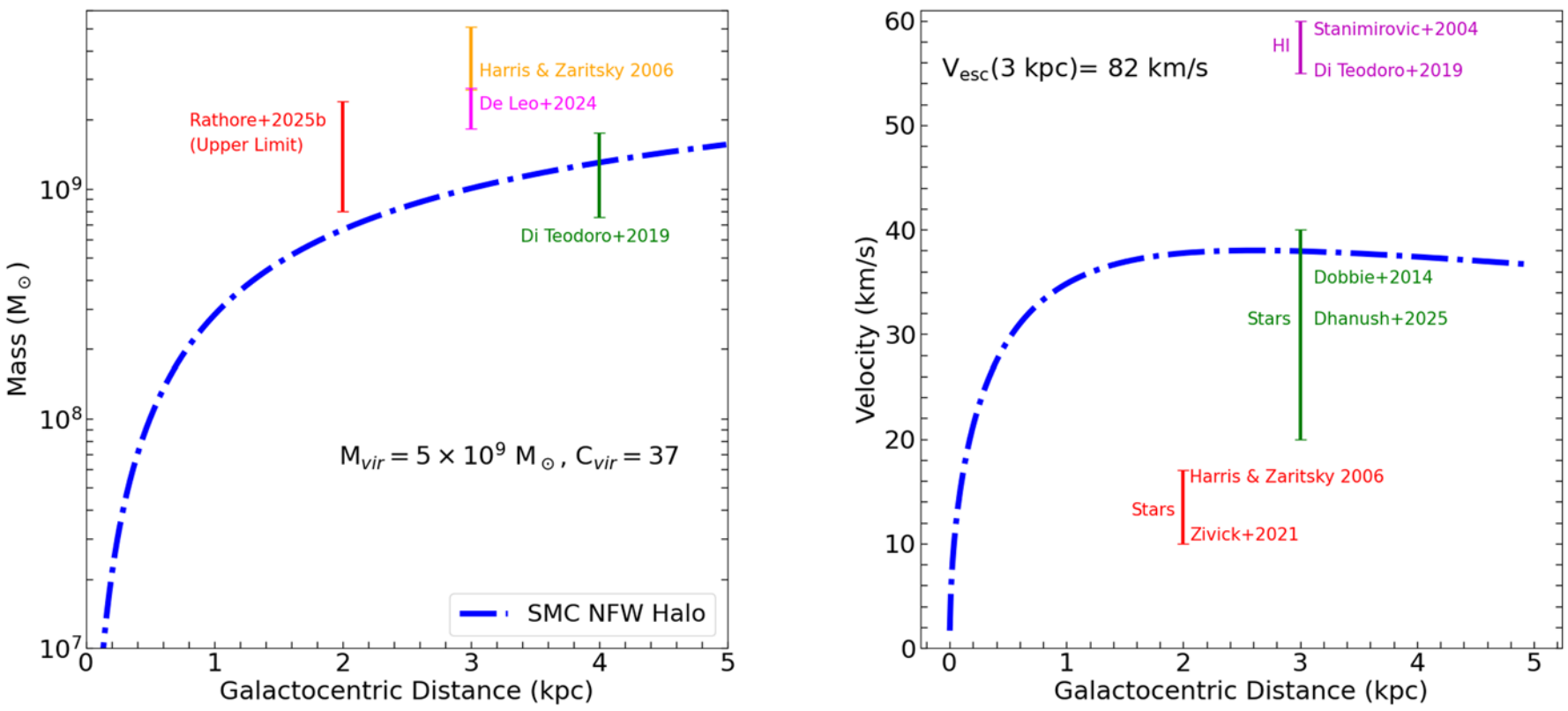

**Fig. 19** Mass profile (left) and rotation curve (right) for the adopted SMC model. The SMC is modeled as an NFW halo with virial mass $M_{\rm vir} = 5 \times 10^9$ M$_\odot$, $R_{\rm vir} = 45$ kpc, and $C_{\rm vir} = 37$. The concentration is selected such that the mass profile approaches observational and theoretical constraints on the mass of the SMC at various radii (Harris and Zaritsky 2006; Di Teodoro et al. 2019; De Leo et al. 2024; Rathore et al. 2025a). There is no gaseous or stellar component included in the SMC model. The corresponding rotation curve does not match recent observational constraints from stars at small radii (Harris and Zaritsky 2006; Zivick et al. 2021) or gas at larger radii (Di Teodoro et al. 2019; Stanimirović et al. 2004). The discrepancy between the gas and stellar kinematics is a known puzzle. The model does agree with observations from old stars at larger radii (Dobbie et al. 2014; Dhanush et al. 2025)

Because of the SMC's low mass, the gravitational force of the SMC acting on the MW is not included. As such, the equation of motion of the MW is the same as in Eq. (11), except that the acceleration imparted by the LMC is now modeled using an NFW halo and Miyamoto-Nagai disk. This approach is similar to that followed by Patel et al. (2020), although they include dynamical friction acting on the SMC within a radius, where the MW and LMC halo densities are comparable, rather than the LMC virial radius.

Orbits are computed using the HI center or Phot center velocities (Table 2), for both the LMC and SMC. In Figs. 20, 21, and 22, the resulting orbits for the SMC and LMC are illustrated in the three fiducial MW models (Table 8). In each case, the LMC is modeled as an NFW halo with virial mass $1.8 \times 10^{11}$ M$_\odot$ and the SMC is the fiducial model with a virial mass $5 \times 10^9$ M$_\odot$ (Table 10). The left panels illustrate the time evolution of the Galactocentric distance of the Clouds, the separation between the Clouds ($\rm R_{S-L}$), and the ratio of their relative speed ($\rm V_{S-L}$) to the escape speed of the LMC's halo at the given separation ($V_{escL}$). The escape speed is defined as

$$V_{escL}(|\mathbf{r}_{\rm S-L}|) = \sqrt{-2\Phi_{\rm L}(\mathbf{r}_{\rm S-L})} \tag{16}$$

where the LMC potential, $\Phi_L$, includes the LMC Miyamoto-Nagai disk and NFW halo potentials defined in Sect. 6.2. If the ratio $V_{escL}/V_{S-L} > 1$, the SMC is not bound to the LMC. Note that this calculation does not account for mass loss as the galaxies are modeled as rigid potentials.

The right panels of each figure show the Galactocentric XY, XZ, YZ projections for the orbits over the past 5 Gyr. In all cases, the LMC and SMC past orbit extends above the MW's disk plane (i.e., the Galactocentric XY plane), consistent with evidence for a northern extension of the gaseous, ionized Stream (Choi et al. 2025). The YZ projection highlights the distinction in the LMC and SMC orbits.

For the least massive MW host (MW1; Fig. 20), both choices of center result in a "first infall" scenario, where the LMC does not complete an orbit about the MW within the past 5 Gyr and is only now making its first close approach (as in the top left panel in Fig. 16). The LMC orbit is not affected significantly by the choise of center. However, the SMC orbit depends sensitively on the choice of center.

Using the HI center velocity, the SMC is bound to the LMC and travels on an eccentric ($e \sim 0.7$), non-circular, decaying orbit. The SMC completes ∼2.5 orbits within the past 5 Gyr. This LMC–SMC relative orbit is consistent with tidal Stream models. If an LMC mass lower than $8 \times 10^{10}$ $M_\odot$ is adopted, the SMC would not be bound to the LMC and would be on a long period orbit about the MW.

Using the Phot center velocity, the relative velocity between the Clouds is almost 25 km/s higher than the HI center velocity solution. The Clouds are currently bound, but the high speeds make it easier for the MW to prevent the binary from being long-lived. The SMC orbits the MW COM for at least 5 Gyr, with an orbital period of ∼ 4 Gyr. In this scenario, the LMC randomly captures the SMC (∼2–3 Gyr ago) as it makes its first approach to the MW. The only close encounter between the Clouds within the past 5 Gyr occurs $\lesssim$ 200 Myr ago. This LMC–SMC relative orbit is inconsistent with tidal models for the Stream as well as the current high gas content of the SMC (see Sect. 10).

For the intermediate MW host (MW2; Fig. 21), the LMC almost completes an orbit about the MW within 5 Gyr. However, in all cases the LMC only reaches its closest approach to the MW today. Again, the HI and Phot center orbits for the SMC differ. Using the HI center, the SMC remains bound to the LMC, completing two orbits about the LMC, but with a larger pericenter (∼50 kpc) than typically considered in models that reproduce the Stream (e.g., Besla et al. (2012); Lucchini et al. (2020)). Using the Phot center velocity, the SMC is unbound from the LMC 1.5 Gyr ago. Instead, the SMC orbits the MW's COM for > 5 Gyr, making very small pericentric approaches to the MW. The LMC only recently captured the SMC upon its recent pericentric approach to the MW.

For a massive MW host (MW3; Fig. 22), the SMC orbits the MW's COM rather than the LMC, regardless of the chosen center. The LMC and SMC's current close proximity on the sky today would be a result of the LMC capturing the SMC ∼1 Gyr ago, while both are in orbit about the MW. The SMC orbits within 200 kpc of the MW center for more than 5 Gyr, making very small pericentric approaches to the MW. This result is similar to the solutions for Model B with the new velocities, which also rely on massive MW models.

Long-lived LMC–SMC binary states, wherein the SMC has repeated close encounters with the LMC on timescales ($\sim$5 Gyr) and separations $< 40$ kpc, which are needed to support the formation of the Stream and influence the star formation history of one or both galaxies (see Sect. 1.3), are not achieved if the MW mass model is $\geq 1.5 \times 10^{12}$ $M_\odot$, regardless of the choice of dynamical center. The longevity of the LMC–SMC binary requires a low mass MW halo and, consequently, a first infall scenario. The impact of LMC mass and the allowed error space on this conclusion will be explored in Sect. 8.

For more massive MW models, the SMC makes multiple orbits about the MW with small pericentric distances; this orbit solution is at odds with the current high gas content of the SMC. The SMC thus presents a unique opportunity to place a firm upper limit on the virial mass of the MW at 1.5 $\times 10^{12}$ $M_\odot$ (see Fig. 30). The Model D framework is necessary to verify/tighten such constraints (Sect. 7.2).

Figure 23 illustrates the resulting orbits for the LMC and SMC in Galactic coordinates over the past 1 Gyr. The projected orbits are similar across all MW and LMC mass models and choice of center - despite the SMC orbit solution being very different depending in 3D Galactocentric coordinates (see the right panel of Fig. 20). As expected from Fig. 18, the LMC's past orbit deviates from the location of the Stream. The SMC's orbit, however, does approach the Stream location in all cases, which suggests a connection between the SMC and the origin of the Stream.

**Summary:** The orbits of the SMC are computed in the Model C: Moving COM framework, using the LMC $M_{\rm vir} = 1.8 \times 10^{11}$ $M_\odot$ model and the mean HI or Phot center velocities. The SMC orbit strongly depends on the choice of center and MW model. Long-lived LMC–SMC binary solutions that are consistent with Stream tidal formation models or the star formation histories of the Clouds are not achieved for high mass MW models (MW2 or MW3) or for the Phot Center velocities in any MW model. The HI Center orbit in MW1 is a binary solution with an orbital period of 1.5 Gyr and an eccentricity of 0.7; the orbit is not circular and decays owing to dynamical friction. In non-binary solutions, the SMC is on a long period orbit about the MW for $>$5 Gyr, making very close pericentric passages with the MW before being randomly captured by the LMC as the LMC makes its most recent pericetric approach to the MW. Such an SMC orbit is at odds with the SMC's current gas content; providing a new method to constrain the MW and LMC virial mass. In all cases, the Clouds experience a close encounter within the past 200 Myr. In non-binary solutions, this recent approach is the only encounter between the Clouds. The SMC's orbit provides a reasonable match to the location of the Stream on the sky, suggesting a connection.

## 7 Model D: distortions to the dark matter halos of the MW and LMC

As discussed in Sect. 3.1, the LMC's infall mass is likely of order $10^{11}$ $M_\odot$. In the Model C: Moving COM framework authors have accounted for the consequent COM motion of the MW owing to the gravitational force from a massive LMC, using rigid analytic models for the MW (e.g., Gómez et al. 2015). However, in a $>$1:10 mass ratio encounter, the dark matter distributions of both the LMC and the MW will be strongly

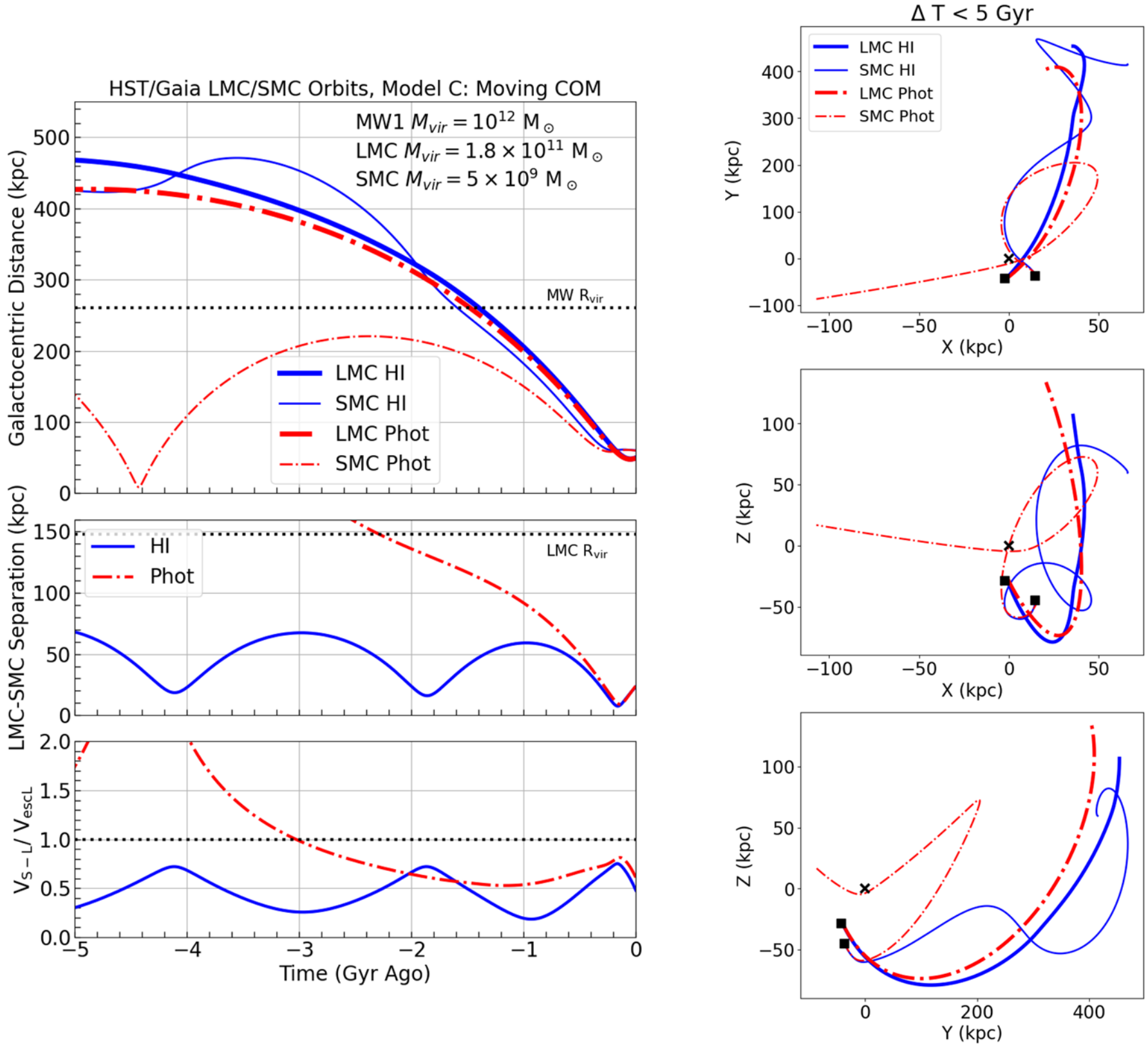

**Fig. 20** Orbit of the SMC and LMC in MW1, computed using the mean HI and Phot velocities/positions (see Table 2) in the Model C: Moving COM framework. In all cases, the LMC is modeled as an NFW halo with $M_{\rm vir} = 1.8 \times 10^{11}$ M$_\odot$ (Table 9) and the SMC is the fiducial model (Table 10). The gravitational force from the SMC on the LMC is included, as is dynamical friction on the SMC owing to its motion through the LMC's halo. The MW COM moves in response to the LMC and exerts dynamical friction on both Clouds. *Top left:* Galactocentric distance of the LMC (thick lines) and SMC (thin lines) as a function of time. *Middle left:* the distance of the SMC relative to the LMC as a function of time. *Bottom left:* the ratio of the relative velocity of the Clouds ($V_{\rm S-L}$) and the escape speed of the LMC at the SMC's separation ($V_{escL}$). *Right panels:* the orbits of the Clouds in the Galactocentric XY, XZ, YZ planes; the MW COM is denoted by the black "X". The LMC orbits are consistent for both dynamical centers; the Clouds are on first infall and their past orbits traverse above the MW's disk plane (positive Z). However, the SMC orbits differ; this is most clearly seen in the YZ plane. Using the HI center the LMC–SMC system can maintain a binary state (completing 2.5 orbits around the LMC in the past 5 Gyr). However, using the Phot center, the SMC does not orbit the LMC COM and instead orbits the MW, with a much small pericenter in the past. The SMC's orbit is strongly perturbed by the LMC's infall

disturbed and change as a function of time. This effect is more complex than can be captured by a rigid MW model with a changing COM.

The Model D framework refers to studies that have modeled the LMC–MW system using fully "live" (N-body) representations for the dark matter distribution of the MW and LMC. In the following, LMC orbit solutions in the Model D framework are discussed, followed by implications for the SMC–LMC binary, which is also a 1:10 mass ratio encounter.

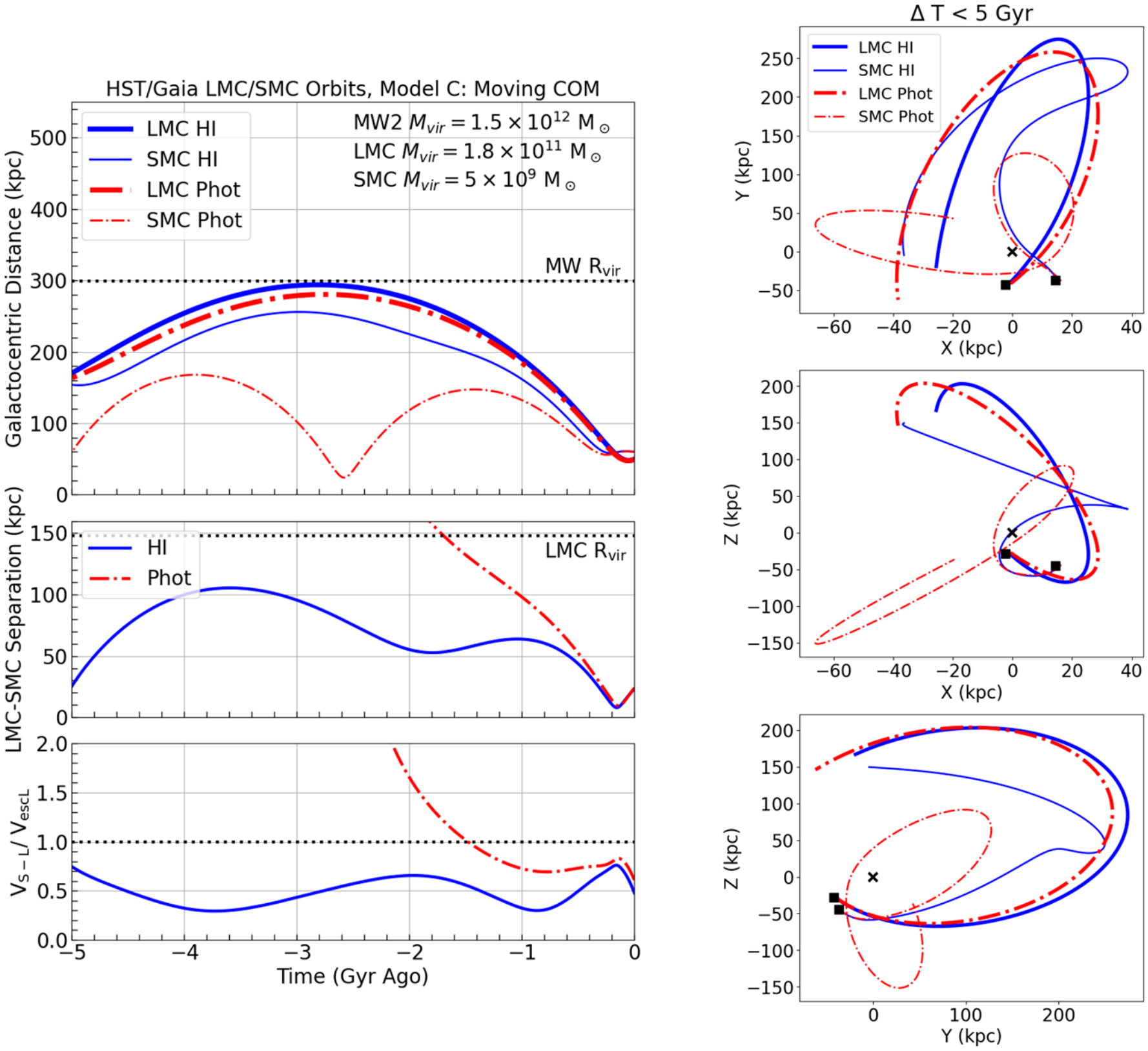

**Fig. 21** Same as Fig. 20, but for MW mass model MW2. The LMC orbits are consistent for both centers; the LMC almost completes an orbit within 5 Gyr. In both cases, the LMC struggles to maintain the SMC as a close binary companion. For both centers, the SMC does not get within 40 kpc of the LMC until the past 0.5 Gyr, which is inconsistent with expectations from their star formation histories and Stream models. For the Phot center, the LMC only captured the SMC upon its recent close approach to the MW; prior to this time the SMC is in orbit about the MW COM (black "x")

## 7.1 The LMC orbit in the model D framework

Garavito-Camargo et al. (2021, 2019) illustrate that significant distortions are expected to the MW's dark matter distribution owing to the orbit of a massive LMC (see Fig. 24). These distortions include a dynamical friction wake that traces the orbit of the LMC (Transient Response) and global over/under densities that result from the displacement of the COM of the inner halo with respect to the outer halo (Collective Response). The resulting combined dark matter distribution is not well-captured by standard spherical, oblate, prolate, or triaxial density profiles. The predicted distortions are also seen in cosmological simulations of LMC–MW analogs (Arora et al. 2025; Darragh-Ford et al. 2025; Mansfield et al. 2026), in different dark matter models (Foote et al. 2023), and have been confirmed by multiple idealized simulations (see review byVasiliev, 2023). Observational evidence for these distortions, such as the stellar counterpart to the dark matter wake, is also mounting (Belokurov et al. 2019; Conroy et al. 2021; Cavieres et al. 2025).

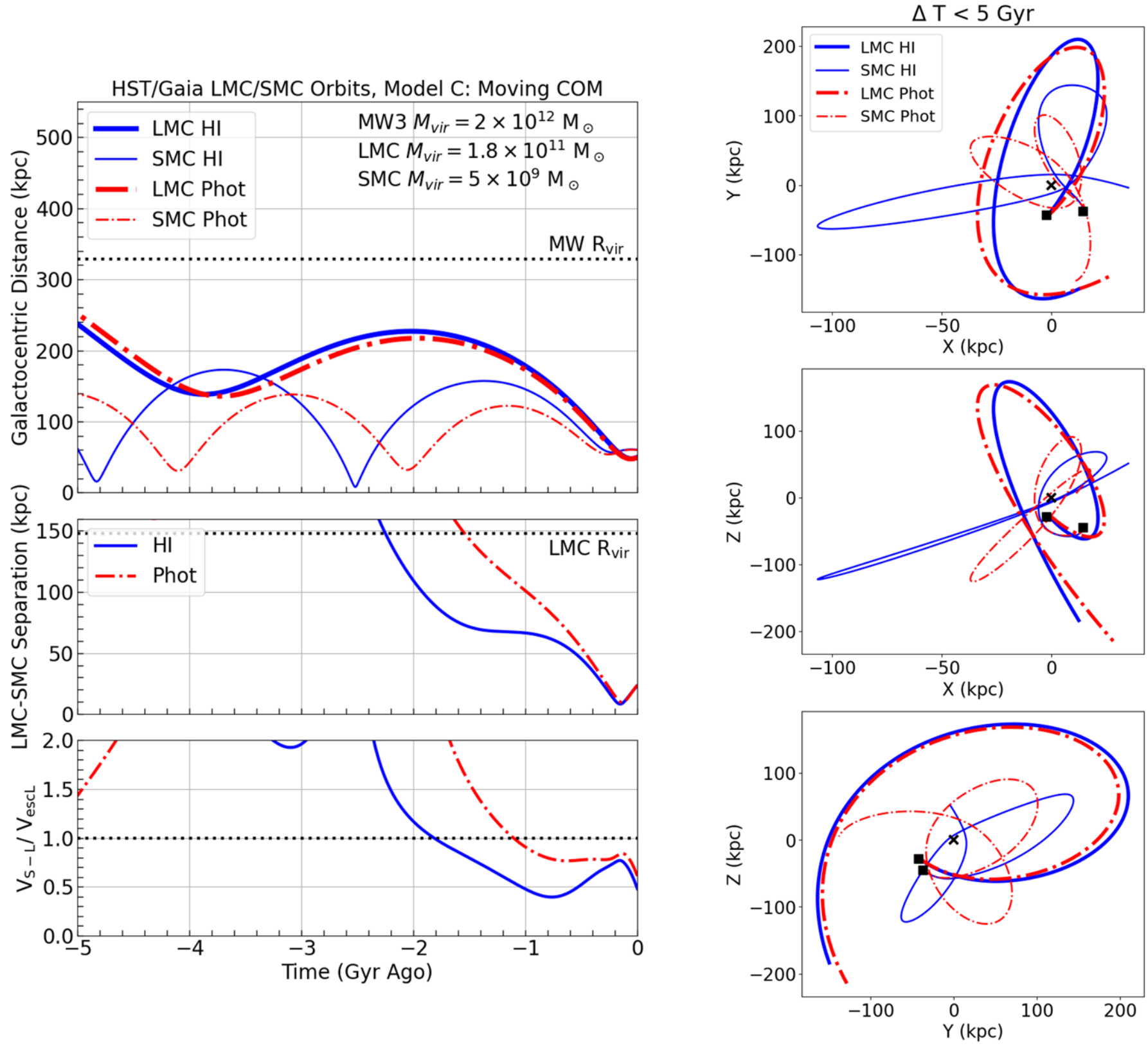

**Fig. 22** Same as Fig. 20, but for MW mass model MW3. Again, the LMC orbits are consistent for both centers, where the LMC completes an orbit within 3.5 Gyr. The LMC's previous pericenter reaches a distance of 150 kpc, meaning the LMC has only recently made its closest approach to the MW. In both cases the LMC–SMC system is not a long-lived binary. The LMC is only recently in proximity to the SMC. The SMC is bound to the MW and completes two orbits about the MW COM (black "x") within the past 5 Gyr

Earlier studies illustrated that the LMC can induce resonant perturbations in the MW's dark matter distribution. These studies typically explored low mass LMC models, where the LMC made multiple orbits around the MW (of order $10^{10}$M$_\odot$,Weinberg, 1998, 2000). However, Vesperini and Weinberg (2000) illustrated that significant perturbations in the halo can be generated even through fly-by encounters with massive satellites.

In general, recent Model D studies employ high mass LMC models and high LMC:MW mass ratios (1:5–15, see Table 6), resulting in first infall orbits. The distortions induced in the MW halo in such an encounter are not fully captured in the analytic Model C: Moving COM framework.

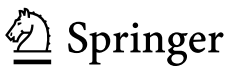

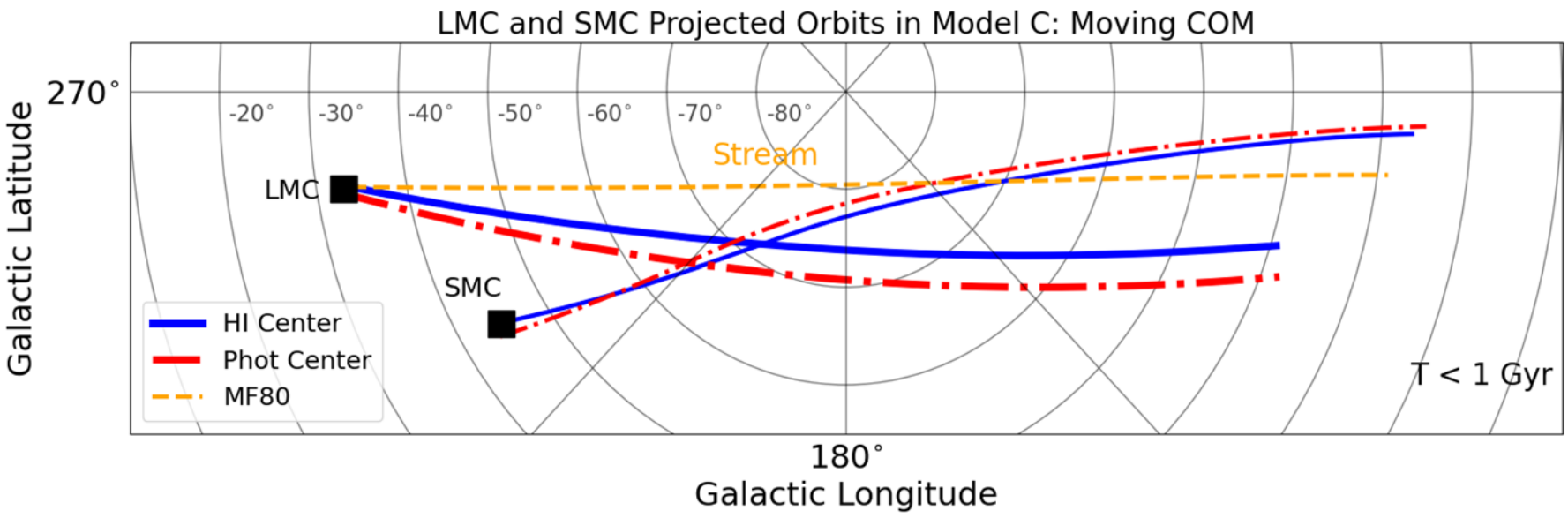

**Fig. 23** Orbital path of the LMC and SMC on the sky, projected in Galactic Coordinates. Orbits are computed over the past 1 Gyr in the Model C: Moving COM framework using the MW1 and $1.8 \times 10^{11}$ $M_{\odot}$ LMC mass models. Results are similar for all MW and LMC models. The LMC orbits are the same as in the top panel of Fig. 18. The SMC HI and Phot orbits differ greatly in 3D Galactocentric coordinates (see the right panels in Fig. 20; the SMC–LMC system is not a binary in the Phot center solution). However, in projection, the SMC HI and Phot Center orbits are similar. Regardless of the choice of center, LMC or MW mass, the SMC's orbit is a better tracer to the location of the Stream on the sky than the LMC orbit

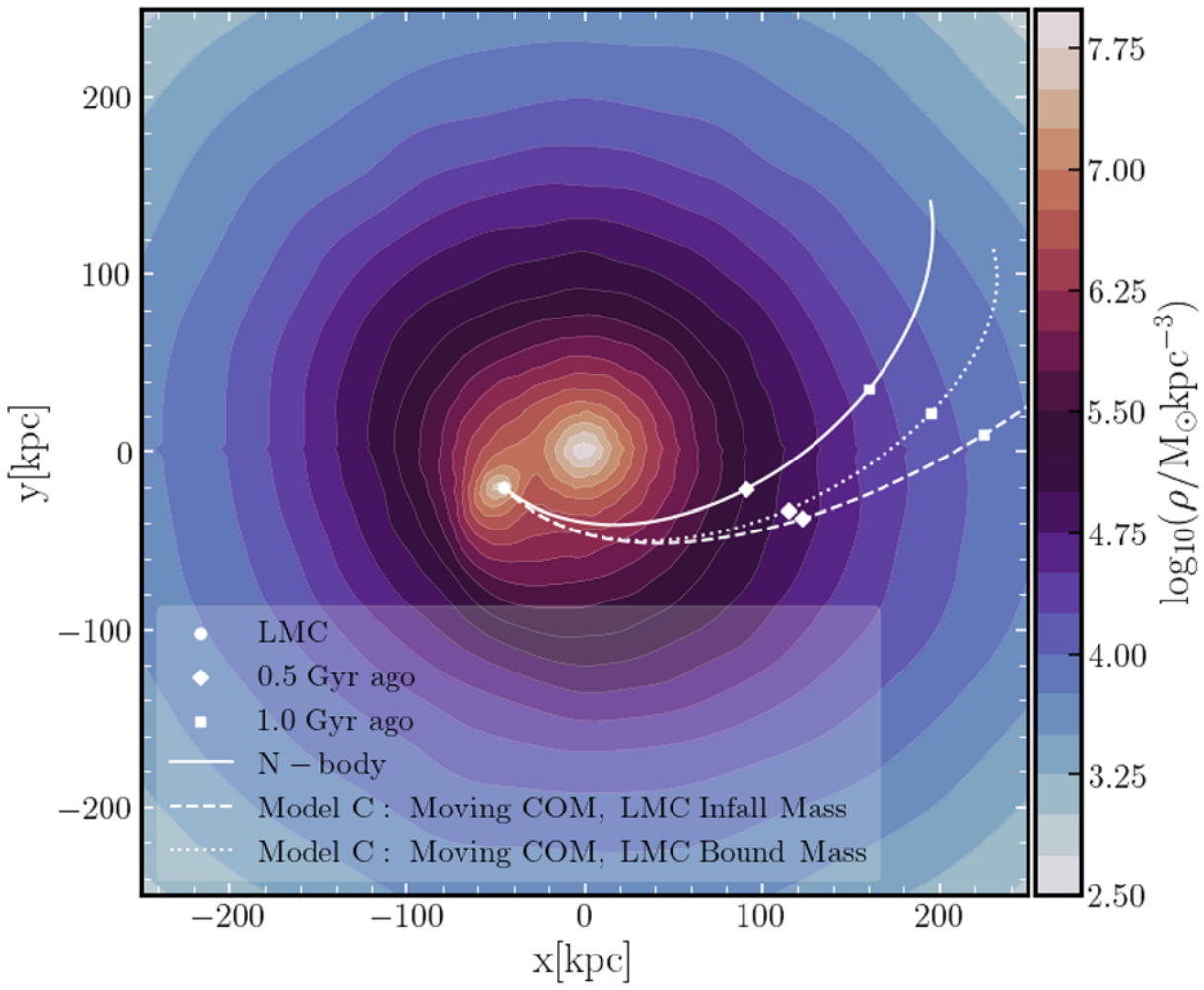

**Fig. 24** Dark matter density distribution of the combined LMC+MW system in the Model D framework in a first infall scenario. The LMC and MW are modeled as N-body systems following Hernquist density profiles. This simulation is from Garavito-Camargo et al. (2021), their fiducial simulation #7, where the LMC mass at infall is $1.8 \times 10^{11}$ $M_{\odot}$ and the MW's virial mass is $1.2 \times 10^{12}$ $M_{\odot}$. The solid line denotes the LMC's orbit in the simulation over the past 2 Gyr. The dashed line is the LMC's analytic orbit constructed backward in time using the Model C: Moving COM framework and assuming that the present-day LMC mass is the infall mass. The dotted line is the same but assuming the LMC mass is the bound mass at present day ($6 \times 10^{11}$ $M_{\odot}$); the match to the N-body orbit is improved. *Image credit N. Garavito-Camargo and E. Patel (2026)*

In particular, in an N-body MW–LMC halo encounter, the COM motion of the MW halo varies as a function of radial distance (Garavito-Camargo et al. 2021). Furthermore, the LMC loses mass, which is not accounted for in the Model C framework. Consequently, an analytic orbit using the Model C: Moving COM framework will differ from the true N-body orbit. This is demonstrated by the dashed line in Fig. 24, where the LMC mass is kept fixed at the infall mass of $1.8 \times 10^{11}$ M$_\odot$.

With an appropriate mass model for the LMC, the orbital differences between the Model C and Model D frameworks can be improved over short timescales. Using the LMC's present-day bound mass ($6 \times 10^{11}$ M$_\odot$; dotted line in Fig. 24) instead of the infall mass results in a better match to the N-body solution for the first infall orbit. This means that the general trends for the LMC orbit with MW mass illustrated in Fig. 16 using the Model C: Moving COM framework are reasonable approximations to the Model D solutions over the past 1–2 Gyr, assuming the LMC mass refers to its current bound mass. However, discrepancies are expected to be substantial over longer timescales, particularly for massive MW models, where the LMC orbits within the MW's halo for a long period of time.

While a first-infall LMC orbit may be reasonably reconstructed in an analytic framework, precision orbit reconstruction for other MW halo tracers (satellites, stars, streams, etc.) needs to account for the range of MW halo distortions induced by the LMC. This requires that these distortions be accurately quantified and understood.

The halo response to the passage of the LMC alone has been quantified in N-body simulations of spherical (Vasiliev 2023; Laporte et al. 2018a, b; Gómez et al. 2015) and aspherical (Sheng et al. 2024; Arora et al. 2025) MW dark matter halos. Interestingly, the strength of the response of the MW to the LMC's passage depends on the MW's halo shape and its large scale structure (Arora et al. 2025; Darragh-Ford et al. 2025). Consequently, Sheng et al. (2024) find that the shape of the MW's halo can also change the orbital solution for the LMC.

Several teams have used basis function expansions (Garavito-Camargo et al. 2021; Petersen and Peñarrubia 2020, 2021; Lilleengen et al. 2023) and perturbation theory (Weinberg 1998, 2000; Rozier et al. 2022) to quantify the MW's dark matter halo's response to the LMC's passage. Through such methods, the perturbative effects of the LMC-induced distortions on the structure/kinematics of stellar streams and the MW's stellar halo can be studied in detail and tested against observational data.

**Summary:** Model D studies adopt time evolving halo potentials for the LMC and MW. The orbit of the LMC distorts the halo of the MW, generating a dynamical friction wake and displaces the COM of the inner MW halo with respect to the outer halo. Orbital solutions are typically first infall scenarios. LMC Model D orbits can be mimicked by the analytic Model C: Moving COM framework, provided an appropriate LMC infall mass is selected. However, the Model C approach is not appropriate to study the orbits of other objects in the MW's halo, which may be impacted by the LMC's halo distortions. In the era of precision astrometry, software and toolkits that enable the community to integrate orbits of halo tracers in a time-evolving MW+LMC system are needed.

## 7.2 The SMC orbit in the model D framework

There are now N-body and hydrodynamic simulations of the LMC and SMC that account for live MW, LMC, and SMC dark matter distributions. The Lucchini et al.

(2021) and Tepper-García et al. (2019) hydrodynamic simulations illustrate that the primary extended gaseous structures of the system, such as the Stream, Leading Arm and Bridge, are explainable even in the Model D framework. The N-body only simulations of Garver et al. (2026) demonstrate that the LMC's stellar disk geometry is explainable in the Model D model with a recent SMC collision. Furthermore, the N-body only KRATOS simulation suite (Jiménez-Arranz et al. 2024) explores multiple LMC–SMC–Milky Way orbital configurations and generally reproduce the stellar structure of the Clouds in the Model D framework.

The LMC/SMC+MW orbit solutions explored in recent Model D studies are generally first infall scenarios. A detailed study of changes to the orbital space of both the LMC and SMC owing to their distortions to the MW's halo has not yet been completed. It is challenging to recover the current observed 3D velocity and position vectors of the LMC and SMC in a scenario, where each body is modeled as a live halo (see Sect. 9).

In the previous section, it was illustrated that a first infall orbit of the LMC in the Model D framework (N-body system) can be approximated by the Model C: Moving COM framework. However, the SMC is expected to have orbited the LMC for an extended period of time.

The LMC and SMC have been modeled as N-body systems by multiple authors, even when the MW is a static model (Model C studies like Besla et al. 2010, 2012; Pardy et al. 2018; Tepper-García et al. 2019; Lucchini et al. 2020; Craig et al. 2022). Besla et al. (2012) introduced a Model D-type framework for the LMC and SMC, wherein the Clouds are both modeled as Hernquist N-body systems and allowed to orbit one another in isolation, prior to their infall toward the MW. In this way, the SMC–LMC (mass ratio $\sim$ 1:10) interaction is similar to the LMC–MW (mass ratio $\sim$ 1:10) interaction in the Model D framework.

Using the high-resolution MEGHA N-body simulation, Foote et al. (2026) demonstrate that the combined dark matter distribution of the LMC and SMC in the B12 scenario is distorted owing to their mutual interactions (Fig. 25). The SMC induces dynamical friction wakes in the LMC's halo and displaces the LMC's COM, much like the LMC's impact on the MW's dark matter distribution. A massive LMC also significantly truncates the SMC's mass distribution; the SMC's mass loss history ultimately also affects the orbit of the LMC. Because the Clouds are modeled in isolation, the LMC halo distortions and the bulk of the SMC's mass loss history are expected to occur *before* the Clouds first approach the MW. LMC halo distortions severely complicate efforts to model the orbits of objects around the Clouds using analytic frameworks, such as those commonly adopted to identify other potential satellites of the Clouds (Jethwa et al. 2016; Patel et al. 2020).

As illustrated in Fig. 26, the Model D, N-body orbit of a massive SMC completes $\sim$2 orbits about the LMC prior to infall. The pericentric distances and decay rate of the orbit are designed to explain the observed enhanced star formation rates of the Clouds over the past $\sim$ 3.5 Gyr (Harris and Zaritsky 2009; Weisz et al. 2013; Massana et al. 2022; Burhenne et al. 2026). The N-body orbit for the Clouds is consistently at separations less than 50 kpc within the past 3.5 Gyr. Observations demonstrate that dwarf pairs with separations less than 50 kpc have star formation rates elevated by a factor of $\geq$2.3 relative to isolated dwarfs (Stierwalt et al. 2015). For a massive

LMC (infall mass $\sim 10^{11}$ $M_{\odot}$), B12 argue that an N-body solution for the LMC–SMC binary requires a decaying orbit with a high eccentricity (e$\sim$ 0.7); otherwise the binary would merge rapidly, preventing the observed sustained period of elevated star formation. Multiple authors have adopted a similar LMC–SMC orbital history to explain the properties of the gaseous Stream (Pardy et al. 2018; Lucchini et al. 2020, 2021; Craig et al. 2022; Garver et al. 2026).

The N-body orbital history of a massive SMC about the LMC (1:10) is not well-reproduced by the standard Model C: Moving COM framework, which assumes a low SMC mass encounter at all times and a backward integration (here from the epoch of infall). However, the recent N-body orbit can be reasonably recovered in a backward integration scheme using the analytic dynamical friction formalism of Van Der Marel et al. (2012b), which is calibrated to an N-body simulation of the M33-M31 system (1:10 mass ratio; see F.2). The reconstructed orbit assumes that the SMC mass at the infall epoch is $\sim 5 \times 10^9$ $M_{\odot}$ (using the fiducial SMC NFW model). In the actual simulation, the SMC's bound mass is $\sim 4 \times 10^9$ $M_{\odot}$ at the epoch of infall. While mass loss is not explicitly accounted for, the Coulomb Logarithm varies along the orbit, mimicking the impact of mass loss.

Model D solutions for SMC–LMC (1:10) binary orbits are substantially more difficult to maintain in the face of MW tides than solutions in the Model C: Moving COM framework (see Sect. 8.2). This further supports the first infall scenario for the Clouds, as a previous approach would have disrupted the wide pair.

Given the expected mass loss of the SMC and the distortions the SMC induces in the LMC halo, understanding the orbits of objects in orbit about the LMC (including the SMC) requires full N-body simulations that are constrained by the present-day structure and kinematics of both the LMC and SMC, rather than analytic orbit reconstructions.

**Summary:** Model D N-body simulations of the LMC/SMC/MW system adopt massive halso for the LMC ($\sim 2 \times 10^{11}$ $M_{\odot}$) and SMC ($\sim 2 \times 10^{10}$ $M_{\odot}$). The Clouds are assumed to be a binary and the SMC's orbit about the LMC is necessarily eccentric ($e \sim 0.7$) to avoid the system merging. The Clouds are typically on first infall, as an eccentric binary is unlikely to survive a previous passage about the MW. The SMC makes $\sim$three orbits about the LMC within the past 6 Gyr with separations $< 50$ kpc within the past 3 Gyr, consistent with enhancements in star formation. The SMC halo mass evolves significantly owing to tides. The SMC also creates pronounced distortions to the LMC's halo. These effects cannot be recovered in the Model C framework - however, these effects increase the orbital eccentricity of the SMC, meaning Model C results regarding the longevity of the LMC–SMC binary are a best-case scenario.

## 8 Orbital statistics

In this section, the observed present-day position and velocity error space for the LMC and SMC is explored to constrain the range of possible orbital solutions for the Clouds.

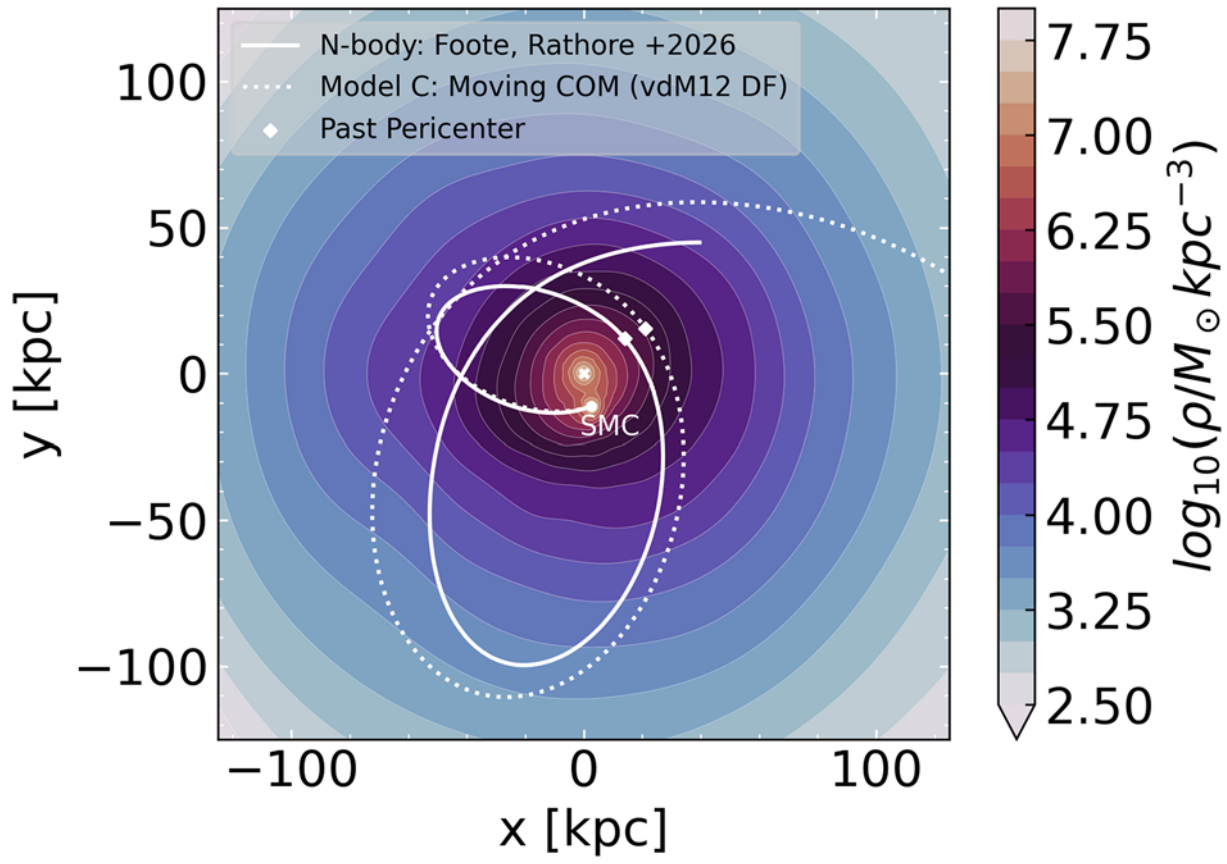

**Fig. 25** N-body simulation of the combined dark matter distribution of the LMC–SMC (1:10 mass ratio) system after 7 Gyr of evolution in isolation (no MW). This figure illustrates a Model D framework for the SMC's orbital history about the LMC. The LMC and SMC are modeled as N-body Hernquist dark matter distributions with initial virial masses of $\sim 1.8 \times 10^{11}$ $M_\odot$ and $2.0 \times 10^{10}$ $M_\odot$, respectively. Plotted is the dark matter distribution of the combined LMC–SMC binary at the approximate epoch of first infall toward the MW. This dark matter distribution is not spherical and evolves with time. The solid line is the N-body orbit of the SMC. The dotted line is the orbit reconstruction using the Model C: Moving COM framework, but with the dynamical friction formalism of (Van Der Marel et al. 2012b, see F.2)assuming the fiducial SMC NFW model (Sect. 6.4.1). This figure is adapted from Foote et al. (2026) and uses the MEGHA simulations created by H. Rathore. *Image credit: H. Foote, H. Rathore, and E. Patel (2026)*

All orbit calculations are performed using the Model C: Moving COM framework (Sect. 6.4.2). As a reminder, the MW halo is modeled using a spherical NFW density profile that is static in shape (Table 8), but the MW's center of mass moves in response to an LMC, depending on the LMC halo mass (NFW models: Table 9). The gravitational force of the SMC on the LMC is also included (fiducial NFW model: Table 10). Dynamical friction from the MW on the LMC and SMC (F.2), and from the LMC on the SMC (F.3) is included.

A large set of possible present-day position and velocity vectors for the LMC and SMC is defined by making ∼4600 Monte Carlo drawings over the $1\sigma$ errors in the measured proper motions (including reported covariances), radial velocities, and distance moduli that result from each choice of LMC/SMC dynamical center (HI or Phot center). These Monte Carlo drawings are then combined into one set of 9200 drawings, referred to in this review as the HI+Phot Center data set. This data set spans the currently reported range of centers for the LMC and SMC (see Figs. 1 and 2).

In the following, 9200 orbits for the LMC, SMC, and MW are computed using the HI+Phot Center data set for the present-day position and velocity vectors of the Clouds. Mean orbital properties from this combined set of orbits are reported, as well as $1\sigma$ errors in order to explore the range of possible values for characteristic properties of the LMC (Sect. 8.1) and SMC (Sect. 8.2) orbits, under the Model C: Moving COM framework.

Note that, as described in Sect. 7, orbit solutions may differ if the MW/LMC/SMC dark matter halos are deformable (e.g., N-body). However, running nearly 10,000

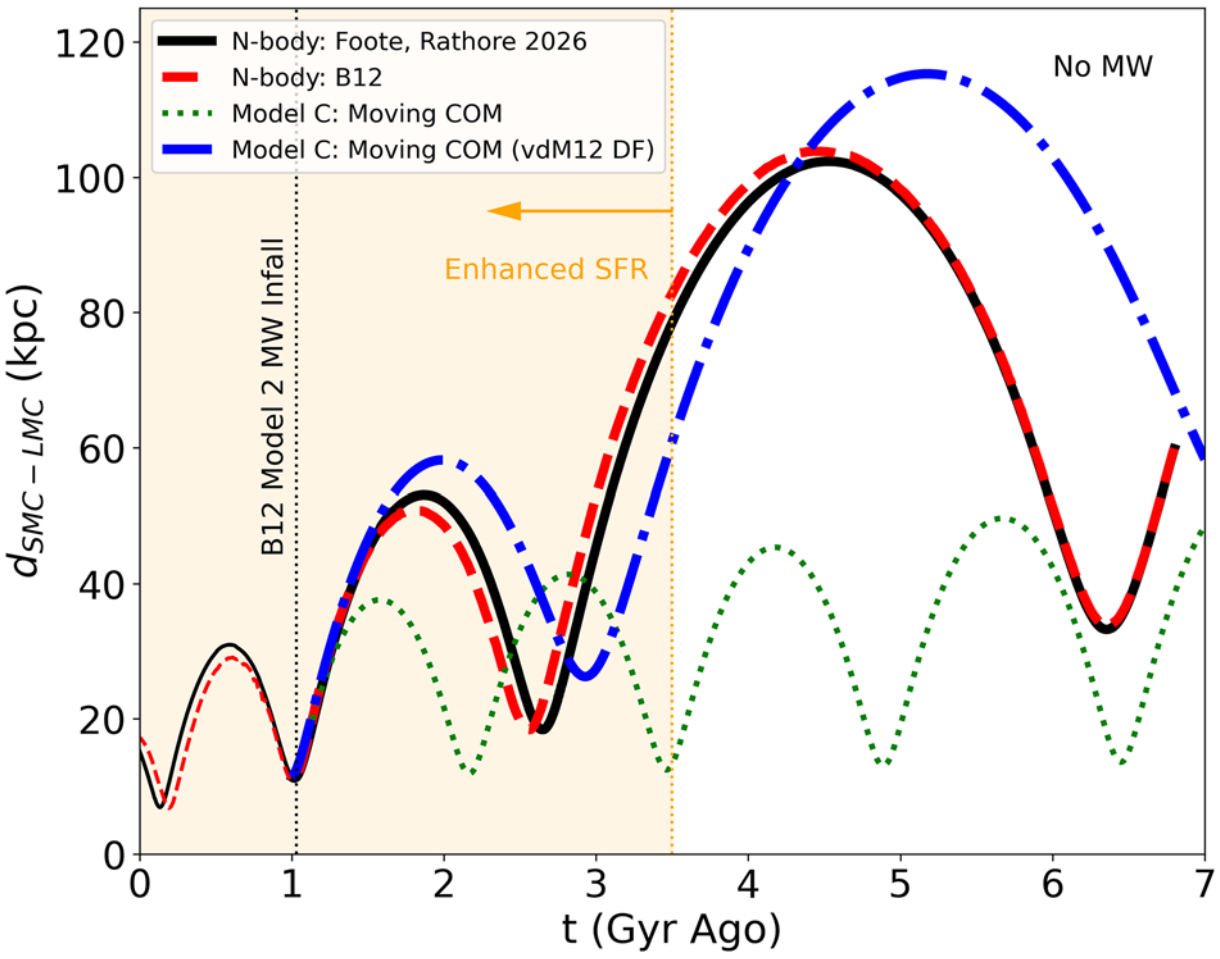

**Fig. 26** Separation of the Clouds as a function of time, where the LMC and SMC evolve in isolation, as in Fig. 25. Today is time=0, and infall is assumed to be ~1 Gyr ago, following Besla et al. (2012). The red line is the N-body orbit from Besla et al. (2012) and the black is from the higher resolution MEGHA simulations (created by H. Rathore Foote et al. 2026). The SMC is on an eccentric orbit ($e = 0.7$) and completes 2 orbits about the LMC before infall to the MW (vertical dashed line). The shaded region denotes the timescale over which the star formation rate is observed to be enhanced in the Clouds ($>$ 2.3 times the average rate, e.g. Massana et al. 2022), and corresponds to separations between the Clouds $<$ 50 kpc. The green dotted line is the orbit reconstruction using the fiducial dynamical friction formalism for the Model C: Moving COM framework, and the fiducial SMC model, starting at the MW infall epoch and going backward in time (t> 1 Gyr ago). The blue dashed line is the same but with the dynamical friction formalism of Van Der Marel et al. (2012b), which is calibrated for a 1:10 mass ratio encounter and provides a better match to the N-body/Model D results. This figure is adapted from Foote et al. (2026) and uses the MEGHA simulations created by H. Rathore. *Image credit: H. Foote, H. Rathore, and E. Patel (2026)*

N-body and hydrodynamic simulations is not feasible for this review. Instead, the analytic Model C: Moving COM framework provides a reasonable approximation for the allowed range of LMC/SMC orbital solutions over a given range of LMC/MW masses.

### 8.1 LMC orbit statistics

The range of orbital solutions possible for the LMC is explored using the HI+Phot Center data set and the Model C: Moving COM framework. The gravitational force of the SMC acting on the LMC is included in this analysis, but SMC orbits are analyzed in the next section.

All orbit solutions (across all LMC and MW mass models and the HI+Phot Center data set) find that the LMC is currently just past its most recent pericentric approach to the MW, which occurred at a distance of 45–50 kpc, 48–78 Myr ago.

Figure 27 illustrates the parameter space of the LMC and MW halo virial masses explored in this review. The virial masses are fixed in time, which means the LMC virial mass is also the infall mass. The color of the marker for each mass combination indicates the fraction of the HI+Phot Center data set, where the LMC completes one

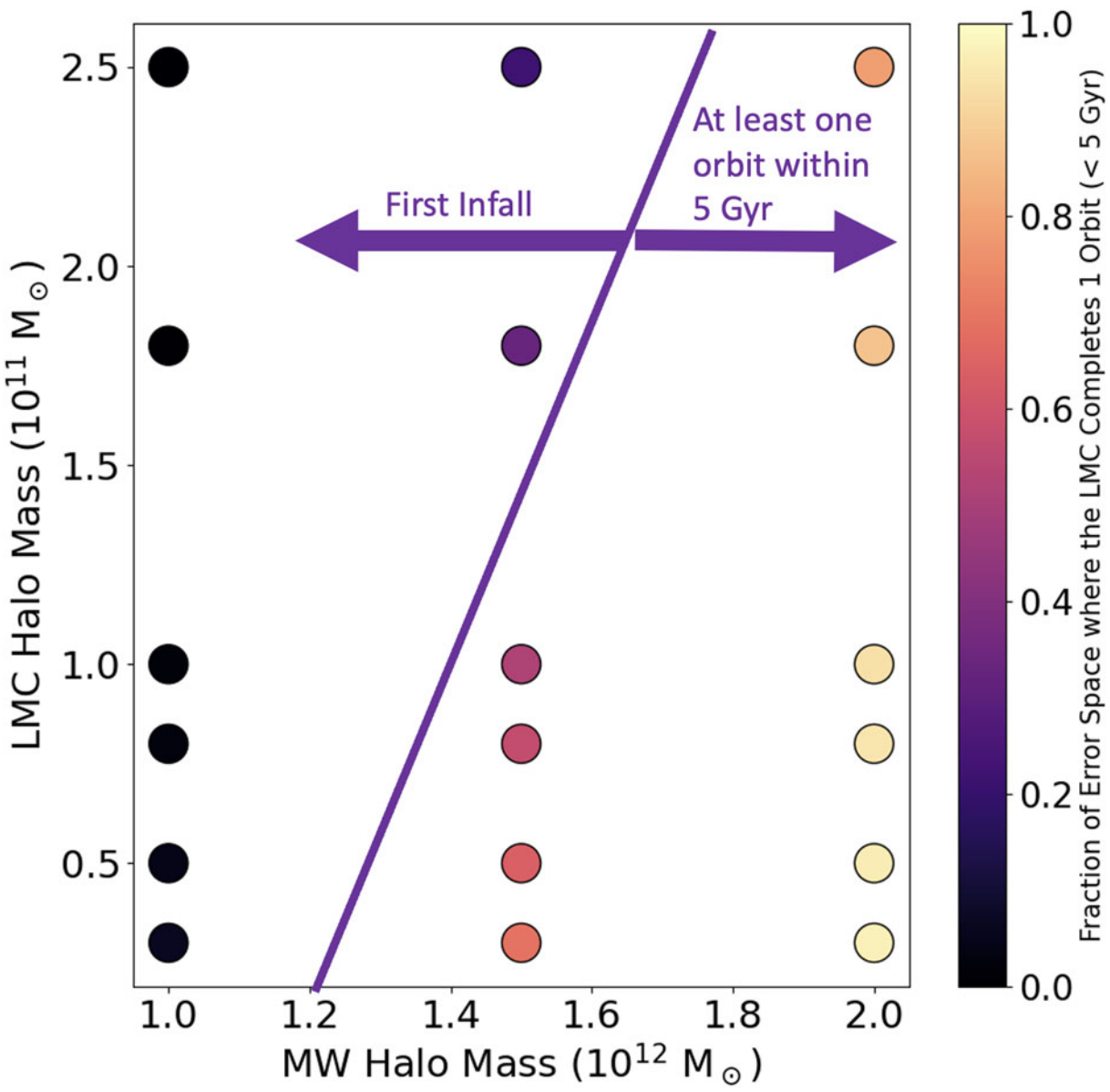

**Fig. 27** Fraction of the HI+Phot Centers data set of allowed present-day LMC position and velocity vectors, where the LMC completes 1 orbit within 5 Gyr (color bar). Orbits are computed using the Model C: Moving COM framework, including the SMC (See Sect. 6.4.2). Orbit solutions are computed using different combinations of fixed halo virial mass for MW and the LMC. The LMC and MW halos are modeled using the fiducial suite of NFW models (Tables 8, 9, 10). The purple diagonal line delimits, where the majority of allowed orbits (>50%) are first infall, in LMC and MW mass space. To the left of this line, the LMC is unlikely to complete an orbit about the MW within 5 Gyr

orbit about the MW within the past 5 Gyr. The diagonal line delimits the parameter space, where 50% of the allowed LMC orbits do not complete an orbit. To the left of this line, the majority of allowed LMC orbits are first infall solutions, wherein the the LMC does not complete an orbit (i.e., two pericentric approaches) about the MW within 5 Gyr.

For the lowest MW mass model (MW1) there are almost no solutions, where the LMC completes an orbit within 5 Gyr. The LMC completes an orbit for $<1\%$ of the error space for high mass LMC models and, at best, $\sim 8\%$ of the error space for the lowest LMC mass model. For the intermediate MW mass model (MW2), if the LMC mass is $\geq 1.8 \times 10^{11}$ $M_{\odot}$, the LMC is on first infall (only 40% of the error space allows an orbit to be completed at this LMC/MW mass scale). For the heavy MW model (MW3) the LMC always completes an orbit within 5 Gyr and is not on its first infall.

In Fig. 28, the range of LMC and MW masses adopted as initial conditions in Models A-D (Table 6) are plotted over Fig. 27. Select studies are highlighted in this parameter space. This figure illustrates the progression in parameter choices for MW and LMC halo masses, and, therefore, of LMC orbit solutions, over time. This progression tends toward “first infall”, high present-day LMC mass, and low MW mass solutions (i.e., evolving from Model A to Model D).

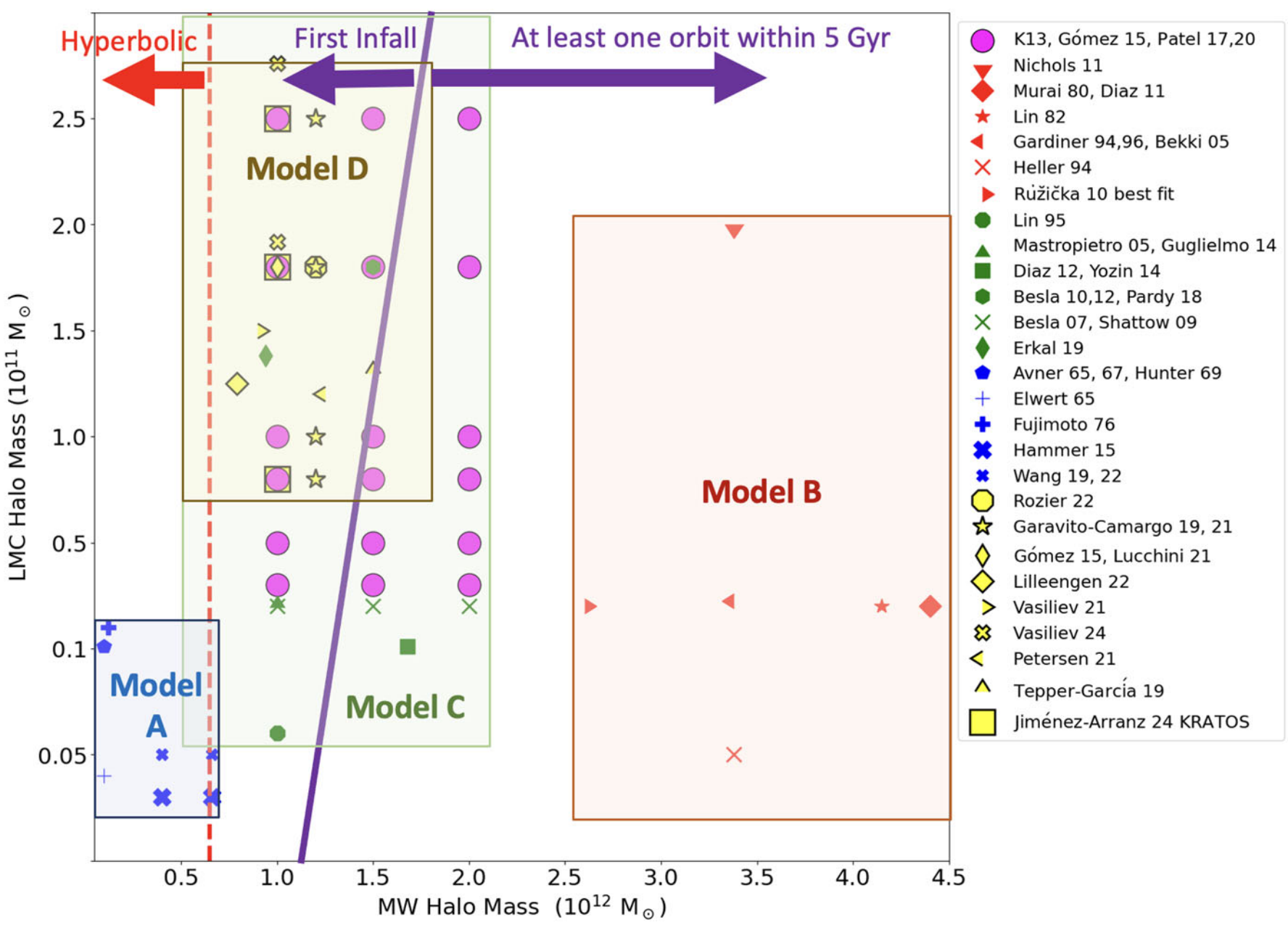

**Fig. 28** Range of LMC and MW masses explored by studies that use the Model A-D frameworks defined in Table 6 are plotted over Fig. 27. The solid purple line again denotes first infall orbits in the Model C: Moving COM framework. The LMC and MW halo masses are the virial mass for Models C and D. For Model A and B studies, the halo mass refers to the MW mass within 300 kpc and LMC mass within 100 kpc. The red dashed line indicates the minimum MW virial mass needed for the LMC to be currently bound to the MW ($M_{vir}$ =6.5 × $10^{11}$ $M_\odot$ NFW halo with all other parameters as in MW1), assuming HI Center or Phot Center velocities ($v_L \sim 320$ km/s, $r_L$= 50 kpc). Model A solutions are largely hyperbolic orbits. Studies from Model C like Kallivayalil et al. (2013); Gómez et al. (2015); Patel et al. (2017); Zivick et al. (2018); Patel et al. (2020) span the range of LMC and MW masses covered by the magenta circle markers. The work of Jethwa et al. (2016) spans the range of values denoted by the size of the Model C box. Studies such as Růžička et al. (2010); Guglielmo et al. (2014) vary the range of the MW mass, rather than LMC mass; marked are their best fit values. Markers for Garavito-Camargo et al. (2019, 2021); Besla et al. (2007); Hammer et al. (2015); Wang et al. (2019); Jiménez-Arranz et al. (2024) are repeated to cover the range of parameter combinations explored in these studies. The Sheng et al. (2024) study consists of many N-body simulations that probes roughly the range covered by the Model D box. Over time, studies have progressed toward first infall, low MW mass, high LMC mass solutions (A to D)

The Model C analytic framework does not account for the LMC's mass loss or the MW's mass growth. In studies where the LMC is treated as an N-body system (e.g., Model D), the present-day halo mass of the LMC will differ from the virial mass/infall mass. As such, the exact orbit solution predicted by the Model C analytic framework will deviate from the N-body solution (see Fig. 24). Instead, the Model D markers in Fig. 28 demonstrate how the LMC and MW initial conditions adopted by N-body studies would map onto the orbit solutions in the Model C framework. More work is needed to understand in detail the delimitation between first infall and multiple orbit solutions in N-body MW+LMC models.

Figure 29 illustrates the mean time (Gyr ago) of the LMC's most recent inward crossing of the MW's virial radius, as a function of the range of LMC and MW halo

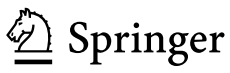

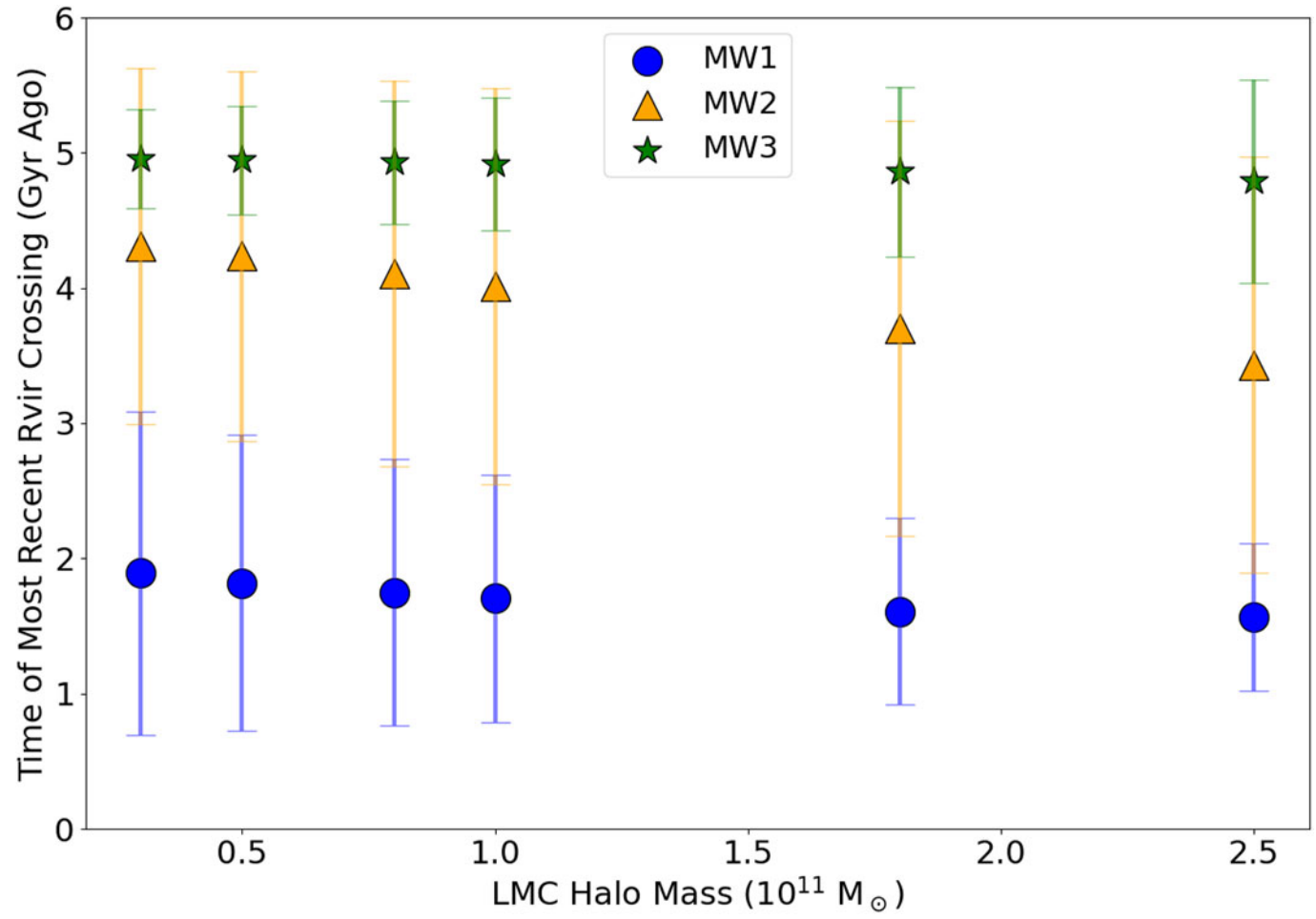

**Fig. 29** Mean time elapsed, since the LMC's most recent inward crossing of the MW's $R_{vir}$ (i.e., time of infall in units of Gyr ago), plotted as a function of LMC halo virial mass. Orbits are computed using the Model C: Moving COM framework (including the SMC) with the fiducial set of LMC, SMC, and MW NFW models. The infall time is computed as the mean across all ∼9200 Monte Carlo drawings in the HI+Phot Center data set. Error bars indicate the standard deviation of the mean and symbols indicate the MW model. For the $1.8 \times 10^{11}$ $M_\odot$ LMC mass model, the mean infall time is: MW1, $1.6 \pm 0.7$ Gyr ago; MW2, $3.7 \pm 1.5$ Gyr ago; and MW3, $4.9 \pm 0.6$ Gyr ago

virial masses explored in this review. This crossing time also corresponds to the LMC's infall time. For the MW1 (MW2, MW3) model, the LMC infall time was 1–3 (3–6, ∼5) Gyr ago. For the most massive MW model (MW3), the majority of the allowed error space supports solutions, where the LMC orbits within the MW's virial radius for the past 5 Gyr.

**Summary:** The range of LMC orbital solutions allowed within 9200 Monte Carlo drawings of the combined HI+Phot Center data set for the present-day position and velocity of the Clouds is explored across all MW and LMC mass models using the Model C: Moving COM framework. All orbits agree that the LMC is just past a pericentric approach to the MW, which occurred at a distance of 45–50 kpc, 48–78 Myr ago. For the lowest MW mass model (MW1) the LMC is on first infall for >90% of the error space across all LMC models (infall 1–3 Gyr ago). In MW2, if the LMC mass is $\geq 1.8 \times 10^{11}$ $M_\odot$, the LMC is on first infall (infall 3–6 Gyr ago). For the heavy MW model (MW3) the LMC always completes an orbit within 5 Gyr and is not on its first infall.

## 8.2 SMC orbit statistics

In this section, the range of orbital solutions for the SMC allowed within the combined HI+Phot Center data set are explored. LMC and SMC orbits are computed using the Model C: Moving COM framework, including the Clouds' gravitational influence on each other and dynamical friction owing to the SMC's motion through the LMC's halo.

All orbit calculations predict a recent close encounter between the Clouds, 115–210 Myr ago at a separation of 5–17 kpc (see Fig. 34). The nature/significance of this recent encounter is discussed in more detail in Sects. 9.2 and 9.3.

Within the HI+Phot Center data set, the relative velocity of the Clouds can exceed the LMC's escape speed, particularly for solutions using the Phot Center. This raises the question of the likelihood that the SMC is presently bound to the LMC. For the lowest mass LMC model ($3 \times 10^{10}$ M$_\odot$; Table 9), the SMC is currently unbound to the LMC for ∼70% of solutions with the HI+Phot Center data set. This is largely driven by the Phot Center relative velocities, which are comparable to the escape speed (Eq. 16) for LMC models with mass $< 8 \times 10^{10}$ M$_\odot$. If the mass of the LMC is $\gtrsim 8 \times 10^{10}$ M$_\odot$, the unbound fraction at present day drops to <5%.

The probability that the SMC is a long-lived binary companion to the LMC is critical for understanding the orbital evolution of the SMC. Figure 30 illustrates the fraction of the combined HI+Phot Center data set that allows for long-lived binary LMC–SMC orbital solutions, plotted as a function of the LMC and MW virial masses discussed in this review.

A long-lived binary orbit is defined, such that the SMC is bound to the LMC ($\mathrm{V_{S-L}} < V_{escL}$) for the past 5 Gyr, maintaining separations of $\mathrm{R_{S-L}} \lesssim 100$ kpc (see Eq. 16 and the bottom panel of Figs. 20, 21, 22). As discussed at the end of Sect. 1.3, there is significant evidence that the Clouds have maintained a binary state for at least 5 Gyr.

Binary LMC–SMC solutions are delimited by the red line in Fig. 30. To the left of this line, over 50% of orbital solutions allow for a binary LMC–SMC orbital solution that lasts for at least 5 Gyr.

In the case of non-binary solutions in MW2 and MW3, the SMC is always bound to the MW, making multiple orbits over a Hubble time within 200 kpc of the MW COM with an orbital period of 2–3 Gyr (see Figs. 21 and 22). This scenario is at odds with the SMC's current high gas fraction and current theoretical models for satellite quenching. Furthermore, in all cases, non-binary solutions require the SMC to be captured by the LMC, while the LMC makes its most recent approach to the MW (see, e.g., Figs. 21 and 22).

The requirement that the SMC is bound to the LMC for the past 5 Gyr significantly constrains the range of plausible LMC and MW mass combinations. As demonstrated in Sect. 6.4.2, non-binary SMC orbits make very close pericentric passages with the MW, which is at odds with its current high gas fraction. Binary LMC–SMC solutions favor high mass LMC models ($> 10^{11}$ M$_\odot$) and low mass MW models ($< 1.5 \times 10^{12}$ M$_\odot$), which maximizes the relative strength of the LMC's gravitational force on the SMC. These long-lived binary solutions are also all "first infall" orbits. This is because if the Clouds entered the halo as a binary and made two orbits about the MW, the tidal field of the MW would split the binary apart on the *first* pericentric approach (see Sect. 10).

In the Model D framework, the SMC is modeled with a relatively massive N-body halo (1:10, LMC:SMC) that has suffered mass loss and has induced significant perturbations in the LMC's dark matter halo (Foote et al. 2026). As illustrated in Fig. 25, the Model D SMC orbit about the LMC is consequently more eccentric than the standard Model C: Moving COM solutions (where dynamical friction is modeled with Eq. F33). This means that the boundary for LMC–SMC binarity in the MW–LMC mass parameter space (red line in Fig. 30) is a *lower* bound on the LMC mass and an *upper* bound on the MW mass, assuming the SMC was at least $2 \times 10^{10}$ $M_{\odot}$ when it was captured by the LMC. In other words, the Model D framework will only make it *harder* to maintain an LMC–SMC binary (and potentially place even smaller upper limits on the mass of the MW).

In the Model C framework, the fiducial SMC model is relatively low in mass (∼2–17% the mass of the LMC), making the mass loss experienced by the SMC less of a problem. To reproduce the SMC orbit in the Model C framework using the more massive SMC models of Model D, mass loss can be mimicked by adopting a different dynamical friction formalism. We follow the Appendix of (van der Marel et al. 2012a) and adopt a formalism for the Coulomb Logarithm that is calibrated for a 1:10 mass ratio N-body simulation (see Appendix Eq. F.2). As shown in Fig. 25, this formalism can reasonably recover the SMC N-body orbit. As in Model D solutions, the new Model C SMC orbit for a 1:10 encounter is significantly more eccentric than that predicted by the standard Model C dynamical friction (see Fig. 25). A more eccentric orbit would make it *harder* for the LMC–SMC binary to survive, decreasing the fraction of binary solutions illustrated in Fig. 30.

Figure 31 plots the mean orbital eccentricity and period of the SMC's most recent orbit. Means are computed for the fraction of the HI+Phot Center data set, where an orbit is completed within the past 5 Gyr, for the given MW model (symbol shape) and LMC model (symbol color).

If the SMC completed at least one orbit about the LMC within 5 Gyr, its orbit was highly eccentric ($e \sim 0.5 - 0.7$) with an orbital period ranging from 1–3 Gyr. This is the time frame of relevance for episodic star formation that is correlated with the tidal forces between the Clouds. Roughly circular orbits of the SMC about the LMC are unlikely, which is in contrast to the assumptions made in most Model B studies.

**Summary:** The range of SMC orbital solutions allowed within 9200 Monte Carlo drawings of the combined HI+Phot Center data set for the present-day position and velocity of the Clouds is explored across all MW and LMC mass models using the Model C: Moving COM framework. The SMC is presently bound to the LMC ($V_{\mathrm{S-L}} < V_{escL}$) for LMC models with mass $\gtrsim 8 \times 10^{10}$ $M_{\odot}$. Long-lived binary LMC–SMC solutions (where the SMC is bound to the LMC over the past 5 Gyr) require that: the LMC is on first infall, the virial mass of the MW is less than $1.5 \times 10^{12}$ $M_{\odot}$, and the LMC's mass is $> 10^{11}$ $M_{\odot}$ at present day. For binary solutions, the SMC's orbital eccentricity is ∼0.5–0.7 and orbital period ranges from ∼1–3 Gyr. If the SMC is not bound to the LMC, the SMC orbits within the MW halo and the LMC randomly captures the SMC while making its most recent approach to the MW.

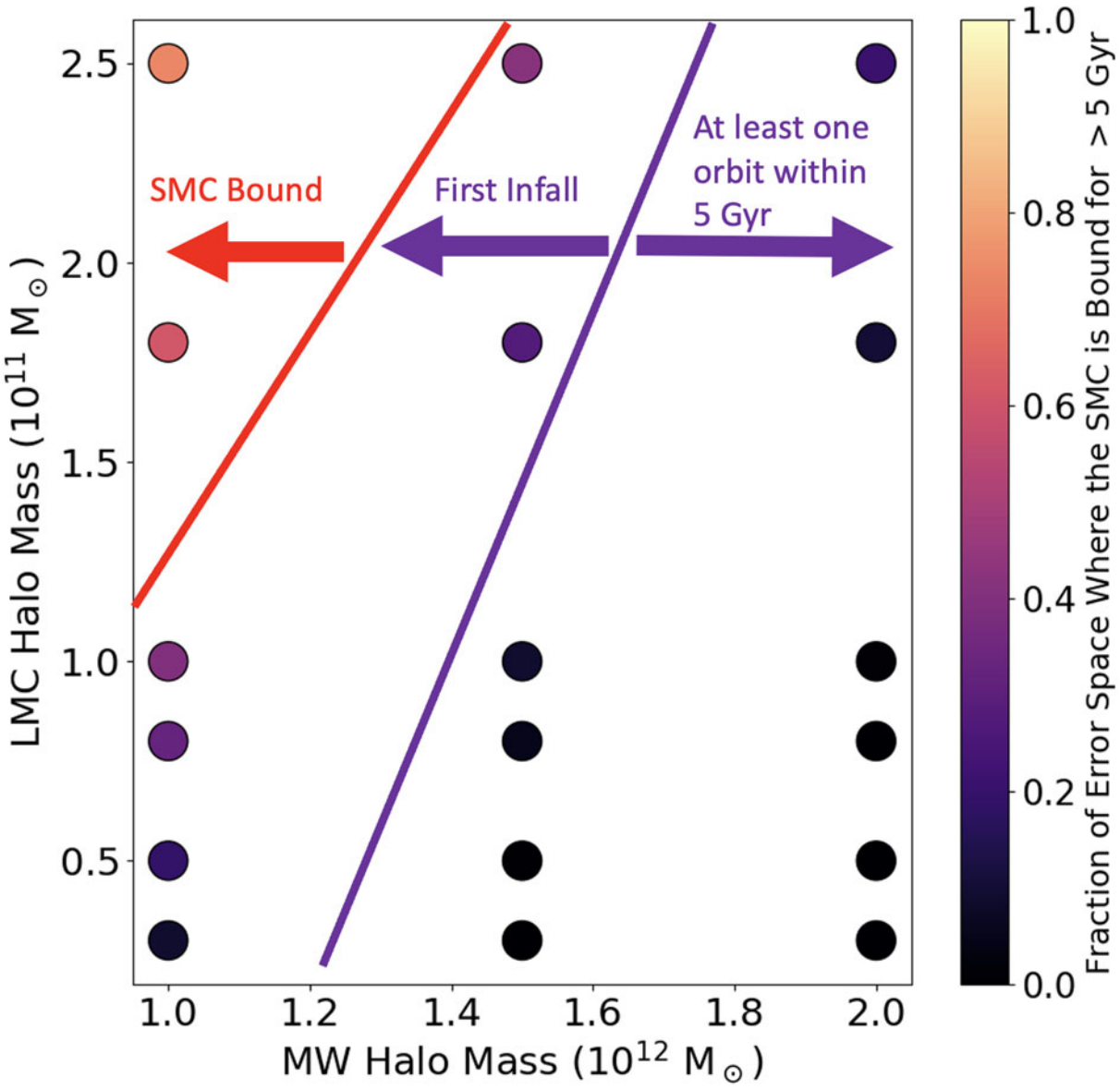

**Fig. 30** Fraction of the HI+Phot Center combined error space, where the SMC is bound to the LMC over the past 5 Gyr, plotted as a function of fixed LMC and MW halo virial mass. The color scale indicates the fraction of the ∼9200 Monte Carlo drawings, where the relative speed of the SMC with respect to the LMC remains less than the local LMC escape speed for 5 Gyr ($V_{S-L}/V_{escL} < 1$). Orbits are computed using the Model C: Moving COM framework. Galaxies are modeled using the fiducial suite of NFW halos for the MW, LMC, and SMC. The purple diagonal line delimits first infall orbits, in LMC and MW mass space (from Fig. 27). The red diagonal line roughly delimits, where the SMC is bound to the LMC for the majority (>50%) of orbits allowed within the error space. The requirement that the SMC remain bound to the LMC for at least 5 Gyr favors first infall orbits, where the LMC mass is high and MW mass is low. For MW1, the LMC must be $\geq 10^{11}$ M$_\odot$ today to maintain a binary state for the majority of the error space. There are no long-lived binary state solutions for the majority of the error space using the MW3 or MW2 models. In the Model D framework, the SMC–LMC orbit will be more eccentric, since the SMC's mass loss is accounted for (Fig. 25), this will make it *harder* for the LMC to keep the SMC as a binary, requiring even higher LMC masses or lower MW masses than indicated by Model C

## 9 Orbital geometry of the LMC–SMC binary and recent interaction

In the previous section, it was determined that long-lived (> 5 Gyr) LMC–SMC binaries favor low mass MW and high mass LMC models. If an SMC orbit is completed about the LMC, the orbital period is 1–3 Gyr and eccentric. In this section, the geometry of the orbit of the SMC about the LMC is examined, using constraints and insights from observations and theoretical studies.

Table 11 lists the present-day relative position and velocity vectors of the SMC with respect to the LMC in Galactocentric coordinates for a subset of observational and theoretical studies. The theoretical studies included in this table are those that have built N-body and/or hydrodynamic simulations of the Stream. These include Model D studies with live MW halos: Jiménez-Arranz et al. (2024) (KRATOS, their K3 model), Lucchini et al. (2021) (L21), and Tepper-García et al. (2019) (TG19, their standard model). Model C: Fixed COM studies (static MW): Craig et al. (2022) (C22,

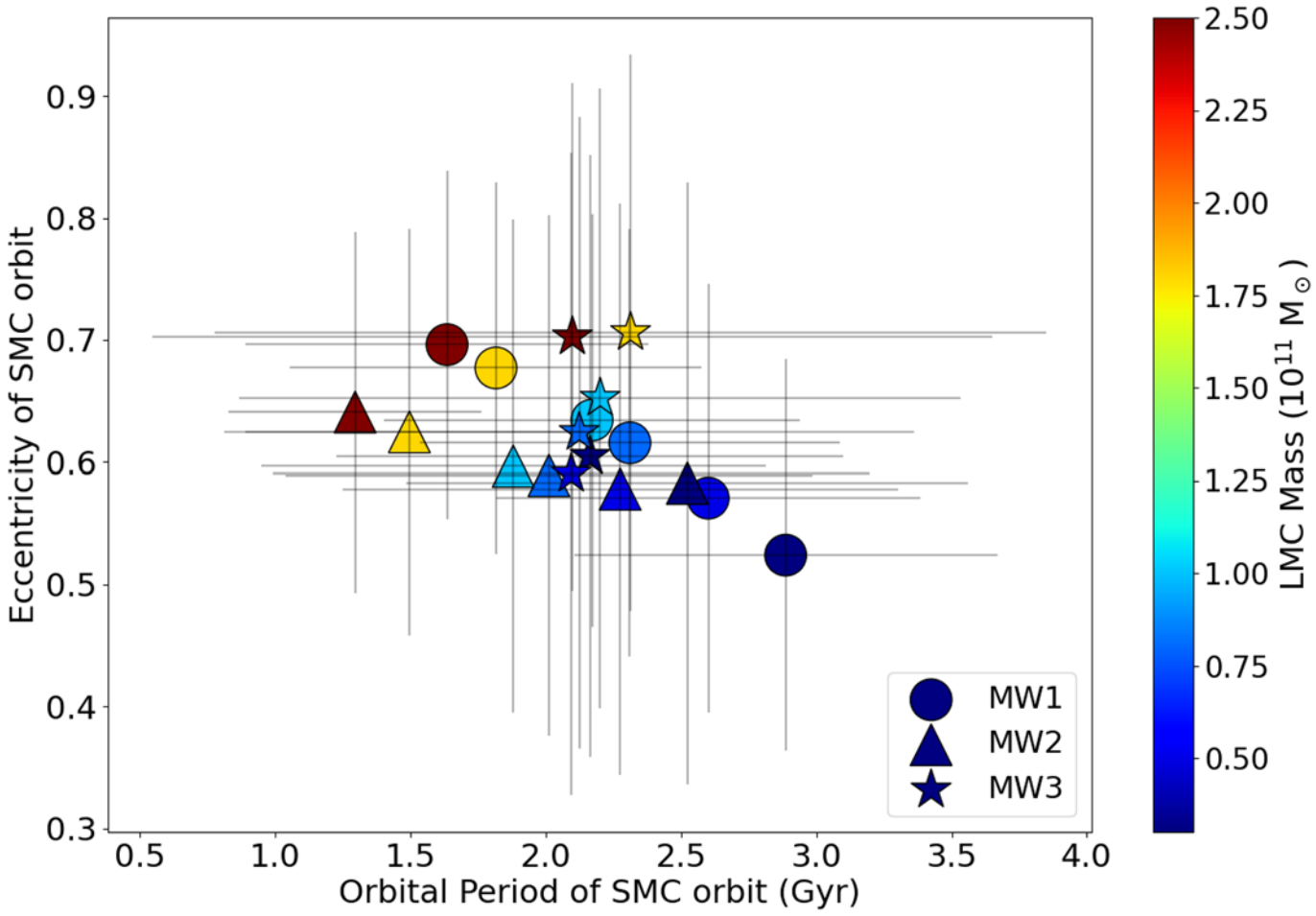

**Fig. 31** Mean orbital eccentricity vs. orbital period of the SMC's most recent orbit about the LMC. Quantities (mean and standard deviations) are computed only for the subset of SMC orbits that complete at least one orbit about the LMC within the past 5 Gyr, within the HI+Phot Center error space. Orbits are computed in the Model C: Moving COM framework. Symbols indicate the MW mass model and the color of the markers indicates the mass model of the LMC. Only high mass LMC models ($> 10^{11}$ M$_\odot$) yield solutions, such that the SMC has completed one orbit about the LMC within 5 Gyr over more than 50% of the HI+Phot Center error space. The mean eccentricity and orbital period for the most recent orbit of the SMC about the LMC is $e \sim 0.5 - 0.7$ and 1–3 Gyr, respectively, across all mass models. Circular orbits for the SMC are unlikely (e.g., Model B solutions)

their M100 model), Pardy et al. (2018) (P18, their 1:9 SMC:LMC mass model), and Besla et al. (2012) (B12, their Model 2). Model B studies: Diaz and Bekki (2012) (hereafter DB12), Gardiner and Noguchi (1996) (hereafter GN96), and Bekki and Chiba (2005) (B05; parameters are equivalent to GN96). Model A study: Wang et al. (2019) (hereafter W19, their model 28).

In all studies listed in Table 11, both the LMC and SMC are live N-body systems, with the exception of DB12, B05, and GN96, where the LMC is modeled as a static potential.

The goal of the next subsections is to understand the orbital geometry of the SMC–LMC binary that has been adopted by different studies, in comparison with the latest proper motion data from Gaia and HST. In particular, the nature of the LMC–SMC's most recent encounter will be discussed in the context of a direct collision or a wider encounter.

### 9.1 The geometry of the SMC's orbit about the LMC

The relative position and velocity of the SMC listed in Table 11 are converted to a coordinate system centered on the LMC disk. Following (Choi et al. 2022), the normal of the LMC disk (last line in Table 11) is transformed to the vector [0,0,1]. Note that the LMC disk rotates clockwise on the sky, and so the LMC Disk Normal is opposite of the LMC disk angular momentum vector in this frame.

**Table 11** Modeled SMC–LMC relative position and velocity vector

| Study | $R_{S-L}$ (kpc) | $V_{S-L}$ (km/s) | Model |
|---|---|---|---|
| C22 | [19,8,0] | [−1,21,−28] | C |
| L21 | [11.3, 1.4, −14.4] | [11.3, −24.3, −60.6] | D |
| TG19 | [0.6, −2.7, −3.0] | [73.4, −38.4, 136.1] | D |
| W19 | [12.6, 3.0,−16.3] | [32.8, 16.0, −57.1] | A |
| P18 | [−1.0, 3.0, −11.0] | [42.0, −33.0, 3.0] | C |
| B12 Model 2 | [7.0, 3.0, −9.0] | [16.0, 5.0, −51.0] | C |
| DB12 | [16.1, 4.7, −16.3] | [46.5, 2.6, −38.3] | C |
| GN96, B05 | [14.6, 6.5, −13.0] | [45.0, 40.0, −23.0] | B |
| LMC Disk Normal | [0.133, 0.963, 0.235] | | Choi et al. (2022) |

Present-day SMC position (velocity) vector relative to the LMC, $R_{S-L}$ ($V_{S-L}$) determined in theoretical studies. All values are in Galactocentric coordinates. Note that the LMC Disk Normal is the opposite of the LMC disk angular momentum vector. W19 (Wang et al. 2019) values for their model 28 were obtained from private communication (Wang, Jan 2025), as were those for TG19 (Tepper-García et al. 2019) (Tepper-Garcia, private communication Jan 2025, standard model, live MW halo), and C22 (Craig et al. 2022) (M100 model, correcting their Table 3 LMC velocity vector to [-47, -179, 265] km/s; Craig, private communication Jan 2025)

The SMC's orbit in the LMC Disk Frame is plotted in Fig. 32. The LMC disk is marked by the orange disk, with a radial extent of 18 kpc (Mackey et al. 2018). In the LMC Disk Frame, an observer at the Sun is located in the positive z direction.

The grey squares indicate the allowed range of positions for the SMC, relative to the LMC, from the combined HI+Phot Center data set. Notably, the SMC is currently located above *or* below the LMC disk plane. The HI+Phot Center data set was randomly sub-sampled to demonstrate examples of plausible SMC–LMC relative velocity vectors (blue and red arrows/squares; the color coding will be discussed in the next section).

The relative orbit for the SMC using the mean HI Center and Phot Center velocities are computed in the Model C: Moving COM framework. The SMC is currently moving in a counter-clockwise orbit about the LMC in all cases. This is consistent with the assumptions made in Model B studies.

In Fig. 33, the range of specific angular momentum vectors of the SMC with respect to the LMC ($J_{S-L}$) from the 9200 Monte Carlo drawings are plotted in the Galactocentric (top panels) and LMC Disk (bottom panels) frames. The color bar indicates the number of Monte Carlo drawings of the HI+Phot Center error space. The resulting $J_{S-L}$ error space encompasses a range of SMC orbital geometries, from polar to coincident with the LMC disk plane.

Circles mark the $J_{S-L}$ derived using the mean HI or Phot Center velocities. In the LMC disk frame, the HI Center relative angular momentum vector points nearly in the LMC disk plane (XY), meaning the corresponding SMC orbit is roughly polar with respect to the LMC disk plane (normalized $J_{S-L}$ = [−0.64, −0.72, −0.26]). Most of the allowed orbits using the HI center are long-lived binary LMC–SMC orbits in low mass MW and high mass LMC models.

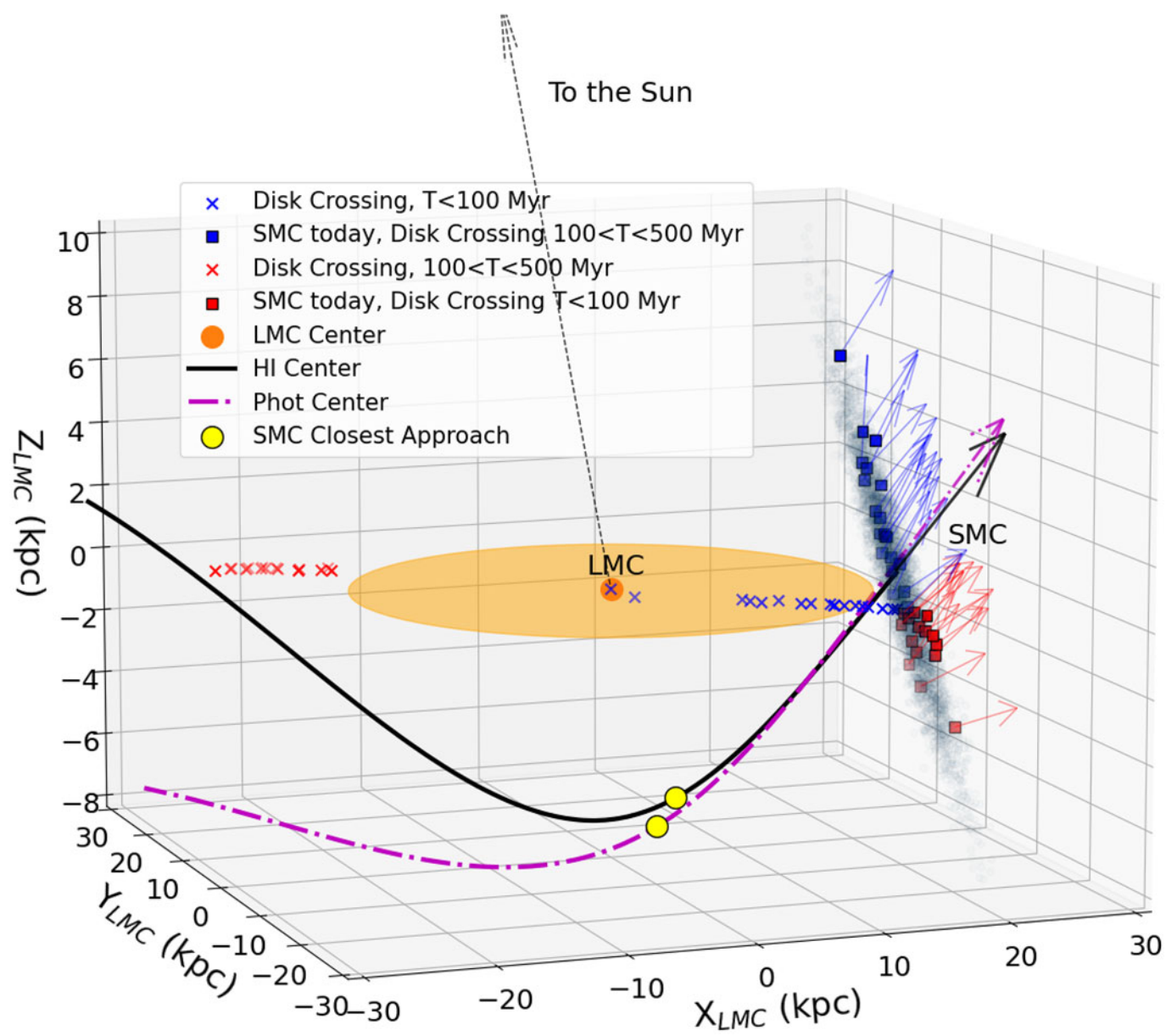

**Fig. 32** 3D visualization of the SMC orbit about the LMC in the LMC's disk frame, including the range of allowed relative positions (grey boxes), and example disk crossings (marked by "x"). The LMC disk frame is defined following Choi et al. (2022), where the normal in the Galactic frame is nLMC = [0.133, 0.963, 0.235]. The LMC's disk angular momentum vector is the negative of this vector in this frame. The LMC's stellar disk is denoted by the orange circle (radius 18 kpc), and is in the XY plane in this coordinate frame. The line-of-sight vector to the Sun is indicated by the black dashed arrow. The SMC's orbit is plotted for the past 500 Myr, using the HI Center (black, solid) and the Phot Center (magenta, dashed-dotted) mean velocities and positions. The orbits computed using the Model C: Moving COM framework, assuming, MW1, an LMC mass of $1.8 \times 10^{11}$ $M_\odot$, and SMC mass of $5 \times 10^9$ $M_\odot$. In all cases, the SMC is moving in a counter-clockwise direction in this frame (moving to the upper right, as indicated by the arrows). The yellow circles denote the location of the closest approach between the Clouds (7.4 kpc the HI Center and 8.9 kpc for the Phot Center). The closest approaches occur below the LMC disk plane (Cullinane et al. 2022a). The grey squares span the range of SMC positions relative to the LMC in the HI+Phot data set. The SMC can be above or below the disk plane of the LMC, leading to a different disk crossing distance and time. A subset of the parameter space is explored, where the SMC is above the disk plane today (blue squares) or below (red squares). The location where the corresponding orbit crosses the LMC's disk plane is indicated by the blue or red "x". If the SMC is currently above the LMC disk plane, the disk crossing time is $<$ 100 Myr, with disk crossing distances that range from the center to the outskirts of the LMC's stellar disk. These are "Direct Impact" solutions. If the SMC is currently below the disk plane, the disk crossing distance occurs $>$ 100 Myr ago, at distances much larger than the extent of the LMC's disk (e.g. Cullinane et al. 2022a)

For the Phot Center, the mean relative angular momentum vector points in the -X,-Y,-Z octant, in the LMC disk frame. Meaning the SMC's orbit is not polar. The normalized $J_{S-L} = [-0.53, -0.63, -0.57]$. The discrepancy with the HI Center result is in $JZ_{S-L}$, whereas $JX_{S-L}$ and $JY_{S-L}$ agree reasonably well.

In the Galactocentric frame (bottom panels), the present-day relative angular momentum vector of the SMC adopted by existing N-body simulations of the Clouds is compared against observations. The K3 KRATOS simulation does not have a MW disk with which to define a Galactocentric coordinate system and could not be straightforwardly included in this comparison.

None of the listed SMC models from the literature have present-day angular momentum vectors that agree with the observations within the $1\sigma$ error space. This is a result of the challenging nature of modeling a live 3-body system. This challenge is illustrated in Fig. 2 from Rathore et al. (2025a), which demonstrates, using the B12 Model 2 simulation, that the sense of the SMC's most recent orbit about the LMC is altered by the tidal field of the MW. In the simulation, the MW causes the SMC to collide with the LMC in a sense opposite to its previous orbit. Consequently, the observed relative angular momentum vector may not define the long-term orbital plane and orbital sense of the SMC–LMC binary. N-body simulations are required to properly understand the geometry of the recent orbit of the SMC about the LMC in the face of mass loss, halo deformations, and the MW's tidal field.

**Summary:** For long-lived, binary LMC–SMC orbits (HI Center solutions), the SMC is in a roughly polar orbit with respect to the LMC's disk plane. Using Phot Center velocities, the mean relative angular momentum vector points in the -X,-Y,-Z octant in the LMC Disk Frame. The uncertainty in the observed 3D Galactocentric position of the SMC implies that the SMC's center can be above or below the LMC disk plane. Reproducing the observed, present-day relative angular momentum vector between the Clouds is a challenge for all simulations, pointing to the complicated dynamics of time-evolving, three-body systems.

### 9.2 The most recent encounter between the clouds

The majority of studies discussed in this review, observational and theoretical, agree that the LMC and SMC have recently ($< 200$ Myr) had a close approach ($< 40$ kpc). This close approach likely resulted in the formation of the Bridge of gas that connects the two galaxies (e.g., Gardiner and Noguchi 1996).

However, the relative distance (impact parameter), timing, and location relative to the LMC's disk plane, of this close encounter are uncertain. The uncertainty stems from two issues.

First, the Clouds have recently approached each other well within their virial radii. As such, orbits will not be well traced using analytic backward integration schemes that do not account for the severe deformation of the dark matter distributions of the LMC, SMC, and MW (Garavito-Camargo et al. 2021; Foote et al. 2026).

Second, the SMC's center of mass position has a large uncertainty (Sect. 2), with significant implications for the SMC's recent trajectory with respect to the LMC's disk plane.

Addressing the first issue (a deforming 3-body system) requires new N-body simulations. The second issue can be examined within the analytic Model C: Moving COM framework using the HI+Phot Center data set.

In Fig. 34, the mean impact parameter for the most recent LMC–SMC encounter is plotted as a function of the timing of the encounter. These values are computed

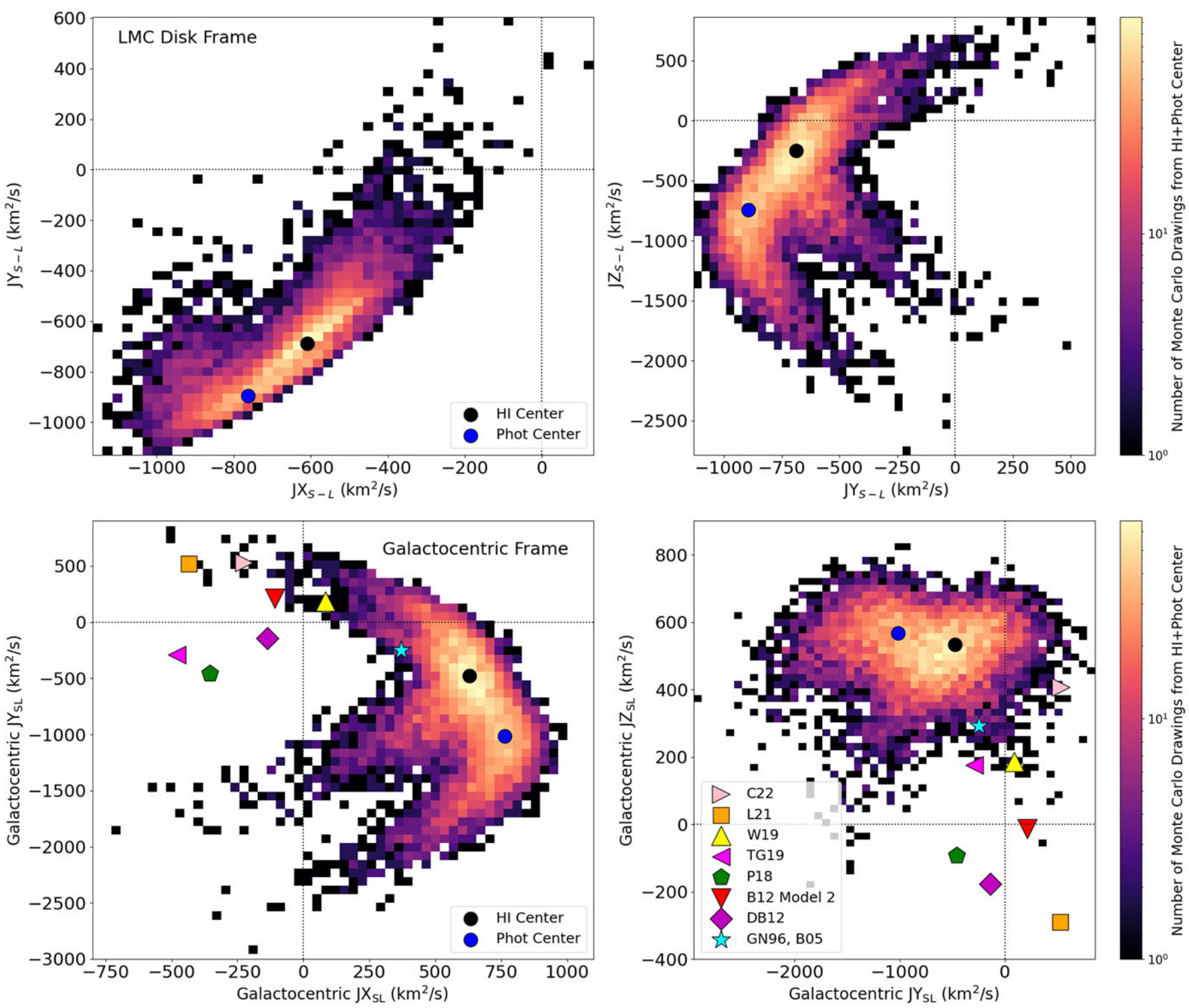

**Fig. 33** Specific orbital angular momentum vector of the SMC relative to the LMC, $J_{S-L}$ plotted in the LMC Disk (top panels, see Fig. 32) and Galactocentric (bottom panels) frames (X,Y,Z). The $J_{S-L}$ vector also denotes the normal to the SMC's orbital plane about the LMC. The colored points represent the values of $J_{S-L}$(X,Y,Z) in the observed HI+Phot Center data set. Yellow circles indicate the means for the HI Center and Phot Center data, separately. In the LMC Disk frame, the observed orbital pole of the SMC about the LMC is predominantly in the −X, −Y, direction and can be near polar ($JZ_{S-L}$ close to 0; HI Center) or coincident with the LMC's disk plane ($JY_{S-L}$ close to 0). In the Galactocentric frame, additional markers indicate $J_{SL}$ for specific N-body studies: Model B: (Gardiner and Noguchi 1996, GN96); Model C: Fixed COM (Diaz and Bekki 2012, DB12) (Besla et al. 2012, B12Model2) (Pardy et al. 2018, P18); and Model D (Lucchini et al. 2021, L21). The fiducial KRATOS K3 simulation from (Jiménez-Arranz et al. 2024, ModelD) is marked in the LMC disk frame, as this study does not include a MW disk to define a Galactocentric reference frame. None of the more recent N-body models are within 1 $\sigma$ of observations

for each combination of MW and LMC mass model explored in this review, using all velocities allowed within the HI+Phot Center data set. There is remarkably little spread in the resulting impact parameter and timing, illustrating why most studies in the literature agree that a recent encounter occurred. Accounting for all MW/LMC models and the HI+Phot Center error space, the most recent LMC/SMC encounter likely occurred 115–210 Myr ago at a separation (impact parameter) of 5–17 kpc.

As discussed in Zivick et al. (2018), the mean impact parameter of the recent LMC/SMC encounter is smaller than the the size of the LMC's stellar disk (radius of $\sim$ 18 kpc (Mackey et al. 2018)). This guarantees that the SMC tidal field had a non-negligible impact on the LMC's current disk structure and internal kinematics (Choi et al. 2022).

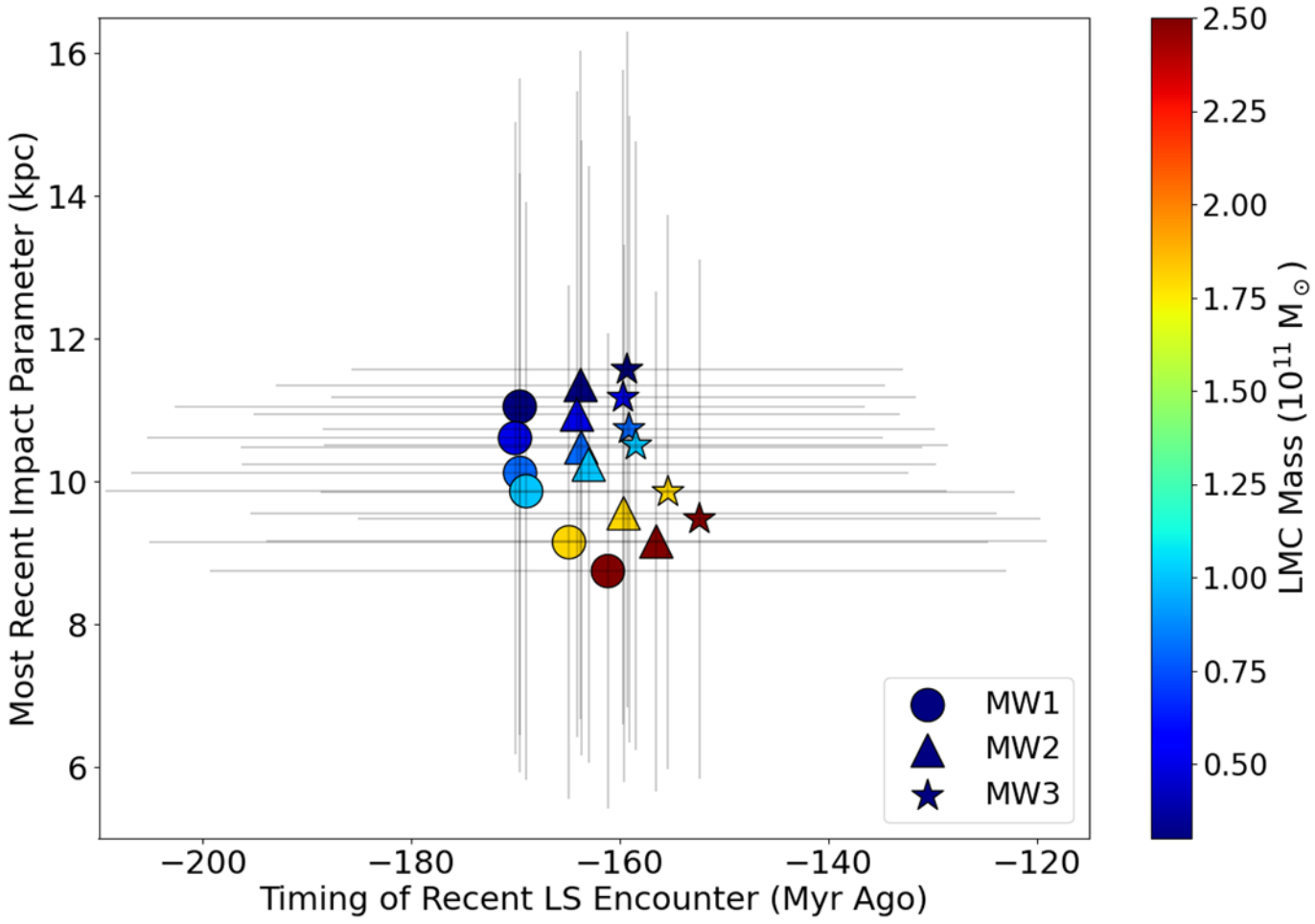

**Fig. 34** Mean impact parameter (distance of closest approach) and timing of the most recent LMC/SMC encounter. The color scale denotes the virial mass of the LMC and the symbol indicates the MW mass model. The orbits are computed using the Model C: Moving COM framework. Plotted are the mean (symbol) and $1\sigma$ error bars of 9200 orbits derived using each LMC/MW mass combination and present-day velocities and positions defined in the HI+Phot Center data set. The recent LMC/SMC impact parameter (5–17 kpc) and timing (115–210 Myr ago) are well-constrained, even when accounting for uncertainties in the centers and mass models. As concluded in Zivick et al. (2018), it is highly improbable that the recent impact parameter is larger than the size of the LMC's stellar disk (radius of $\sim$ 18 kpc, Mackey et al. 2018)

However, as discussed in Cullinane et al. (2022a), a small impact parameter does not necessarily imply that the SMC directly collided with the LMC's disk. Indeed, these authors report that the closest approach occurs *below* the LMC disk plane.

This issue is illustrated in Fig. 32, where the orbit of the SMC with respect to the LMC is plotted in the LMC Disk frame. The mean HI and Phot Center orbits do not encounter the disk directly. In other words, the distance between the LMC's disk center of mass and where the SMC's orbit crosses the LMC's disk plane ("disk crossing") is larger than the radius of the LMC's stellar disk. The closest approach of the SMC to the LMC is ∼7 kpc (HI Center) and ∼9 kpc (Phot Center), and is marked by the yellow circles. However, this minimum separation occurs below the disk plane in both cases (Cullinane et al. 2022a).

The question remains; does the entire observational error space rule out the possibility of a "Direct Impact" between the Clouds, wherein the most recent LMC disk crossing of the SMC occurs *within* the LMC's disk extent.

In Fig. 35, the expected distances and timings of the SMC's most recent LMC disk crossing are computed as a function of the corresponding present-day relative velocity of the SMC with respect to the LMC ($V_{S-L}$), as constrained by the HI+Phot Center data set.

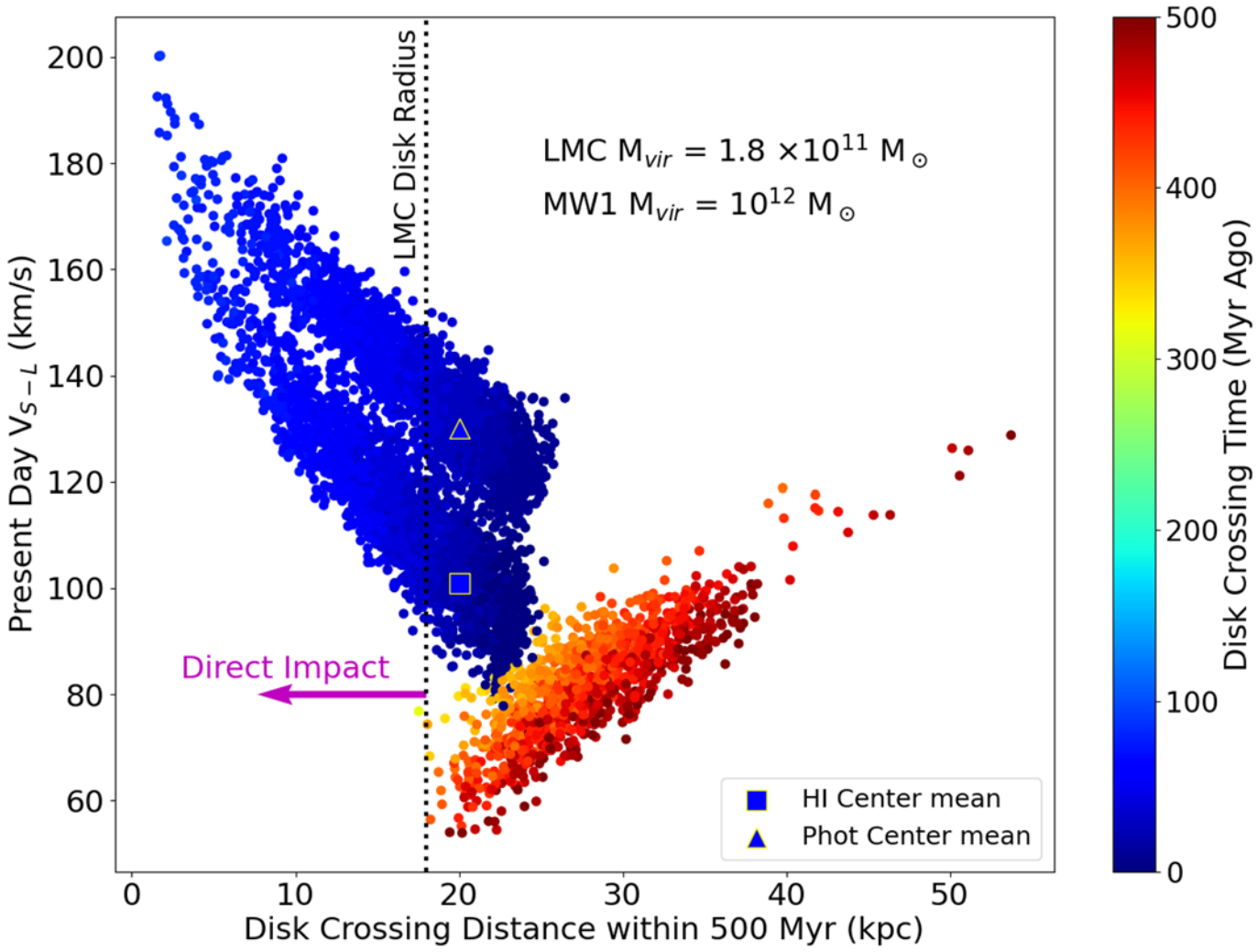

**Fig. 35** Distance from the LMC's disk center at which the SMC most recently crossed the LMC's disk plane (disk crossing), plotted against the present-day relative velocity of the SMC–LMC system ($V_{S-L}$). Orbits are computed using the HI+Phot Center data set in the Model C: Moving COM framework, with MW model MW1 and LMC mass model $1.8 \times 10^{11}$ $M_\odot$. Results are similar for all mass models. Color indicates the time of the disk crossing; there are two families of solutions, where disk crossings occur $< 100$ Myr ago (57%) or $> 350$ Myr ago (31%). Using either the mean HI Center (triangle) or Phot Center (square) velocities, the disk crossing distance is ∼20 kpc, occurring ∼20–30 Myr ago (see Fig. 32). The different centers have different mean $V_{S-L}$, resulting in the two vertically shifted tracks of blue points. Solutions with disk crossings within 18 kpc of the LMC center (<∼100 Myr ago; "Direct Impact" orbits) are found in 26% of the HI+Phot Center data set. Direct Impacts occurs if the present-day LMC–SMC relative velocity is $\gtrsim$100 km/s and the SMC center is currently located above the LMC disk plane (blue squares in Fig. 32). A Direct Impact within 10 kpc is of low probability (∼4%), but analytic calculations cannot capture the detailed evolution of live systems that undergo such close encounters. Note that the Phot Center solutions occupy a distinct space in $V_{S-L}$ than the HI center solutions

Orbit calculations assumed the MW NFW model MW1 and an LMC NFW model with an infall mass of $1.8 \times 10^{11}$ $M_\odot$. The following results are similar in all combinations of MW and LMC models explored in this study.

Solutions for the properties of the SMC's most recent disk crossing divide into two possibilities. First, if the present-day $V_{S-L}$ is large ($\gtrsim$ 100 km/s), disk crossings will occur recently (within ∼100 Myr ago), and at low separations $< 25$ kpc. Such solutions (blue points) occur for ∼54% of the HI+Phot Center data set. A Direct Impact scenario (disk crossing distance $< 18$ kpc) occurs within 26% of the HI+Phot Center data set; all Direct Impacts occur within 100 Myr.

Second, for slower $V_{S-L}$ ($\lesssim$ 100 km/s), the disk crossing occurs at distances larger than the radius of the LMC's stellar disk and at earlier epochs ($> 350$ Myr ago). These solutions correspond to 31% of the HI+Phot Center data set. There are no disk crossings within 5 Gyr for 12% of the HI+Phot Center data set.

To better understand the origin of the dual disk crossing solutions ($< 100$ or $> 350$ Myr ago), return to Fig. 32, where the location of the SMC's most recent LMC disk crossing is marked in a 3D rendering of the LMC's disk plane. Red "x" markers indicate disk crossings that occurred >100 Myr ago, while the blue "x" markers correspond to more recent crossings (<100 Myr ago). The latter are also "Direct Impact" solutions.

The origin of the dual disk crossing solutions is the assumed present-day location of the SMC's center of mass relative to the LMC's disk plane. If the SMC is currently above the LMC's disk plane, it will have had a recent "Direct Impact" with the LMC's disk. If the SMC is currently below the LMC's disk plane, the disk crossing will occur at much earlier times and larger distances.

**Summary:** All HI+Phot Center orbit solutions require a recent (115–210 Myr ago), close encounter (5–17 kpc) between the Clouds, regardless of whether the Clouds are a long-lived binary. However, not all close encounters are collisions. If the present-day $V_{S-L}$ is large ($\gtrsim$ 100 km/s), the SMC will cross the LMC disk plane $\lesssim$100 Myr ago at low separations (< 25 kpc). A Direct Impact scenario (disk crossing distance within the extent of the LMC disk; < 18 kpc) occurs within ~26% of the HI+Phot Center data set; all Direct Impacts occur within 100 Myr. Direct Impacts are thus allowed by the error space, using the Model C: Moving COM framework.

### 9.3 Theoretical models favor a recent direct impact between the clouds

The analytic calculations presented in the previous section cannot capture the detailed evolution of live systems that undergo close encounters. As such, the presented distribution of disk crossing properties (timings, distances) should be considered a general proof of concept that a recent, direct impact of the SMC with the LMC disk is plausible.

However, theoretical models can break the degeneracy in the disk crossing solutions. Numerical studies have identified several observed features of the Clouds that strongly favor a recent (< 200 Myr), close (disk crossing < 10 kpc) "Direct Impact" scenario.

1. *The properties of the LMC's bar and spiral arm.* The LMC has a one-armed spiral and an off-center bar (de Vaucouleurs and Freeman 1972). The bar is absent in all tracers of the ISM, has a potentially slow pattern speed (Jiménez-Arranz et al. 2024), and is tilted out of the disk plane (Choi et al. 2018 and references there in). Theoretical studies have shown that a Direct Impact between the LMC–SMC can simultaneously explain all these properties. For example, Berentzen et al. (2003) illustrated that a vertical collision between a barred spiral and its satellite can induce offset bars and one-armed spirals. Similarly, Bekki (2009) invoked a collision between the LMC and a dark satellite to explain the LMC bar's properties (see also Yozin and Bekki 2014b).
   B12 model 2 was the first to simulate a very close Direct Impact (impact parameter ~ 2 kpc, occurring 100 Myr ago) between the Clouds using hydrodynamic simulations of the LMC–SMC–MW system (Model C). As shown in Rathore et al. (2025a), the SMC torques during the impact result in a simulated LMC stellar bar with a changed bar pattern speed, off-set location from the outer disk isophotes, tilt out of the disk plane, and absence of a gaseous counterpart. In contrast, B12 model 1, which does not invoke a Direct Impact but does include the tidal field of the SMC and MW, cannot explain the LMC's bar properties. The LMC bar's properties requires a strong perturbation.

2. *The star formation and kinematics of the Bridge* The properties of the Bridge of gas and stars that connect the LMC and SMC are consistent with numerical models of a Direct Impact. The Bridge is star forming (Demers and Battinelli 1998; Harris 2007; Chen et al. 2014; Ficara et al. 2026), whereas the Stream is not (Matthews et al. 2009). This is naturally explainable if the gas density of the Bridge is higher than the Stream, owing to ram pressure stripping of the gaseous SMC by the LMC's gaseous disk during a Direct Impact. B12 model 2 results in a star forming, tidal+ram pressure formed Bridge and a non-star forming, tidally formed Stream. In B12 model 1, neither Bridge nor Stream is star forming, as the two structures are both purely tidal.
   These distinct bridge-forming scenarios result in different predictions for the dynamics of the stellar bridge. In B12 model 2, the stellar dynamics should point toward the LMC along the bridge, whereas in B12 model 1, the stars would also have motions perpendicular to the bridge. The B12 model 2 predictions were confirmed with GaiaDR2 (Zivick et al. 2018; Oey et al. 2018).

3. *The internal stellar dynamics of the LMC.* In Choi et al. (2022), the internal in-plane stellar dynamics of the LMC were analyzed. After subtracting a model for a rotating disk, these authors found proper motion residuals that were significantly larger than expected from the tidal field of the MW or the SMC in a non-Direct Impact scenario. Comparison with the B12 model 2 Direct Impact model resulted in much better agreement, suggesting that the LMC is disrupted to a level that requires significant perturbation, consistent with a Direct Impact, $< 10$ kpc, occurring $< 250$ Myr ago. Navarrete et al. (2023) similarly find that a Direct Impact ($\sim$4.5 kpc) is needed to explain the kinematics of the LMC's stellar periphery. Using data from the VMC survey, Vijayasree et al. (2025) also find that the LMC's internal kinematics are consistent with a recent ($\sim$ 150 Myr ago) encounter with the SMC.

4. *The internal gas and stellar dynamics of the SMC are not in equilibrium.* The SMC gas distribution displays a pronounced velocity gradient of order 60–100 km/s (Stanimirović et al. 2004; Di Teodoro et al. 2019). In contrast, the old stars are dispersion dominated with a rotation amplitude $\lesssim$ 10 km/s (Harris and Zaritsky 2006; Zivick et al. 2021) and kinematics consistent with expansion from the LMC's tidal field (Zivick et al. 2018, 2021; Niederhofer et al. 2021; Vijayasree et al. 2026). Rathore et al. (2026) use N-body hydrodynamic simulations to illustrate that the discrepancy between the gas and stellar kinematics naturally arises after a direct, hydrodynamic collision between the Clouds, with impact parameter $\sim$2 kpc. During a collision, the SMC stellar and gaseous components are both subject to the strong tidal field of the LMC causing pronounced radially outward motions. However, the gas is also subject to ram pressure stripping, which is sufficient to destroy the inner rotation and offset the gas and stellar kinematic centers, explaining the discrepancy.

5. *The SMC is presently a starbursting galaxy.* The current star formation rate of the SMC is more than three times higher than its average rate (Weisz et al. 2013; Massana et al. 2022). Given the high gas content of the SMC (1:1 mass ratio gas:stars) an elevated star formation rate is theoretically consistent with a recent collision (Hernquist 1989). Using simple ISM sub-grid models and star formation prescriptions that do not include stellar feedback, the hydrodynamic simulations of B12 model 2 find a strong burst of star formation is expected at the time of Direct Impact. More modern ISM treatments are needed, but the B12 model 2 results indicate that the SMC is a prime candidate for detailed numerical studies of star formation and stellar feedback induced by a galactic collision.

**Summary:** Theoretical arguments favor a recent ($< 200$ Myr), Direct Impact (disk crossing distance $< 10$ kpc) between the Clouds, even though such solutions comprise 5% of the observed HI+Phot Center data set (analytic orbits computed using the Model C: Moving COM framework). Studying a direct impact scenario with high-resolution N-body and hydrodynamic simulations, rather than analytic methods, is critical to interpret observations of the ISM, star formation, and geometry of the Clouds.

## 10 Are the LMC and SMC on first infall to the MW?

A "first infall" orbital solution is an orbit, where the LMC is currently just past its first approach *toward* the MW. This wording allows for the LMC/SMC to originate in the same overdensity that formed the MW and M31, putting it on a long period orbit (Alar Toomre, private communication ca. 2006).

In this review, the definition of "first infall" is further constrained by limiting the timescale over which the LMC has first approached the MW to be within the past 5 Gyr. In other words, the LMC has not had a second pericenter about the MW within the past 5 Gyr. This definition was chosen for two reasons. First, studies have illustrated that satellite orbits cannot be accurately reconstructed over long timescales without knowledge of the full cosmological assembly history of the MW, LMC, and Local Group (Lux et al. 2010; D'Souza and Bell 2022; Santistevan et al. 2024). Second, known massive perturbers (Gaia-Enceladus Structure, Sagittarius dSph, M31) have not been accounted for in this review, each of which can alter the mass growth history of the MW and the orbital solution of the LMC and SMC over timescales $> 5$ Gyr.

As illustrated by several studies, if the LMC's orbit is integrated further back in time despite these caveats, it is possible that another pericentric approach could occur earlier than 5 Gyr ago (Besla et al. 2007; Shattow and Loeb 2009; Kallivayalil et al. 2013) with a separation greater than 50 kpc (Patel et al. 2020). This can occur even in areas of the parameter space marked as "first infall" in Fig. 27, i.e., low MW mass and high LMC mass (Vasiliev 2024).

Figure 36 illustrates this idea. The orbit of the LMC and SMC about the MW is computed by integrating backward over the past 13 Gyr, using the Model C: Moving COM framework, assuming MW mass model MW1 and a heavy LMC ($2.5\sim 10^{11}$ $M_\odot$). The MW mass does not evolve over time; if it did, the Galactocentric orbits would become more eccentric backward in time (Kallivayalil et al. 2009). The orbits are computed using both the mean HI Center and mean Phot Center present-day velocities/positions for the Clouds.

Both sets of orbits illustrate the “first infall” scenario defined in this review; the Clouds are a binary that has only made one pericentric approach to the MW within the past 5 Gyr. However, in both cases, there is an earlier pericentric approach to the MW that occurred ∼9–10 Gyr ago. The timing of this earlier close approach is model dependent. Vasiliev (2024) find a previous close approach 6–8 Gyr ago, but their scenario is similar to that illustrated here. This also means that the second approach solution in Vasiliev (2024) would also be defined a “first infall” scenario in this review.

The orbital space marked as “first infall” in Fig. 27 is thus a statement that, within the past 5 Gyr, orbits derived using the majority of the HI+Phot Center error space do not allow for a second close approach between the LMC and the MW. Without a second close MW–LMC encounter within 5 Gyr, the internal structural and ISM properties of the LMC and SMC cannot be dominated by the MW’s tides. Instead, the tidal and hydrodynamic interaction history between the Clouds must dictate the present-day internal structure and kinematics of the LMC and SMC (Besla et al. 2010, 2012).

Given that a second pericenter >5 Gyr ago cannot explain the current structural/dynamical state of the Clouds, does it matter if the Clouds made an earlier approach, at timescale larger than 5 Gyr? Significantly, yes it does.

The Clouds are an ideal laboratory for precision tests of cosmological structure formation models. As discussed in this review, at infall, the LMC’s virial mass was likely ≥10–25% of the MW’s virial mass. The timing of the LMC’s infall thus impacts the inferred cosmological mass growth history of the MW. This growth history in turn impacts how MW-mass analogs are identified in cosmological simulations, a necessary step to place the MW in a cosmological context.

For example, if there is currently $\sim 10^{12}$ $M_\odot$ of dark matter within the virial radius of the MW (i.e., including the LMC), in a first infall scenario it would be more appropriate to identify MW-mass analogs in cosmological simulations as centrals with lower virial masses of 0.75–0.9$\times \sim 10^{12}$ $M_\odot$, i.e., subtracting a massive LMC $1 - 2.5 \times 10^{11}$ $M_\odot$ (see, e.g., Erkal et al. 2020; Kravtsov and Winney 2024). Consequently, a first infall or second pericenter scenario changes the severity of tensions between the observed properties of the MW and cosmological expectations. The expected number and mass function of current MW satellite galaxies and dark subhalos, the total baryonic mass of the MW’s CGM, and the dark matter density profiles of MW satellites, all scale with the mass of the host.

Given the uncertainties in analytic orbit integrations over long timescales, distinguishing whether the LMC is just past its first or second pericentric approach to the MW requires examining properties of the Clouds that are incompatible with a previous approach. The strongest arguments against a second pericentric approach scenario are as follows:

1. *Statistics of LMC mass satellites in MW-like hosts in $\Lambda$CDM cosmological simulations disfavor a second approach scenario.*

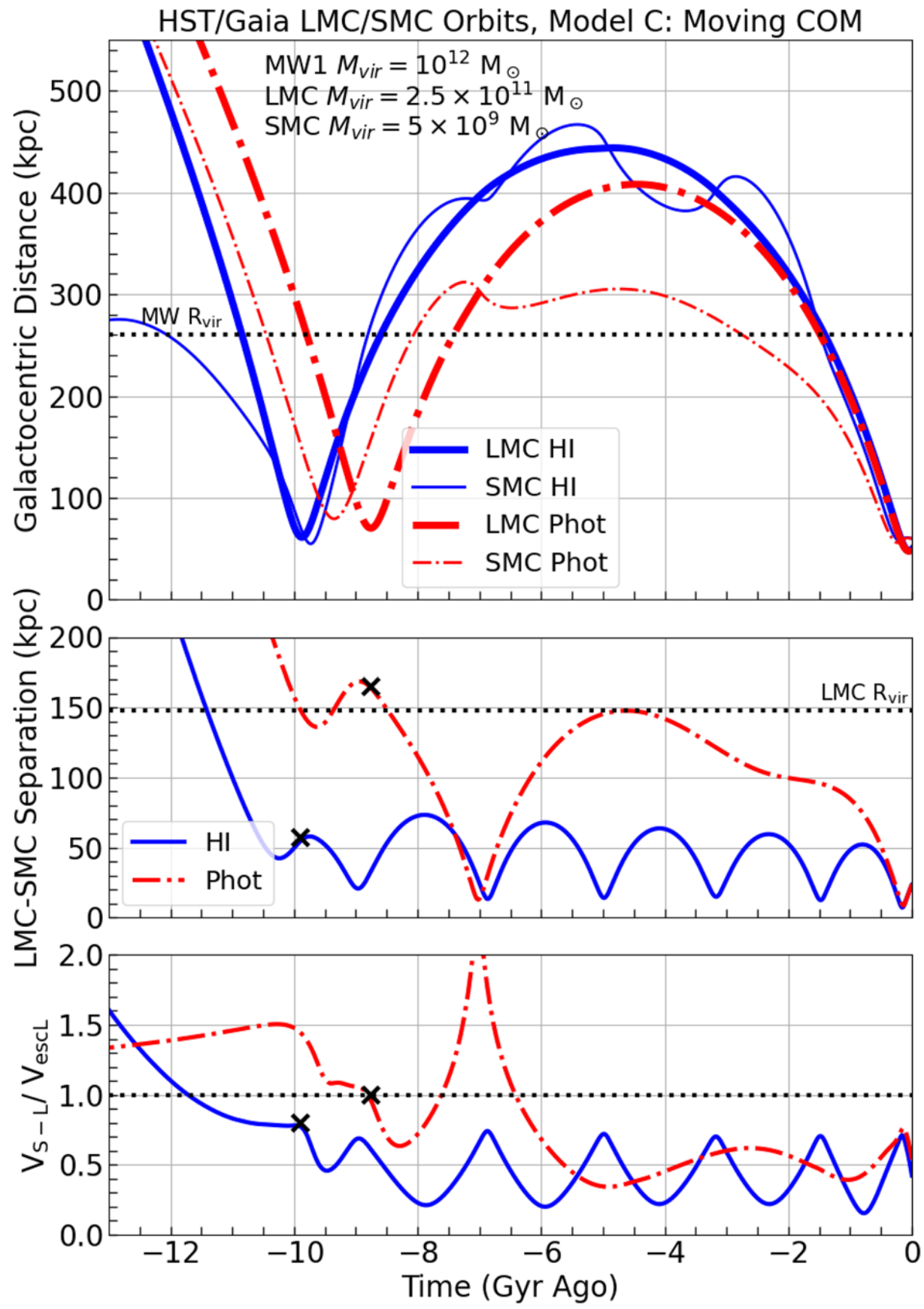

**Fig. 36** Galactocentric distance (top) and relative separation (bottom) of the LMC and SMC as a function time in the past. Orbits are constructed using the Model C: Moving COM framework for MW1 and the most massive LMC model ($2.5 \times 10^{11}$ M$_\odot$). This figure is the same as Fig. 20, except for the higher LMC mass and the orbits are computed over the past 13 Gyr, much longer than the typical 5 Gyr considered in this review. The MW mass is static and other external perturbers are ignored in this calculation (Sag DSph, M31, Gaia-Enceladus). This is an illustration to demonstrate an example "second passage" scenario. In both cases, the orbits are "first infall" as defined by this review; the LMC only has one pericentric approach to the MW within the past 5 Gyr. Yet, the LMC has an earlier infall ∼10–11 Gyr ago, and its first pericentric approach to the MW was ∼9–10 Gyr ago. For such a high mass LMC, the SMC is a long-lived binary in both the Phot and HI center orbits (with very different orbital periods). *However*, at the time of the LMC's first pericenter, the LMC–SMC binary is always destabilized (marked by an x in the middle and bottom panels). In both orbit solutions, a second LMC passage requires that the SMC was randomly captured by the LMC as it makes its first *pericentric* passage of the MW

Statistics from ΛCDM cosmological simulations of structure formation identify LMC-mass analogs about MW-type hosts at z∼ 0 in 10–35% of systems (Boylan-Kolchin et al. 2011; Busha et al. 2011; González et al. 2013; Patel et al. 2017). Meaning, MW–LMC mass analogs are not cosmologically rare; this is also confirmed observationally (Robotham et al. 2012). Statistically, such ΛCDM LMC analogs are typically accreted at late times (Boylan-Kolchin et al. 2011; Busha et al. 2011). In addition, surviving LMC analogs accreted at early times are typically on circular orbits today, which are strongly discrepant with the observed angular momentum of the LMC (Boylan-Kolchin et al. 2011; Patel et al. 2017). The high speed of the LMC and its present day close proximity to the MW favor a first infall scenario in $\Lambda$CDM.

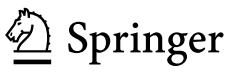

Future theoretical and observational studies of the dynamical response of the MW's stellar halo to the passage of the LMC may provide further proof that the Clouds did not complete an earlier pericentric approach to the MW. In particular, Sheng et al. (2024) find that the magnitude of the mean motion of the MW's inner halo in response to the LMC's passage should be weaker than observed if the LMC is on second infall, favoring instead a first infall scenario. At the same time, the orbital poles of many MW satellites are known to be aligned with that of the LMC over long timescales, which may favor a second LMC pericenter (Martínez-García and del Pino 2025). As such, whether the LMC is on first or second infall will impact efforts to reconstruct the assembly history of the MW using the kinematics of halo tracers (e.g., Helmi 2008; Johnston et al. 2008).

2. *The present-day gas content and star formation histories of the SMC and LMC are inconsistent with quenching.*

   The strongest observational argument against the second pericenter scenario is the present-day gas content and star formation histories of the Clouds. The Clouds are the only two gas-rich dwarf Irregulars at close Galactocentric distance to the MW or M31, leading Van Den Bergh (2006) to describe them as interlopers. The SMC has as much gas as the LMC and a stellar:gas mass ratio of $\sim$1:1. There are numerous environmental factors that work to quench star formation and deplete gas after galaxies become satellites of massive hosts. Indeed, the Clouds are undergoing active ram pressure stripping today, as the LMC's HI disk is asymmetrically truncated (Salem et al. 2015). Ram pressure from the MW's CGM must contribute to the formation of the Stream (Mastropietro et al. 2005; Lucchini 2024), and also modify the structure of the LMC's CGM (Krishnarao et al. 2022; Carr et al. 2025), where the LMC CGM is unlikely to survive a second pericentric approach to the MW (Lucchini et al. 2026).

   Upon accretion, satellites are cut off from their gas supply and thus their star formation rates should decline over time (Wetzel et al. 2015). However, the global star formation rates of the LMC and SMC have been steadily increasing over the past 5 Gyr, with a rapid increase in the past few hundred Myr (Weisz et al. 2013; Massana et al. 2022). In fact, the LMC is bluer in color than even field galaxies of comparable stellar mass (Tollerud et al. 2011) and the SMC is presently starbursting (Massana et al. 2022). It is improbable that a gas-rich, starburst galaxy like the SMC has been in orbit within the MW's CGM for the past 6–10 Gyr (Engler et al. 2023; Pathak et al. 2025).

   Invoking alternative gravity models, like MOdified Newtonian Dynamics (MOND), does not help the scenario. Owing to the high relative speeds of the Clouds, the Clouds are not a binary in MOND models. Instead the LMC is on first infall, while the SMC orbits about the MW for a Hubble time and only happens to approach the LMC at the LMC's pericentric passages about the MW (e.g., Zhao et al. 2013). The SMC should be significantly quenched in this scenario.
A second pericenter vs. single pericenter/first infall orbit is an ideal test case for numerical models of galaxy quenching and key for understanding the nature of quenched/star forming satellites of MW analogs, such as those found in the SAGA survey Mao et al. (2021, 2024)

3. *The tidal field of the MW will disrupt the LMC–SMC binary upon its first pericentric approach.*
The only way for the SMC to avoid the quenching influence of the MW's environment is if it were in orbit with the LMC as a long-lived binary (e.g., Fig. 36). Binarity would also naturally explain the posited synchronization in timing of bursts of star formation in both galaxies (Harris and Zaritsky 2009; Massana et al. 2022; Burhenne et al. 2026; Besla et al. 2012).
As outlined in Fig. 30, the preferred configuration for a long-lived LMC–SMC binary is a low mass MW + a high mass LMC. This is the *opposite* requirement for orbital solutions, where the LMC makes a previous passage about the MW within 5 Gyr. The high present-day relative velocity between the Clouds implies that the SMC is on an eccentric orbit about the LMC (see Fig. 31). Such binary configurations are easily disrupted by MW tides, meaning that even one previous pericentric passage is sufficient to have destroyed the binary (González et al. 2013). This is illustrated in Fig. 36, where at the previous close approach ~9–10 Gyr ago, the LMC–SMC binary is disrupted, even with the most massive LMC model and lightest MW model considered in this review. This means that a second approach scenario requires the LMC to have randomly captured the SMC while making its first pericentric approach to the MW.
If mass loss from the SMC were accounted for in this calculation the scenario would be even worse, as dynamical friction would make the SMC's orbit more eccentric in the past. *The LMC–SMC binary is unlikely to survive two pericentric passages about the MW*. This is the same intuition that led Avner (1965) to argue that the Clouds are on a hyperbolic orbit passing by the MW for the first time (see Sect. 4).

The idea that the Clouds are a binary whose first infall occurred recently is consistent with observations and ΛCDM theory. Only 3.5–0.4% of observed MW-like galaxies host both an LMC and SMC stellar mass analog (Liu et al. 2011; Tollerud et al. 2011; James and Ivory 2011; Robotham et al. 2012; Mao et al. 2021). Similarly, statistics from cosmological simulations find that only ~2.5% of MW type dark matter halos host both LMC and SMC mass analogs (Boylan-Kolchin et al. 2011; see also Busha et al. 2011; González et al. 2013; Patel et al. 2017; Haslbauer et al. 2024). The MW system is thus rare in that it hosts two massive satellites in close proximity to each other. However, Fig. 36 illustrates that this rare configuration can be understood if the Clouds have only recently passed their first pericentric approach to our MW. Only now are MW tides operating to disrupt this configuration. Congruently, LMC–SMC analogs become more common in isolated environments (more than 1.5 Mpc from the tidal field of a massive host, Besla et al. (2018)). Such analogs are also more common at earlier redshifts (Chamberlain et al. 2024). There is thus no tension between ΛCDM theory and the existence of an MW–LMC pair or an MW–LMC–SMC triple, if the Clouds are a long-lived binary on their first infall to the MW.

**Summary:** A scenario where the LMC is on its second approach about the MW, with a previous approach ∼9–10 Gyr ago, is allowed within the observational error space in the Model C: COM Moving Framework. There are several arguments against this scenario. 1) Statistics of LMC mass satellites with similar z=0 kinematics in MW-like hosts in CDM cosmological simulations disfavor a second approach scenario. 2) The present-day gas content and star formation histories of the SMC and LMC are inconsistent with significant quenching, requiring the Clouds to have spent limited time within the MW's CGM. 3) The tidal field of the MW will disrupt the LMC–SMC binary upon its first pericentric approach. For a second approach scenario, this means that the LMC is required to have randomly captured the SMC at its first pericentric approach. The LMC–SMC binary is unlikely to survive two pericentric passages about the MW. Accordingly, LMC–SMC–MW systems are extremely rare, both cosmologically and observationally.

## 11 Conclusion

This review describes the changing observational and theoretical landscapes that have dramatically altered our understanding of the orbital history of the LMC and SMC about the MW and about each other.

Observationally, space-based proper motion measurements (HST vs. Gaia) now show acceptable agreement and convergence, for consistent choices in dynamical center. In conversion to Galactocentric velocities, there is also consistency if solar velocity and distance values are held constant across studies. Regardless of center choice, the LMC $\mu_N$ component in the measurements is non-zero, causing the orbital trajectory of the LMC to deviate from the Stream when projected on the sky. The remaining debate centers around the choice of appropriate dynamical center for the LMC and SMC. The consequences are most extreme for the SMC and the binarity of the LMC/SMC as the choice of center can change the relative LMC–SMC velocity vector ($V_{\rm LS}$) by $\sim 25$ km/s. This difference ($V_{\rm LS} > 100$ km/s vs. $< 100$ km/s) is large enough to prevent the SMC from completing an orbit about the LMC.

Theoretically, our understanding of the distribution of dark matter in galaxies has changed significantly over time. Early studies assumed point mass MW models (Model A), which then changed to extremely massive Isothermal Sphere models for the MW (Model B). With improvements in understanding of cosmological structure formation in ΛCDM models, rigid MW halos with steep density profiles, like NFW profiles, were adopted (Model C). More recently, time-evolving MW halos with cosmologically motivated density profiles (Model D) have been used. The change to time-evolving models is a direct result of mounting evidence that the LMC's infall mass was of order $10^{11}$ $M_\odot$. This infall mass is more than a factor of 10 more massive than assumed by Model A and Model B studies and requires the SMC to be on a highly eccentric orbit to avoid merging with the LMC.

While the Model C: Moving COM framework can reasonably approximate the orbital history of the MW–LMC system, recovering the complex MW–LMC–SMC 3 body interaction history requires the Model D framework. In addition to N-body simulations, the high gas content of the LMC/SMC requires high resolution hydrodynamic simulations. Such simulations are particularly pressing given evidence that the LMC and SMC have directly collided within the recent 200 Myr. This recent collision will significantly impact the determination of the correct kinematic center of the LMC and SMC.

The orbital history of the Clouds profoundly impacts our understanding of the current dynamical state and mass assembly history of the MW system. Because the LMC is expected to deposit a significant amount of mass in the MW's halo (1:4–1:10 mass ratio), the LMC's infall time impacts the appropriate mass range within which cosmological MW analogs should be selected. If the LMC were accreted recently (first infall) the appropriate cosmological MW mass analog should be lower in mass, potentially easing tensions with known cosmological challenges (Missing Satellites, Too Big to Fail, Missing Baryons problems).

This review concludes from the existing theoretical frameworks and the latest observational data from Gaia and HST, that the Clouds are a long-lived binary pair and are just past their *first* close approach to the MW. This first infall scenario is most consistent with existing observational constraints, expectations from $\Lambda$CDM models, and theoretical studies of galaxy quenching. A previous approach of the Clouds to the MW would disrupt the LMC–SMC binary; the LMC would have to randomly capture the SMC while making its first close approach to the MW. All non-binary orbital solutions require the SMC to have orbited within the MW's CGM for more than 5 Gyr, which is incompatible with its high present-day gas fraction and recent elevated star formation history. The requirement of a long-lived ($>$5 Gyr) LMC–SMC binary requires a massive LMC ($M_{\rm vir} > 10^{11}$ M$_\odot$) and low mass MW ($M_{\rm vir} < 1.5 \times 10^{12}$ M$_\odot$).

### 11.1 Outlook to the future

High precision proper motions of the LMC and SMC utilizing HST and Gaia, coupled with new insights into dark matter halo models, now favor a picture in which the LMC and SMC are on recent infall. The configuration of a MW mass galaxy with two massive satellites infalling at present day is cosmologically rare, giving us unique opportunities to study galaxy interactions and the growth of structure in our Universe. The proper motion measurements further indicate that the LMC and SMC have had a recent strong encounter with each other. Coupled with the proximity of the LMC/SMC system, this evolutionary picture further affords a detailed view of how strong interactions between galaxies (the LMC and SMC) impact their stellar, gaseous and dark matter structures, as well as their chemical evolution.

At the time of this review, we have reached the ability to target proper motions of millions of stars in both galaxies with surveys, such as Gaia and VMC. However, per star errors at these distances are still large (except for intrinsically bright stars in Gaia which are not representative of the Clouds as a whole) and thus high precision is achieved by averaging large numbers of sources in bins or per region. This leads to varying precision as a function of source density or population, for example, proper motions for the faint and sparse outskirts of the Clouds are much less well-determined than the inner regions.

A deep, homogeneous survey of the Clouds using *Roman* and *Rubin* would enable the determination of more precise and higher spatial resolution proper motion coverage in the outskirts of the Clouds. A near-IR *Roman* survey would also contribute to our understanding of the more extincted inner regions, such as the bar. Higher precision for faint stellar populations throughout the Clouds would, in concert with dynamical models, allow us to better understand the impact of repeated interactions on their internal structures, and specifically the relationship between the stellar, HI and dark matter centers.

While this review has focused on proper motions, it is important to note the obvious importance of spectroscopic surveys, and the powerful complementary probes of interaction history provided by star formation histories, metallicity and age gradients. Building on the pioneering works from, e.g., the SMASH and Scylla surveys, a uniform, deep, multi-band survey of the Clouds in LSST and *Roman* filters would allow new insights into the star formation histories of the Clouds, especially the ancient SFH, which is presently not well-known.

The VISCACHA survey includes dedicated spectroscopic follow-up, and also cross-matches with the SMASH survey to obtain multi-band CMDs to investigate the age metallicity relation and chemical evolution history of SMC clusters. This reveals metallicity dips in the SMC clusters which could coincide with the formation epoch for the Stream and the Bridge (Oliveira et al. 2023), as well as a metallicity dip dated at an older epoch (6 Gyr ago) interpreted as the result of a major merger with a metal-poor gas cloud or dwarf galaxy (Saroon et al. 2025), and first put forward by Tsujimoto and Bekki (2009) based on a compilation of SMC clusters in the literature at that time.

Upcoming are the DELVE-MC survey, covering an area of $\sim$135 sq. degrees around the Clouds down to $g, r, i \sim 24.5$ mag (Massana et al. 2025); the metallicity-sensitive DECam Mapping the Ancient Galaxy in CaHK (MAGIC) survey, which will image $>$ 5000 sq. degrees in the southern hemisphere including the Clouds (Chiti et al. 2026); and the SDSS-V Magellanic Genesis Survey which will map the kinematic and chemical structure of the Clouds using APOGEE and BOSS spectroscopy (Nidever et al. 2026).

While not focused solely on the area of the Clouds, photometric and spectroscopic surveys aimed at bright tracers in the outer MW halo are uncovering evidence for the response of the MW to the infalling LMC (Garavito-Camargo et al. 2019). This is a rapidly developing field that underscores complexity in interpreting what may be the large scale response of the MW vs. more small-scale, not yet fully phase-mixed substructure within it (Byström et al. 2025). However, there is overall agreement of a clear effect of the LMC. Works focused on star counts and photometry (e.g., Cavieres et al. 2025; Li et al. 2026) are uncovering differences in the MW halo density profile in the southern hemisphere vs. the northern hemisphere. Spectroscopic works (e.g., Petersen and Peñarrubia 2021; Erkal et al. 2021; Chandra et al. 2025; Byström et al.

2025) show that the MW is moving at roughly 20 − 40 km/s toward a location along the LMC's past orbit, with general agreement in amplitude but less agreement in direction.

These works show the power of future analyses that will combine all of these complementary probes in conjunction with numerical simulations.

**Acknowledgements** We are grateful to Maria-Rosa Cioni, Hayden Foote, Nico Garavito-Camargo, Paul McMillan, Himansh Rathore, Ekta Patel, Andrew Pace and Eugene Vasiliev for their help with this review. We thank the anonymous referee for feedback that helped to improve clarity and presentation of the work. GB is supported by NSF CAREER AST-1941096 and NASA ATP award 80NSSC24K1225. The Theoretical Astrophysics Program (TAP) at the University of Arizona provided resources to support this work. NK is supported by a Dean's Research Fellowship, and acknowledges fruitful discussions within the Galaxy Evolution and Cosmology (GECO) Initiative at the University of Virginia. We respectfully acknowledge the University of Arizona is on the land and territories of Indigenous peoples. Today, Arizona is home to 22 federally recognized tribes, with Tucson being home to the O'odham and the Yaqui. Committed to diversity and inclusion, the University strives to build sustainable relationships with sovereign Native Nations and Indigenous communities through education offerings, partnerships, and community service. We respectfully acknowledge the University of Virginia is on the ancestral homeland and traditional territory of the Monacan Indian Nation. We pay respect to their elders and knowledge keepers—past and present. We acknowledge and pay respect to the enslaved Africans, enslaved laborers, and free Black laborers who built UVa, as well as their descendants. *Software:* `Astropy` (Collaboration et al. 2013; Price-Whelan et al. 2018, 2022); `Jupyter` (Kluyver et al. 2016; Granger and Pérez 2021); `Matplotlib` (Hunter 2007); `Numpy` (Harris et al. 2020); `Scipy` (Virtanen et al. 2020).

**Author Contributions** G.B. and N. K. wrote the manuscript text and prepared the figures (except where otherwise noted).

## Declarations

**Competing interests** The authors declare no competing interests.

## Appendix A Singular isothermal sphere potential

The Singular Isothermal Sphere is typically used to characterize the dark matter halo of the MW in the Model B framework. In this review, this profile is referred to as "Isothermal Sphere", for short. The density profile ($\rho_{\rm Iso}$) for the isothermal sphere is

$$\rho_{\rm Iso}(\mathbf{r}) = \frac{V_C^2}{4\pi G|\mathbf{r}|^2}, \tag{A1}$$

where $V_C$ is the circular speed at the distance of the Sun. Studies have adopted a range of values for $V_C$, which dictates the mass enclosed within a given radius (see Sect. 5). In this review, both the IAU standard of $V_C = 220$ km/s and the more recent value of $V_C = 239$ km/s (McMillan 2011) are considered.

The corresponding mass profile ($M_{\rm Iso}$) increases linearly with distance, resulting in very heavy mass models, often exceeding observational limits at large distances. The mass profile is computed as

$$M_{\rm Iso}(<|\mathbf{r}|) = \frac{V_C^2|\mathbf{r}|}{G} \tag{A2}$$

The resulting potential ($\Phi_{\rm Iso}$) and acceleration ($\mathbf{a}_{\rm Iso}$) felt by a satellite at a distance $\mathbf{r}_{MW}$ from the MW are

$$\Phi_{\rm Iso}(\mathbf{r}) = -V_C^2 ln(|\mathbf{r}|), \tag{A3}$$

$$\mathbf{a}_{\rm Iso}(\mathbf{r}) = \frac{\partial}{\partial \mathbf{r}}\phi_{\rm Iso}(\mathbf{r}) = \frac{-V_C^2\mathbf{r}}{|\mathbf{r}|}, \tag{A4}$$

In Sect. 6.3.3, Fig. 18, orbits in a prolate Isothermal halo are examined to illustrate that the motion of the MW center of mass can complicate efforts to infer the MW's halo shape from the orbital motion of objects in the MW halo.

The gravitational potential for an Isothermal Sphere with axial ratio dependence is

$$\Phi_{\rm qIso}(\mathbf{r}) = -\frac{V_C^2}{2} ln(|R + \frac{z^2}{q^2}|), \tag{A5}$$

where $R = \sqrt{x^2 + y^2}$ and $q < 1$ denotes an oblate halo, $q > 1$ denotes a prolate halo, and $q = 1$ is spherical. The corresponding acceleration is

$$\mathbf{a}_{\rm qIso}(\mathbf{r}) = \frac{\partial}{\partial \mathbf{r}}\phi_{\rm qIso}(\mathbf{r}) = \frac{-V_C^2\mathbf{r}}{(R^2 + z^2)}[1, 1, q^{-2}], \tag{A6}$$

## Appendix B Plummer sphere potential

In Model B studies, and many Model C studies, the dark matter distribution of the LMC and SMC are modeled using Plummer (Plummer 1911) potentials.

The Plummer density ($\rho_{\rm Plummer}$), mass ($M_{\rm Plummer}$), potential ($\Phi_{\rm Plummer}$) and acceleration ($\mathbf{a}_{\rm Plummer}$) profiles are defined at a distance $r$ as

$$\rho_{\rm Plummer}(\mathbf{r}) = \frac{3M_{\rm sat}k^2}{4\pi(|\mathbf{r}|^2 + k^2)^{2.5}} \tag{B7}$$

$$M_{\rm Plummer}(<|\mathbf{r}|) = \frac{M_{\rm sat}|\mathbf{r}|^3}{(|\mathbf{r}|^2 + k^2)^{1.5}} \tag{B8}$$

$$\Phi_{\rm Plummer}(\mathbf{r}) = -\frac{GM_{\rm sat}}{\sqrt{|\mathbf{r}|^2 + k^2}}, \tag{B9}$$

$$\mathbf{a}_{\rm Plummer}(\mathbf{r}) = \frac{\partial}{\partial \mathbf{r}}\phi_{\rm Plummer}(\mathbf{r}) = -\frac{GM_{\rm sat}\mathbf{r}}{(|\mathbf{r}|^2 + k^2)^{1.5}} \tag{B10}$$

where $k$ is the scale length of the potential. Most Model C studies follow the assumptions in Murai and Fujimoto (1980), where the LMC and SMC are low mass galaxies (halo masses $\sim 10^{10}$ M$_\odot$) modeled as Plummer potentials with scale lengths of $k_\mathrm{L}$ =3 kpc and $k_\mathrm{S}$=2 kpc, respectively. Although higher LMC masses are explored in Model C, the LMC and SMC are often still modeled as Plummer potentials with larger scale lengths chosen to match dynamical constraints on the mass profile (Patel et al. 2020, see Table 7).

## Appendix C Hernquist potential

The MW's bulge is typically modeled using a Hernquist potential (Hernquist 1990, H90). In several studies the LMC dark matter halo is also modeled using a H90 potential (e.g. Besla et al. 2012; Garavito-Camargo et al. 2019, 2021), as the mass profile converges and does not need to be artificially truncated.

The Hernquist density ($\rho_\mathrm{H90}$), mass ($M_\mathrm{H90}$), potential ($\Phi_\mathrm{H90}$), and acceleration ($\mathbf{a}_\mathrm{H90}$) profiles are defined at a distance $r$ as

$$\rho_\mathrm{H90}(\mathbf{r}) = \frac{M\, r_a}{2\pi\, |\mathbf{r}|(|\mathbf{r}| + r_a)^3} \tag{C11}$$

$$M_\mathrm{H90}(< |\mathbf{r}|) = \frac{M|\mathbf{r}|^2}{(|\mathbf{r}| + r_a)^2} \tag{C12}$$

$$\Phi_{H90}(\mathbf{r}) = -\frac{GM}{(|\mathbf{r}| + r_a)^2}, \tag{C13}$$

$$\mathbf{a}_{H90}(\mathbf{r}) = \frac{\partial}{\partial \mathbf{r}}\phi_\mathrm{H90}(\mathbf{r}) = -\frac{GM\,\mathbf{r}}{|\mathbf{r}|(|\mathbf{r}| + r_a)^2} \tag{C14}$$

where $r_a$ is the scale radius of the halo or bulge, and $M$ is the total mass of the halo or bulge. Note, $M$ is not the same as the virial mass.

## Appendix D Miyamoto-Nagai potential

Exponential disks do not have analytic prescriptions for their acceleration, making them difficult to utilize in backward integration schemes. Instead, studies have approximated the MW's exponential stellar disk using the Miyamoto-Nagai potential (Miyamoto and Nagai 1975).

The Miyamoto-Nagai density ($\rho_\mathrm{MN}$), potential ($\Phi_\mathrm{MN}$), and acceleration ($\mathbf{a}_\mathrm{MN}$) profiles are defined at a Galactocentric distance $\mathbf{r} = [\mathbf{x}, \mathbf{y}, \mathbf{z}]$ as

$$\rho_\mathrm{MN}(\mathbf{r}) = \frac{M_\mathrm{disk}\, z_d^2}{4\pi}\, \frac{r_d\, R^2 + (r_d + 3A)(r_d + A)^2}{A^3(R^2 + (r_d + A)^2)^{5/2}} \tag{D15}$$

where $z_d$ is the disk scale height, $r_d$ is the disk scale length, $R = (x^2 + y^2)^{0.5}$, and $A = (z^2 + z_d^2)^{0.5}$. Values for the MW's disk parameters used in this review the MW are given in Table 8. The disk mass profile is obtained by numerically integrating Eq. (D15):

$$\Phi_{\rm MN}(\mathbf{r}) = -\frac{GM_{\rm disk}}{R^2 + (A + r_d)^2}, \tag{D16}$$

$$\mathbf{a}_{\rm MN}(\mathbf{r}) = -\frac{GM_{\rm disk}}{(R^2 + B^2)^{3/2}} \left[\mathbf{x}, \mathbf{y}, \frac{B\,\mathbf{z}}{B - r_d}\right] \tag{D17}$$

where $B = r_d + A$.

## Appendix E NFW potential

The MW's dark matter halo is often characterized using NFW (Navarro et al. 1996) profiles in Model C and D studies. Many studies also use the NFW profile to model the dark matter distribution of the LMC and SMC.

The NFW density ($\rho_{\rm NFW}$), mass ($M_{\rm NFW}$), potential ($\Phi_{\rm NFW}$), and acceleration ($\mathbf{a}_{\rm NFW}$) profiles are defined at a distance $r$ as

$$\rho_{\rm NFW}(\mathbf{r}) = \frac{M_{\rm vir}}{4\pi f(C_{\rm vir}) r_s^3} \frac{1}{a(1+a)^2} \tag{E18}$$

where $a = |\mathbf{r}|/r_s$ and $r_s = R_{\rm vir}/C_{\rm vir}$ is the scale radius.

$C_{\rm vir}$ is the virial concentration parameter. $C_{\rm vir}$ is typically a free parameter in Model C and D that is chosen, such that the modeled MW's rotation curve reasonably approximates observations. Assumed values of $C_{\rm vir}$ for each MW model adopted in this review are listed in Table 8.

$f(C_{\rm vir})$ is a function defined as

$$f(x) = \ln(1 + x) - \frac{x}{1 + x}. \tag{E19}$$

$M_{\rm vir}$ is the virial mass of the halo, defined as the mass enclosed within $R_{\rm vir}$. Following the Appendix of Van Der Marel et al. (2012b), the virial radius, $R_{\rm vir}$, is defined as the radius, where the enclosed average dark matter density of the halo is

$$\rho(R_{\rm vir}) = \frac{\Delta_c}{\Omega_m} \Omega_m \rho_{\rm crit} = \Delta_{\rm vir} \rho_{\rm avg}, \tag{E20}$$

where $\rho_{\rm crit}$ is the critical density of the universe and $\rho_{\rm avg} = \Omega_m \rho_{\rm crit}$ is the average dark matter density of the universe. The overdensity $\Delta_c$ is defined by Bryan and Norman (1998) as

$$\Delta_c = 18\pi^2 + 82x - 39x^2, \tag{E21}$$

where $x = \Omega_m(z) - 1$. Assuming $z = 0$ and cosmological parameters $h = 0.7$, $\Omega_m(0) = 0.27$, and $z = 0$, then $\Delta_c = 97$ and the virial overdensity, $\Delta_{\rm vir} = \Delta_c/\Omega_m = 359$.

Note that $M_{\rm vir}$ is not the same as $M200$, which is defined as the mass enclosed within $R200$, which is the radius, where the density is $200\rho_{\rm crit}$. In general, $200 < R_{\rm vir}$, where the virial radius $R_{\rm vir}$ is

$$R_{\rm vir} = 206\, h^{-1}{\rm kpc} \left(\frac{\Delta_c}{97.2}\right)^{1/3} \left(\frac{M_{\rm vir}}{10^{12} h^{-1} M_{\odot}}\right)^{1/3} \tag{E22}$$

where in this review, $h = 0.7$. The resulting mass, potential, and acceleration for the NFW profile are defined as

$$M_{\rm NFW}(< |\mathbf{r}|) = M_{\rm vir}\frac{f(a)}{f(C_{\rm vir})}, \tag{E23}$$

$$\Phi_{\rm NFW}(\mathbf{r}) = -\frac{GM_{\rm vir}}{f(C_{\rm vir})}\frac{ln(1+a)}{|\mathbf{r}|}, \tag{E24}$$

$$\mathbf{a}_{\rm NFW}(\mathbf{r}) = \frac{\partial}{\partial \mathbf{r}}\phi_{\rm NFW}(\mathbf{r}) = -GM_{\rm vir}\frac{f(a)\,\mathbf{r}}{f(C_{\rm vir})|\mathbf{r}|^3} \tag{E25}$$

The peak of the halo rotation curve occurs at $r_{\rm max} = 2.163\, r_s$:

$$V_{\rm max} = 0.465 V_{\rm vir}\frac{C_{\rm vir}}{f(C_{\rm vir})}, \tag{E26}$$

where $V_{\rm vir} = GM_{\rm vir}/R_{\rm vir}$ is the circular speed at $R_{\rm vir}$.

## Appendix F Dynamical friction

### F.1 Satellite orbiting within an isothermal sphere potential

The Model B class of studies typically invoke dynamical friction acting on a satellite in orbit about the MW, where the MW is modeled as an Isothermal Sphere. Following Binney and Tremaine (1987), dynamical friction acting on a satellite is computed as

$$\mathbf{a}_{\rm DF,Iso,MW}(\mathbf{r}, \mathbf{v}) = -0.428\, {\rm GM}_{\rm sat}\Lambda_C\frac{\mathbf{v}}{|\mathbf{r}|^2|\mathbf{v}|}, \tag{F27}$$

where $\mathrm{M_{sat}}$ is the mass of the satellite, $|\mathbf{r}|$ is the magnitude of the relative distance between the satellite and the MW, and $\mathbf{v}$ is the relative velocity vector of the satellite with respect to the MW. $\Lambda_C$ is the Coulomb logarithm:

$$\Lambda_C = \ln(b_{\rm max}/b_{\rm min}), \tag{F28}$$

where $b_{\rm max}$ is the separation between the satellite and the host at any given point in time, and $b_{\rm min}$ is the impact parameter. Most Model B studies that account for dynamical friction adopt a constant value of $\Lambda_C = 3$ (Gardiner and Noguchi 1996; Bekki and

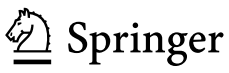

Chiba 2005). An exception is Murai and Fujimoto (1980), where they assume that $b_{\rm min}$ is the scale length of the satellite halo.

### F.2 Satellite orbiting within an NFW potential

To describe dynamical friction for a host and satellite system modeled with NFW halos, the generalized Chandrasekhar formula (Chandrasekhar 1943) is used:

$$\mathbf{a}_{\rm DF,NFW,MW}(\mathbf{r},\mathbf{v}) = -\frac{4\pi G^2 M_{\rm sat} \ln(\Lambda)\rho_{\rm NFW}(r)}{|\mathbf{v}|^2}\left[{\rm erf}(X) - \frac{2X}{\sqrt{\pi}}\exp(-X^2)\right]\frac{\mathbf{v}}{|\mathbf{v}|}, \tag{F29}$$

where $\rho_{\rm NFW}(r)$ (Eq. E18 is the density of the host NFW halo at the Galactocentric distance ($|\mathbf{r}|$) of a satellite with virial mass $M_{\rm sat}$. $\mathbf{v}$ is the orbital velocity vector of the satellite, and $X = |\mathbf{v}|/(\sqrt{2}\sigma_{\rm NFW})$. For the NFW halo, the one-dimensional halo velocity dispersion, $\sigma_{\rm NFW}$, is computed using the approximation provided in Eq. 6 of Zentner and Bullock (2003):

$$\sigma_{\rm NFW}(x) \approx V_{\rm max}\frac{1.4393\,x^{0.354}}{1 + 1.1756\,x^{0.725}} \tag{F30}$$

where $V_{\rm max}$ is the peak of the NFW halo rotation curve, and x is $|\mathbf{r}|/r_s$, where $r_s = R_{\rm vir}/C_{\rm vir}$ is the NFW scale radius for the MW. See relevant equations in Appendix E.

As discussed in Sect. F.1, the Coulomb logarithm, $\ln(\Lambda)$, has been approximated in several different ways by different authors. In order to describe a general set of LMC orbit solutions for Model C (cosmological halos), the Coulomb logarithm is chosen to vary as a function of the satellite-host distance. The analysis of van der Marel et al. (2012a) (their Appendix A) is followed, where these authors calibrated the analytic dynamical friction equations to an N-body simulations of the M33:M31 (1:10 mass ratio) encounter, which is a system with a similar mass ratio to the LMC:MW system. The calibration yields the following Coulomb logarithm:

$$\ln(\Lambda_{\rm L}) = \max\left[10^{-5}, \log\left(\frac{|\mathbf{r}|}{a\,1.22}\right)\right], \tag{F31}$$

where $a$ is the relevant size scale for the satellite, which in this review will be the NFW scale radius, $r_s$, for the LMC (in orbit about the MW) or the SMC (in orbit about the LMC). This formalism provides a reasonable match to N-body simulations of the SMC's orbit about the LMC, as seen Fig. 25.

To describe the dynamical friction experienced by the SMC within a MW NFW halo, the same formalism is used but with a different Coulomb Logarithm. The mass ratio between the SMC:MW is much smaller than 1:10. In this case the Coulomb Logarithm is defined as in Patel et al. (2020) and Hashimoto et al. (2003):

$$\ln(\Lambda_{\rm S}) = \ln\left(\frac{|\mathbf{r}|}{1.4a}\right) \tag{F32}$$

where $a$ is the NFW scale radius for the SMC.

In both cases, dynamical friction is only assumed to act on the satellite, while they orbit inside the virial radius of the host.

This prescription for dynamical friction most closely matches the LMC orbital studies of Gómez et al. (2015); Patel et al. (2017, 2020), except that the satellite is modeled as an NFW halo rather than a Plummer or H90 profile. This prescription is used in this review for orbit calculations in the Model C framework to compute dynamical friction acting on the Clouds by the MW, while each satellite is in orbit within the virial radius of the MW.

### F.3 SMC orbiting the LMC

Starting with Bekki and Chiba (2005), studies include the impact of dynamical friction on the SMC as it orbits within the LMC's dark matter distribution.

Typically, dynamical friction is included, while the SMC is within some limiting radius of the LMC. Studies like Bekki and Chiba (2005) take this limiting radius to be the LMC's tidal radius owing to the MW, which for their LMC halo masses is 13 kpc. Besla et al. (2007) adopt a value of 15 kpc for the LMC tidal radius.

The deceleration experienced by the SMC owing to dynamical friction from the LMC halo is computed by these studies as

$$\mathbf{a}_{\mathrm{DF,Iso,L}}(\mathbf{r},\mathbf{v}) = -0.428\,\mathrm{GM_S}\,\Lambda_{\mathrm{L}}\frac{\mathbf{v}_{\mathrm{S-L}}}{|\mathbf{r}_{\mathrm{S-L}}|^{2}|\mathbf{v}_{\mathrm{S-L}}|}, \tag{F33}$$

where $\mathbf{v}_{\mathrm{S-L}}$ and $\mathbf{r}_{\mathrm{S-L}}$ are the relative velocity and position vectors of the SMC with respect to the LMC. The Coulomb logarithm is $\Lambda_{\mathrm{L}} = 0.2$ (Bekki and Chiba 2005) and the mass of the SMC is denoted $\mathrm{M_S}$. The above model for dynamical friction is used in this review for orbits in the Model B framework.

For the Model C framework (Sect. 6), the same dynamical friction prescription is used as above, but with $\Lambda_{\mathrm{L}} = 0.3$, following Patel et al. (2020). Furthermore, dynamical friction acts only when the SMC is within the virial radius of the LMC (which differs from the Patel et al. 2020). The resulting orbits with this dynamical friction formulation are illustrated in Sect. 6.4.2.

In the Model D framework (Sect. 7.2), the SMC is usually initialized with a more massive halo than in other frameworks ($\sim 2\times10^{10}\ \mathrm{M_\odot}$). This means that the SMC:LMC mass ratio is ∼1:10, like the LMC:MW. As such, the dynamical friction formalism described in the previous section (Eq. F29), where the Coulomb Logarithm is calibrated to N-body simulations of a 1:10 merger by van der Marel et al. (2012a), can be used to reconstruct the SMC orbit in the Model C: Moving COM framework. In an analytic, *backward* integration scheme, where mass loss is not captured, an N-body orbit of the SMC can be reconstructed using the bound mass of the SMC at the starting time of the integration. The resulting orbits are reasonable approximations for the N-body results for at least one orbit (see Fig. 25).